\documentclass[12pt]{report}

\newcommand{\RomanNumeralCaps}[1]
    {\MakeUppercase{\romannumeral #1}}
\usepackage{amsmath, graphicx, amssymb, tikz, pgfplots, color, float,bm,caption, mathtools,lscape,ragged2e,booktabs,indentfirst}
\usepackage{pgfgantt}

\usepackage[margin = 25mm]{geometry}

\usepackage{fancyhdr}
\usepackage{colortbl}
\usepackage{listings}
\usepackage{float}
\usepackage[toc,page]{appendix}

\usepackage[onehalfspacing]{setspace}
\usepackage[style=british]{csquotes}
\usepackage{multirow,multicol}
\usepackage{xcolor,colortbl}
\usepackage{pdfpages}
\usepackage[natbibapa]{apacite}
\usepackage[font=scriptsize]{caption}
\usepackage{algorithm}
\usepackage[noend]{algpseudocode}

\pgfplotsset{compat=1.15}
\author{Zhengkun LI}   
\begin{document}

\begin{titlepage}
\newcommand{\HRule}{\rule{\linewidth}{0.3mm}} 
\center 
\includegraphics[width=\linewidth]{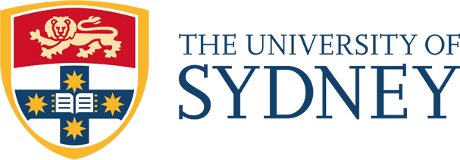}\\[1cm] 
\textsc{\LARGE University of Sydney}\\[0.5cm] 
\textsc{\Large Business School}\\[0.5cm] 
\textsc{\large Discipline of Business Analytics}\\[0.5cm] 
\HRule \\[0.5cm]
{ \Large \bfseries A Joint Value at Risk and Expected Shortfall Combination Framework and its Applications in the Cryptocurrency Market}\\[0.3cm] 
\HRule \\[0.5cm]

A dissertation submitted in partial fulfilment\\
of the requirements for the award of Honours\\
{\vspace{2cm}}
\begin{minipage}{0.4\textwidth}
\begin{flushleft} \large
\emph{Author:}\\
Zhengkun \textsc{Li}\\
450018729
\end{flushleft}
\end{minipage}
~
\begin{minipage}{0.5\textwidth}
\begin{flushright} \large
\emph{Supervisors:} \\
Prof. Richard \textsc{Gerlach}\\
Dr. Laurent \textsc{Pauwels}
\end{flushright}
\end{minipage}\\[2cm]
{\large \today}\\[3cm] 
\vfill 
\end{titlepage}
\begin{center}
\section*{Certificate of Originality}    
\end{center}

I hereby declare that I have read and understood the Academic Honesty in Coursework Policy 2015. \\

I understand that failure to comply with the above can lead to the University commencing proceedings against me for potential student misconduct under Chapter 8 of the University of Sydney By-Law 1999 (as amended). 
I declare that this submission is my own work and to the best of my knowledge it contains no materials previously published or written by another person, nor material which to a substantial extent has been accepted for the award of any other degree or diploma at University of Sydney or at any other educational institution, except where due acknowledgement is made in the thesis. \\

Any contribution made to the research by others, with whom I have worked at University of Sydney or elsewhere, is explicitly acknowledged in the thesis.
I also declare that the intellectual content of this thesis is the product of my own work, except to the extent that assistance from others in the project’s design and conception or in style, presentation and linguistic expression is acknowledged.\\

\begin{center}
\line(1,0){120}\\
Zhengkun LI
\end{center}

\newpage
\begin{center}
\section*{Acknowledgement}
\end{center}
First and foremost, I would like to express my gratitude to my supervisors, Professor Richard Gerlach and Dr. Laurent Pauwels. Richard, you have introduced me to the academic world of quantitative financial risk management and time series modelling with your expert knowledge, patient guidance and support. Laurent, the suggestions, guidance and time series knowledge from you are invaluable. I greatly appreciate my supervisory team, as without their unselfish attitude and profound knowledge, this thesis and other works would not have been possible in this challenging honour year.

I wish to thank the BA honour coordinator Dr. Dmytro Matsypura and the lecturer Professor Junbin Gao. 
Dmytro, thanks for your continuous support and kind suggestions that have guided us through the honour year. Junbin, I cannot understand most journal articles without the statistical knowledge from your lectures, and the knowledge you shared has introduced me to the statistical world. Additionally, I would like to thank the 2018 business school honour coordinator Associate Professor Diane van den Broek: the knowledge and skills from your lectures taught me how to conduct research as a new researcher, and thank you for your support during the whole year. 

I also wish to acknowledge the supports of all the staff in the discipline of Business Analytics, especially the comments, suggestions and insights of Associate Professor Artem Prokhorov, Associate Professor Daniel Oron and Dr. Chao Wang. Your advice related to the fields of cryptocurrency and financial time series modelling have guided me to improve my research quality. Additionally, thanks American Journal Experts (AJE) for English language editing. 

Thanks also to the research lab crew with whom I have shared the majority of my honour time. Firstly, to the 2018 BA cohort - Haonan, Jiarui, Kaiming, Siyu and Zhenhao - it has been my pleasure to spend my honour year with this incredible group of people. To the PhD candidates - Jessica Leung and Henry Cheung - the enjoyable and humorous moments shared with you are unforgettable. 

Finally, I would like to convey my gratefulness to my parents and girlfriend Chen TAN for continuously supporting me in my studies and daily life, which always motivates me to keep going. I owe all my sanity to you. Thank you.

\newpage
\begin{center}
\section*{Abstract}    
\end{center}

Value at risk and expected shortfall are increasingly popular tail risk measures in the financial risk management field. Both academia and financial institutions are working to improve tail risk forecasts in order to meet the requirements of the Basel Capital Accord; it states that one purpose of risk management and measuring risk accuracy is, since extreme movements cannot always be avoided, financial institutions can prepare for these extreme returns by capital allocation, and putting aside the appropriate amount of capital so as to avoid default in times of extreme price or index movements.  Forecast combination has drawn much attention, as a combined forecast can outperform the individual forecasts under certain conditions. We propose two methodology, one is a semiparametric combination framework that can jointly produce combined value at risk and expected shortfall forecasts, another one is a parametric regression framework named as Quantile-ES regression that can produce combined expected shortfall forecasts. The favourability of the semiparametric combination framework has been presented via an empirical study - application in cryptocurrency markets with high-frequency data where the necessity of risk management application increases as the cryptocurrency market becomes more popular and mature. Additionally, the general framework of the parametric Quantile-ES regression has been presented via a simulation study, whereas it still need to be improved in the future.  The contributions of this work include but are not limited to the enabling of the combination of expected shortfall forecasts and the application of risk management procedures in the cryptocurrency market with high-frequency data. 
\tableofcontents
\newpage
\listoffigures
\newpage
\listoftables
\newpage

\pagenumbering{arabic}
\setcounter{page}{1}

\chapter{Introduction}
Effective risk management is of paramount importance. 
Therefore, an accurate risk measure is needed, as any unexpected movements of the financial market can have a large impact on both investors and regulators. The advantage of measuring risks in an acceptable way is that we can change our behaviour to reduce or avoid risks. Value at risk (VaR) and expected shortfall (ES) are two sensitive risk measures adopted and recommended by the Basel Capital Accord for financial institutions to estimate the minimum capital requirements for possible losses to protect their investments from extreme market events and movements.

VaR, introduced by J.P. Morgan in their RiskMetrics publication \citep{morgan1996riskmetrics}, is formally defined as a threshold value such that, within a given probability, the potential loss on an underlying asset or portfolio over a given period will not exceed the threshold value. 
However, VaR is criticised as it gives no information regarding any possible loss beyond the given threshold value. Also, VaR is not a subadditive measure; that is, the overall VaR of a well-diversified portfolio can be larger than the sum of the individual VaRs, which is not consistent with finance literature. Those shortcomings led to the proposal of ES \citep{artzner1997thinking,artzner1999coherent}, which is defined as the expected loss over a given period conditional on the loss being greater than the VaR threshold. Thus, ES is also called conditional VaR (CVaR) and is a coherent risk measure that can overcome several shortcomings of VaR, such as  non-subadditivity. Financial institutions can allocate their capital efficiently for extreme market risks by jointly using VaR and ES as suggested by the Basel \RomanNumeralCaps3 Capital Accord and \cite{yamai2005value}.
As a result, the need for highly accurate VaR and ES forecasts is increasing dynamically.

To generate accurate VaR and ES forecasts, many models are proposed to estimate tail risk. Those models can be mainly categorised into three groups: parametric, nonparametric and semiparametric. Parametric models such as GARCH family models are highly dependent on the distribution assumptions of the underlying residual series; that is, an adaptable distribution assumption can produce a large improvement in the corresponding forecast performances. On the other hand, nonparametric (including semiparametric) models may be subject to the bias-variance tradeoff, and dependent on whether the corresponding measurement equation can be the true equation of the underlying time series. 

The application of forecast combination has drawn much attention, as it can be benefited by combining different individual forecasts and leveraging the advantages from each forecast. The methodology of forecast combination in point forecast has been well developed, especially in the forecasting of economic indicators such as interest rates and inflation rates. For tail risk measures, the methodology used to combine different VaR quantile forecasts is well developed by adeptly using a series of quantile regression frameworks such as quantile regression \citep{halbleib2012improving} or quantile LASSO regression \citep{li20081}. However, there is a sparse amount (even no) of literature focused on the ES combination puzzle; that is, how to combine several ES forecasts is still questionable.

Recently, the emergence of cryptocurrency markets, especially the Bitcoin market, has drawn much attention from academics, investors and regulators. However, following the large impact of cryptocurrency markets on the financial world, the debates on the maturity of cryptocurrency markets at different levels are never-ending. Nevertheless, several studies, such as \cite{chu2017garch} and \cite{dyhrberg2018investible}, have focused on the investment and forecast properties of cryptocurrency.  
Additionally, there is a research gap concerning tail risk measures in cryptocurrency markets, as the quantile risk measure may require a larger sample size than others and cryptocurrency markets began to emerge in the middle of 2017.

Regarding those research gaps concerning how to combine ES forecasts efficiently to generate accurate ES forecast series, this thesis proposes a new semiparametric joint VaR and ES combination method that allows the combination of a series of VaR forecasts and ES forecasts jointly. The empirical study shows the favourability of the proposed method as well as the application of several tail risk measure estimation methods in five selected cryptocurrencies according to their market capitalisation. Additionally, this thesis also extends the existing quantile regression frameworks to a new parametric regression frameworks (named as Quantile-ES Regression) which can combine different ES forecasts by treating them as VaR forecasts. Consequently, the two proposed methods may start a new discussion on ES forecast combination and can be extended in several aspects to further improve the forecast accuracy.

The remainder of this thesis is organized as follows. Chapter \ref{literature} reviews the literature related to the estimation and evaluation of VaR and ES forecasts as well as the forecast combination puzzle. Chapter \ref{methodology} illustrates the methodology used in this thesis. Chapter \ref{empirical} shows the application of several existing VaR and ES estimation methods as well as the proposed method. Chapter \ref{Simulation} proposes the parametric Quantile-ES regression with a simulation study. Concluding remarks are in Chapter \ref{conclusion}.

\chapter{Literature Review}\label{literature}
\section{Cryptocurrency}
The emergence of the cryptocurrency market since Bitcoin was introduced by \cite{nakamoto2008bitcoin} has posed great challenges and opportunities for investors, entrepreneurs, economists and policymakers. The popularity and the use of cryptocurrencies did not grow substantially until 2017, and the analysis of cryptocurrencies, especially Bitcoin, has received considerably attention from both academia and industry for different purposes. 

The total market capitalisation of cryptocurrency is over 230 billion USD, with more than 1700 cryptocurrencies (CoinMarketCap.com accessed on 8th August, 2018). Bitcoin has a market capitalisation of 112 billion USD, which is nearly 50 percent of the entire market. Ethereum, Ripple, Bitcoin Cash and EOS have the next largest market capitalisations, each with a market capitalisation greater than 5 billion dollars.

However, the cryptocurrency market was criticised for the existence of a potential bubble underlying the prosperity gained \citep{fry2016negative} as well as for its level of maturity \citep{sontakke2017cryptocurrencies}. Many investors continue to enter the market for the large potential profits caused by market immaturity.

Nevertheless, \citet{dyhrberg2018investible} examined the investibility of cryptocurrencies by estimating the transaction costs and intraday trading patterns, which suggested that cryptocurrencies, especially Bitcoin, are investible. Thus,  the application of the risk management procedures and exploration of the volatility properties remain innovative due to the existence of a large number of trading platforms and investors. 
 A comprehensive risk analysis would be a useful tool for both investors and decision makers.    

Currently, the literature on the volatility and other risk properties analysis of cryptocurrencies is limited. GARCH family models were most commonly employed in the existing papers to investigate the volatility properties of cryptocurrencies, especially Bitcoin. 

\citeauthor{dyhrberg2016hedging} (\citeyear{dyhrberg2016bitcoin}, \citeyear{dyhrberg2016hedging}) explored the hedging properties of Bitcoin by employing the GARCH family models with the daily return between 2010 and 2015. She found that Bitcoin has some properties similar to those of gold and can clearly be used as a hedge asset against American stocks and the dollar. However, this result may not reflect the higher volatility exhibited by the cryptocurrency market since the cryptocurrency market became active in the middle 2017.

\citeauthor{chu2017garch} (\citeyear{chu2017garch}) assessed the VaR of 7 cryptocurrencies by applying 12 GARCH family models with the daily return data between 2014 and 2017. However, cryptocurrencies exhibited much larger intraday volatility than traditional financial stocks; therefore, the use of daily data may tend to underestimate the risk properties of the given cryptocurrency.

\citeauthor{chan2017statistical} (\citeyear{chan2017statistical}) conducted a statistical analysis of 7 cryptocurrencies by fitting parametric models with several distribution assumptions to the daily data from 2014 to 2017. They found that no single distribution can fit all the cryptocurrencies. In other words, different cryptocurrencies have different distribution behaviours.

\citeauthor{katsiampa2017volatility} (\citeyear{katsiampa2017volatility}) explored the volatility of Bitcoin price with daily data from 2010 to 2016. Moreover, he applied a goodness-of-fit test to investigate the powers of several GARCH family models. However, no additional evaluation methods are employed to test the performance of each model.

\citeauthor{osterrieder2017statistical} (\citeyear{osterrieder2017statistical}) provided an extreme value analysis of Bitcoin daily returns and compared them to the traditional currency exchange rates. They found that Bitcoin exhibits higher volatility and that the return distribution is heavier tailed and has stronger non-normal characteristics.

\citeauthor{gkillas2018application} (\citeyear{gkillas2018application}) studied the tail behaviour of the daily returns of five major cryptocurrencies according to their market capitalisation by applying extreme value theory. They found that the riskiest cryptocurrency is Bitcoin Cash, whereas the least risky cryptocurrencies are Bitcoin and Litecoin.

\section{Tail Risk Measures}
\subsection{Value at Risk} 
Value at risk (VaR) is a tail quantile of a portfolio's return distribution and is defined as the maximum possible loss over a certain holding period at the given confidence level. VaR has become a standard measure of market risk since J. P. Morgan published the RiskMetrics system in 1994. Furthermore, VaR has been widely used by financial institutions since the Basel Committee on Banking Supervision (BCBS) chose it as an indicator for setting regulatory capital requirements \citep{komitee1996overview}.

\citet{jorion1996risk2}, \citet{duffie1997overview} and \citet{dowd1998beyond} introduced the theoretical framework for VaR, which can be defined as:
\begin{equation}
    \alpha = \mathbf{P}_{t+1}(r_{t+1}\leq \mathbf{VaR}_{t+1|t}(\alpha)|\mathcal{F}_t)\label{VaR}
\end{equation}
or
\begin{equation}
    \mathbf{VaR}_{t+1|t}(\alpha)\equiv F_t^{-1}(\alpha),
\end{equation}
where $\alpha\in(0,1)$ is the quantile or significance level, and $\mathcal{F}_t$ denotes the information set available at time t. $F_t$ is the conditional distribution, and $F_t^{-1}$ its inverse. Therefore, VaR can be regarded as the $\alpha$-level quantile of a given portfolio's loss distribution.

Although VaR is an intuitive risk measure and has been widely used in both industry and academia, the practicability of VaR was heavily criticised, as the measure is incoherent and considers only the distribution quantile while ignoring extreme loss beyond the given quantile level. 

\citet{artzner1999coherent} provided results on the non-subadditivity of VaR; that is, the VaR for a well-diversified portfolio can be even larger than the sum of the individual VaRs under certain conditions. This scenario is inconsistent with finance literature. 
 
Additionally, the Bank of International Settlement Committee on the Global Financial System argued that extreme market movement and the tail behaviour of the financial asset's return distribution cannot be captured and monitored by VaR \citep{chen2012bayesian}. 

Furthermore, \citet{ibragimov2016heavy} provided new results on the non-subadditivity of VaR which showed that diversification does not reduce VaR for a large class of dependent extreme-tailed risks.

Consequently, another measure addressing those limitations, and that can also be regarded as a complement to VaR, is expected shortfall.

\subsection{Expected Shortfall}

\citet{artzner1997thinking} proposed the concept of ES, which gives the expected loss conditional on returns exceeding the corresponding VaR threshold. Therefore, ES is also known as conditional VaR (CVaR). Following \citet{artzner1999coherent}, ES can be defined as:
\begin{equation}
    \mathbf{ES}_{t+1|t}(\alpha) = \mathbf{E}_{t}[r_{t+1}|r_{t+1}<\mathbf{VaR}_{t+1|t}(\alpha)]\label{ES}
\end{equation}

ES combines aspects of VaR with more information about the returns distribution in the tail; thus, it can alleviate the problems inherent in VaR \citep{acerbi2002coherence}. Additionally, a considerable advantage of ES over VaR is the preservation of the convexity property, which allows ES to enter the objective and constraints in optimisation problems under certain conditions \citep{rockafellar2002conditional}.

\citet{yamai2005value} compared the performances of VaR and ES under fat-tailed distributions and found that ES is a better risk measure than VaR in terms of tail risk. They emphasised the facts that VaR can disregard losses beyond the VaR level, which can cause serious problems in the real world, and that the information provided by VaR can mislead the investors.

Both VaR and ES can be used as tail risk measures to provide a more comprehensive risk quantification jointly because ES, in general, can be regarded as the subproduct of VaR estimation, as noted by \citet{yamai2005value} and Basel \RomanNumeralCaps3 Capital Accord.

\section{Volatility and Tail Risk Modelling}
According to \citet{10.2307/2350752}, the distribution of financial returns should be fat-tailed and negative skewed with conditional heteroscedasticity. This observation is sufficiently supported by empirical results. On the basis of these results, different models have been introduced to capture those characteristics of financial time series.

\citet{engle2001value} categorised VaR and ES estimation methods into three groups: parametric, semiparametric and nonparametric. Additionally, they noticed that the main difference between those groups is the method used to address the return distributions \citep{engle2004caviar}.

Volatility is an important indicator in financial time series' risk properties and can be statistically defined as:
\begin{equation}
    \sigma_{t+1} = \sqrt{\mathbf{Var}_{t+1}|\mathcal{F}_t},
\end{equation}
where $\mathbf{Var}_{t+1}$ denotes the conditional variance of the given time series at time t+1. 

Some VaR and ES estimation approaches, especially the parametric class, require modelling of volatility, which has drawn considerable attention from both academia and industry for risk management purposes. 

\citet{engle1982autoregressive} proposed the autoregressive conditional heteroscedasticity (ARCH) model, which has become the most popular and has been widely applied in volatility modelling. The ARCH model allows the conditional variance to change as a function of past residuals while leaving the unconditional variance constant over time.

\citet{bollerslev1986generalized} extended the ARCH model to the generalised ARCH (GARCH) model, which allows a much more flexible lag structure. Until recently, the ARCH and GARCH models are widely used and recognised as essential methods in financial time series analysis because they can capture the changes in variance over time. Furthermore, GARCH (1,1) is commonly used as a benchmark in the literature. 

A group of GARCH models are derived to capture the different volatility features of financial time series. Hence, these models constitute the so-called GARCH family models. \citet{nelson1991conditional} introduced the exponential GARCH (EGARCH) model to capture the leverage effect, that is, negative innovation (bad news) tends to have a larger impact on the future volatility of the given financial asset \citep{black1976stuedies}. Furthermore, \citet{glosten1993relation} proposed the GJR-GARCH model to capture the leverage effect and other stylised features of financial time series, such as volatility clustering. In addition, a large group of extensions, such as the threshold GARCH \citep{zakoian1994threshold} and the double threshold GARCH \citep{li1996double}, are proposed to capture different features and overcome the drawbacks suffered by the GARCH model.

Because parametric modelling requires assumptions about the return distributions, several distributions, such as the normal distribution, Student's t-distribution, asymmetric Laplace (AL) distribution \citep{chen2012bayesian}, and two-sided Weibull distribution \citep{chen2013two}, are employed in parametric models to capture the tail behaviour of the return distributions.

Nonparametric methods, such as the historical simulation approach, are widely applied in industry as they do not make any distribution assumption and are easy to implement. These methods are based on the assumption that past information provides a complete representation of future expected returns, that is, ``history repeats itself''.

Semiparametric methods combine components of both parametric and nonparametric methods. A commonly used semiparametric approach is conditional autoregressive value at risk (CAViaR) proposed by \citet{engle2004caviar}, which provides a framework to estimate VaR directly rather than based on the volatility estimation. However, as the CAViaR model does not provide a framework for ES estimation, \citet{taylor2017forecasting} extended it by incorporating the maximum likelihood based on an AL density, which enables joint modelling of the VaR and ES. Additionally, \citet{taylor2008estimating} proposed the conditional autoregressive expectiles (CARE) models, which incorporated the expectile theory into the VaR and ES estimation.
\section{Forecast Combination}
No model is universally adequate for the estimation of VaR and ES; that is, the best model is unknown and may vary by asset and over time. In this situation, an alternative approach to selecting a specific risk model is to combine the forecasts generated by several individual models.

\citet{timmermann2006forecast} provided an overview of the forecast combination problem. He provided three main arguments to support the belief that combined forecast may be more favourable than individual forecasts. First, combined forecasts tend to be robust to structural breaks. Second, combined forecasts benefit from diversification gains from the combination of individual forecasts with different assumptions, information sets, etc. Third, the impact of misspecification from individual forecasts can be reduced by combining them.

The field of forecast combination has developed several methods for combining point forecasts. Additionally, the empirical facts support the hypothesis that the combined forecast can outperform individual forecasts under certain conditions (see, e.g., \citeauthor{pauwels2012forecast}, 2012; \citeauthor{vasnev2013forecasting}, 2013). The existing well-established methods include simple average \citep{timmermann2006forecast}, median forecast \citep{chan1999dynamic}, Bayesian averaging \citep{liu2009forecasting}, and exponentially weighted averaging \citep{yang2004combining}.

However, less literature has focused on the combination of quantile forecasts. Increasing attention has recently been devoted to the combination of quantile forecasts to improve forecast accuracy, especially the estimation of VaR in the financial risk management field. Moreover, no framework for how to combine ES forecasts has been proposed in the literature.

Because true quantile prediction will always minimise the quantile loss function, which is a linear piecewise function, \citet{halbleib2012improving} combined VaR forecasts by employing the quantile regression introduced by \citet{koenker1978regression}. Additional extended quantile regression frameworks were introduced to avoid overfitting and improve forecast accuracy by incorporating different penalty terms, for example, quantile lasso regression \citet{li20081}, quantile ridge regression and quantile elastic net regression \citep{bayer2017combining}. Those combination methods are called optimal weighted averages as they were estimated by optimising the given tick loss function.
\section{Forecast Evaluation}
Following the needs for accurate VaR and ES forecasts in financial risk management, various commonly used tests are proposed to assess the VaR and ES forecasts.

The evaluation procedures for VaR quantile forecasts are well-developed; VaR forecasts can be directly assessed by applying the existing tests.

\citet{kupiec1995techniques} proposed the violation ratio and unconditional coverage (UC) test to assess whether the actual number of VaR violations over the testing period equals the theoretically expected number.

\citet{christoffersen1998evaluating} extended the UC test and developed a likelihood ratios statistic to assess the joint assumptions of UC and the independence of failures, that is, whether the probability of an exception or violation in any step depends on the outcome of the previous steps. This test, called the conditional coverage (CC) test, can reject VaR quantile forecasts with either more or less clustered violations.   

\citet{engle2004caviar} introduced the dynamic quantile (DQ) test. Similar to the CC test, DQ test is used to assess the UC, CC and independence properties. The DQ test has been proved to be more powerful than the CC and UC tests. Moreover, the quantile loss function from \citet{koenker1978regression} was regularly applied as a criterion because the best set of VaR quantile forecasts should minimize the quantile loss function.

However, optimal assessment of ES forecasts remains an open issue under investigation. \citet{chen2012bayesian} illustrated how to treat ES forecasts as VaR quantile forecasts, which allowed the ES forecasts to be directly assessed.

\citet{fissler2016higher} introduced a series of joint score functions to evaluate VaR and ES forecasts jointly based on the theoretical fact that the true ES forecasts may not be able to minimise the quantile loss function or any other loss function \citep{weber2006distribution, gneiting2011making}. Furthermore, \citet{taylor2017forecasting} developed the AL log score function based on the joint framework and AL distribution. Additionally, the model confidence set (MCS) proposed by \cite{hansen2011model} can produce the model set that are statistically favoured based on the given loss function. 
\chapter{Methodology}\label{methodology}
\section{Individual Model Specification}
This section specifies the individual models that will be considered and applied to generate the one-step-ahead forecasts of both VaR and ES. Moreover, these models will be categorised into three groups: parametric, nonparametric and semiparametric.
\subsection{Parametric Models}

\subsubsection*{Exponential Weighted Moving Average}
The exponential weighted moving average (EWMA) method, developed by J. P. Morgan \citep{morgan1996riskmetrics}, computes the conditional variance at time $t+1$ by assigning unequal weights to past returns. The EWMA assumes that the residuals of financial returns follow a normal distribution with the most recent returns tending to have a larger impact on the return at time $t+1$ and thus having higher weights. The weights exponentially decrease from recent returns to returns that are far in the past. Thus, the conditional variance at time $t+1$ is defined as:
\begin{align*}
    \hat{\sigma}^2_{t+1|t}=(1-\theta)r_t^2+\theta{\sigma}_t^2,
\end{align*}
where $\theta$ is a decay factor, with recommended values of $0.94$ for daily data and $0.97$ for monthly data. When $\theta$ is set to $1$, the weights are assigned equally, and the EWMA method is equivalent to the normal distribution method, which assigns equal weights to all past information.

\subsubsection*{GARCH (1,1)}

According to \citet{bollerslev1986generalized}, the GARCH (1,1) model is defined as:
\begin{align*}
    &\sigma_{t+1}^2=\omega+\alpha\epsilon^2_{t}+\beta\sigma_{t}^2\\
    &r_{t+1} = \mu_{t+1} + \epsilon_{t+1}\\
    &\epsilon_{t+1} = \sigma_{t+1}z_{t+1}\\
    &z_{t+1}\sim\mathcal{D}(0,1)\\
    &\alpha,~\beta\geq0,
\end{align*}
where $\{z_t\}$ is a sequence of independent and identically distributed (i.i.d.) random variables. $\mu_{t+1}$ is the conditional mean at time t+1.
\subsubsection*{EGARCH (1,1)}

In contrast to the GARCH model, the EGARCH model proposed by \citet{nelson1991conditional} makes no restriction on the parameters and considers the leverage effect. The EGARCH (1,1) model is defined as:
\begin{align*}
    &\ln({\sigma_{t+1}^2}) = \alpha_0 + \alpha_1\frac{|\epsilon_{t}|+\gamma_1\epsilon_{t}}{\sigma_{t}}+\beta_1\ln(\sigma^2_{t})\\
    &r_{t+1}=\mu_{t+1}+\epsilon_{t+1}\\
    &\epsilon_{t+1} = \sigma_{t+1}z_{t+1}\\
    &z_t\sim\mathcal{D}(0,1),
\end{align*}
where the term $|\epsilon_{t-i}|+\gamma_i\epsilon_{t-i}$ is used to capture the leverage effects in the financial market. When $\epsilon_{t-i}$ is positive, which indicates positive innovation, the term for the positive innovation is $1+\gamma_i$, whereas for a negative innovation, this term will be $1-\gamma_i$.
\subsubsection*{GJRGARCH (1,1)}
To capture other stylised characteristics of financial time series, such as volatility clustering, \citet{glosten1993relation} proposed an asymmetric GARCH model named according to the authors' initials, GJRGARCH. The model is formally defined as:
    \begin{align*}
    &\sigma_{t+1}^2 = \alpha_0+\alpha_1\epsilon_{t}+\beta_1\sigma^2_{t}+\gamma_1\mathbf{I}_{t}\epsilon_{t}^2\\
    &r_{t+1}=\mu_{t+1}+\epsilon_{t+1}\\
    &\epsilon_{t+1} = \sigma_{t+1}z_{t+1}\\
    &z_t\sim\mathcal{D}(0,1)
    \end{align*}
and:
  \[
  \mathbf{I}_{t}=
  \begin{cases}
    1 & \text{if $\epsilon_{t}<0$} \\
    0 & \text{if $\epsilon_{t}\geq0$}
  \end{cases}
\]
where the indicator function, $\mathbf{I}_{t}$, is used to denote the leverage effect. Clearly, a negative innovation will have a different impact on conditional volatility from a positive innovation with the same magnitude under the GJRGARCH framework with the indicator function.

\subsubsection*{VaR and ES Estimators}
This section defines the corresponding estimators of both VaR and ES under different distributions as the parametric models require an assumption on the residuals distribution $\mathcal{D}$. This thesis will consider three commonly used distributions: the standard normal distribution, the standardised Student's t-distribution and the standard skewed Student's t-distribution.

\paragraph{Standard Normal Distribution} \citet{mcneil2005quantitative} show that the corresponding $1-\alpha$ confidence level VaR and ES estimators are defined as:
\begin{equation*}
\begin{split}
    &\mathbf{VaR}_{t+1}(\alpha) = \mu_{t+1} + \sigma_{t+1}\Phi^{-1}(\alpha)\\
    &\mathbf{ES}_{t+1}(\alpha) = \mu_{t+1} + \sigma_{t+1}\frac{\phi(\Phi^{-1}(\alpha))}{\alpha},
\end{split}
\end{equation*}
where $\Phi^{-1}$ and $\phi$ are the inverse cumulative distribution function (CDF) and probability density function (PDF) of the standard normal distribution, respectively. $\mu_{t+1}$ and $\sigma_{t+1}$ are the corresponding one-step-ahead conditional mean and standard deviation forecasts from GARCH family models, respectively. 

\paragraph{Standardised  Student's t-Distribution} According to \citet{mcneil2005quantitative}, the corresponding $1 - \alpha$ confidence level VaR and ES estimators are defined as:
\begin{equation*}
    \begin{split}
        &\mathbf{VaR}_{t+1}(\alpha) = \mu_{t+1} + \sigma_{t+1}t^{-1}_{v}(\alpha)\sqrt{\frac{v-2}{v}}\\
        &\mathbf{ES}_{t+1}(\alpha) = \mu_{t+1} + \sigma_{t+1}\frac{t_{v}(t_{v}^{-1}(\alpha)}{\alpha}\Big(\frac{v+(t_v^{-1}(\alpha))^2}{v-1}\Big)\sqrt{\frac{v-2}{v}},
    \end{split}
\end{equation*}
where $v$ is the degree of freedom (DOF) of the standardised Student's t-distribution, which is estimated from the empirical sample distribution. $t_v$ and $t_v^{-1}$ are the PDF and inverse CDF of the standardised Student's t-distribution, respectively.

\paragraph{Standard Skewed Student's t-Distribution} \citet{contino2017bayesian} derive the VaR and ES estimators based on the standard skewed Student's t-distribution from \citet{hansen1994autoregressive}; the corresponding $1-\alpha$ confidence level VaR and ES estimators are defined as:
\begin{equation*}
    \begin{split}
        &\mathbf{VaR}_{t+1}(\alpha) = \mu_{t+1} + \sigma_{t+1}skt_{\alpha,v,\lambda}^{-1}\\
        &\mathbf{ES}_{t+1}(\alpha) = \mu_{t+1} + \sigma_{t+1}E[\epsilon|\epsilon<\mathbf{VaR}_{t+1}(\alpha)],
    \end{split}
\end{equation*}
where:
\begin{equation*}
\begin{split}
    &E[\epsilon|\epsilon<\mathbf{VaR}_{t+1}(\alpha)] = \frac{c(1-\lambda)^2}{b\times skt_{v,\lambda}(\mathbf{VaR}_{t+1}(\alpha))}\frac{v-2}{1-v}\Big(1+\frac{1}{v-2}\Big(\frac{b\mathbf{VaR}_{t+1}(\alpha)+a}{1-\lambda}\Big)^2\Big)^{\frac{1-v}{2}} - \frac{a}{b}\\
\end{split}
\end{equation*}
The term $skt_{v,\lambda}$ is the CDF of the standard skewed Student's t-distribution, which is defined as:
\begin{equation*}
    skt(\epsilon|v,\lambda) =  \left\{
\begin{array}{ll}
      (1+\lambda)t_v\Big(\sqrt{\frac{v}{v-2}}\Big(\frac{b\epsilon+a}{1-\lambda}\Big),\lambda\Big) & \text{if}~\epsilon<-\frac{a}{b} \\
      \frac{1-\lambda}{2}(1+\lambda)\Big[t_v\Big(\sqrt{\frac{v}{v-2}}\Big(\frac{b\epsilon+a}{1-\lambda}\Big),v\Big)-0.5\Big] & \text{if}~\epsilon\geq-\frac{a}{b}
\end{array} 
\right.
\end{equation*}
$v, \lambda$ are the kurtosis parameter with a range of $2<v<\infty$ and the skewness parameter with a range of $-1<\lambda<1$, respectively. Furthermore, a, b and c in the CDF function above are defined as:
\begin{equation*}
    \begin{split}
        &a = 4\lambda c(\frac{v-2}{v-1})\\
        &b = \sqrt{1+3\lambda^2-a^2}\\
        &c = \frac{\Gamma(\frac{v+1}{2})}{\sqrt{\pi(v-2)}\Gamma(\frac{v}{2})},
    \end{split}
\end{equation*}
$\Gamma(*)$ denotes the gamma function.

\subsection{Nonparametric Models}

\subsubsection*{Historical Simulation}

Instead of making a distribution assumption, the historical simulation (HS) method simply takes the corresponding $\alpha$-level VaR quantile of the empirical return distribution based on only the assumption that past information can be a good representation of future returns. Similarly, the HS method computes ES by taking the average of past empirical losses that exceed the corresponding VaR threshold. Thus, the VaR and ES estimators are defined as:
\begin{align*}
    \begin{split}
    &\mathbf{VaR}_{t+1}(\alpha) = \textbf{Quantile}\{\{r_t\}_{t=1}^n,\alpha\}\\
    &\mathbf{ES}_{t+1}(\alpha) = \frac{1}{n}\sum_{t=1}^n r_t\mathbf{I}(r_t<\mathbf{VaR}_{t}(\alpha)),
    \end{split}
\end{align*}
where $\{r_t\}_{t=1}^n$ is the past return series, and the indicator function $\mathbf{I}(r_t<\mathbf{VaR}_{t}(\alpha))$ is used to denote the scenarios in which the losses exceed the corresponding VaR threshold.

\subsection{Semiparametric Models}
\subsubsection*{Conditional Autoregressive Value at Risk and Expected Shortfall}
\citet{engle2004caviar} proposed the CAViaR framework, which specifies the evolution of the quantile over time by using an autoregressive process and estimates the parameters with the regression quantiles introduced by \citet{koenker1978regression}. The general CAViaR specification is defined as:
\begin{align*}
    \mathbf{VaR}_{t}(\alpha) = \beta_0 + \sum_{i=1}^q\beta_i\mathbf{VaR}_{t-i}(\alpha) + \sum^{r}_{j=1}\beta_j\boldsymbol{l}(\mathbf{x}_{t-j}),
\end{align*}
where $p = q+r+1$ is the dimension of parameter vector $\{\boldsymbol{\beta}\}$, and the $\boldsymbol{l}(*)$ is a function of a finite number of lagged values of observables. The second term, $\beta_i\mathbf{VaR}_{t-j}$, is used to ensure that the quantile can change `smoothly' over time, and the function $\boldsymbol{l}(\mathbf{x}_{t-j})$ is used to connect the quantile to the past observable variables $\mathbf{x}_{t-j}$ that belong to the information set $\mathcal{F}_{t-j}$.

Furthermore, with the general CAViaR specification, \citet{engle2004caviar} introduced four examples of CAViaR processes defined as follows:

\textbf{Adaptive (ADA)}:
\begin{equation*}
    \begin{split}
        \mathbf{VaR}_{t}(\alpha) = \mathbf{VaR}_{t-1}(\alpha) + \beta_1\Big\{\Big[1+\text{exp}\Big(G[r_{t-1}-\mathbf{VaR}_{t-1}(\alpha)]\Big)\Big]^{-1}-\alpha\Big\},
    \end{split}
\end{equation*}
where G is a positive finite number. The ADA model incorporates the following rule: whenever the empirical losses exceed the corresponding VaR threshold, the model immediately increases the VaR, whereas the VaR is slightly decreased when the empirical losses do not exceed the threshold. The purpose of this strategy is to reduce the probability of sequences of exceedances. Moreover, the ADA model has a unit coefficient the lagged VaR. Additionally, three other alternatives are available and are defined as:

\textbf{Symmetric Absolute Value (SAV)}:
\begin{flalign*}
    ~~~~~~~~~~\mathbf{VaR}_{t}(\alpha) = \beta_1 + \beta_2\mathbf{VaR}_{t-1}(\alpha) + \beta_3|r_{t-1}|,&&
\end{flalign*}

\textbf{Asymmetric Slope (AS)}:
\begin{flalign*}
    ~~~~~~~~~~\mathbf{VaR}_{t}(\alpha) = \beta_1 + \beta_2\mathbf{VaR}_{t-1}(\alpha) + \beta_3(r_{t-1})^{+} + \beta_4(r_{t-1})^{-},&&
\end{flalign*}
where $(x)^{+} = \text{max}(x,0)$ and $(x)^{-} = -\text{min}(x,0)$.

\textbf{Indirect GARCH(1, 1) (IG)}:
\begin{flalign*}
    ~~~~~~~~~~\mathbf{VaR}_{t}(\alpha) = \sqrt{\beta_1+\beta_2\mathbf{VaR}_{t-1}^2(\alpha)+\beta_3r^2_{t-1}},&&
\end{flalign*}
where the parameters set $\{\boldsymbol{\beta}\}$ should be restricted to be $> 0$ to ensure a positive value under the square root. The SAV and IG models respond symmetrically to the past returns, and the AS model responds differently to the past returns.

\textbf{CAViaR Parameters Estimation}

The parameters of CAViaR models were estimated by incorporating the regression quantiles by \citet{koenker1978regression}. They extended the notion of the sample quantile to the linear regression model. The parameter set $\{\boldsymbol{\beta}\}$ can be estimated by solving the following optimisation problem:
\begin{equation*}
    \underset{\beta}{\text{min}}\frac{1}{T}\sum^{T}_{t=1}[\alpha - \mathbf{I}(r_t<\mathbf{VaR}_{t}(\alpha))][r_t-\mathbf{VaR}_{t}(\alpha)]
\end{equation*}

One special case of the regression quantiles is the least absolute deviation (LAD), which is a more robust estimator than the ordinary least squares (OLS) estimator, especially when the errors have a fat-tailed distribution.

However, CAViaR models do not provide a framework for ES estimation, which limits the use of CAViaR models for financial risk management purposes. Nevertheless, \citet{taylor2017forecasting} extended the CAViaR framework by incorporating the AL log score function and adding the connection between VaR and ES.

Based on the CAViaR framework, \citeauthor{taylor2017forecasting} specified the additional ES component by considering the crossing between VaR and ES, as by definition, ES must be lower than the corresponding VaR threshold. \citeauthor{taylor2017forecasting} introduced two examples to connect the ES component to the corresponding VaR from the CAViaR framework. 

The first formulation models the ES component as the product of the VaR quantile and a constant multiplicative factor \citep{gourieroux2012converting}. To avoid crossing, the constant was restricted to be greater than 1 by expressing it in terms of an exponential function. The connection function is defined as:
\begin{equation}
    \mathbf{ES}_{t}(\alpha) = (1+\text{exp}(\gamma_0))\mathbf{VaR}_{t}(\alpha)\label{ES-EXP}
\end{equation}

Expression \eqref{ES-EXP} correctly describes the relationship between ES and VaR but may excessively restrict the ES component as ES will have the same dynamics as the corresponding VaR under this restriction. Therefore, an alternative approach that incorporates the autoregressive process is defined as:
\begin{align*}
\begin{split}
    &\mathbf{ES}_{t}(\alpha) = \mathbf{VaR}_{t}(\alpha) - x_t\\
    &x_t = \left\{
    \begin{array}{cc}\gamma_0+\gamma_1(\mathbf{VaR}_{t-1}(\alpha)-r_{t-1})+\gamma_2x_{t-1} & \text{if}~ r_{t-1}\leq\mathbf{VaR}_{t-1}(\alpha)\\
    x_{t-1} & \text{otherwise}
    \end{array}
\right.
\end{split}
\end{align*}
where the parameter set $\{\boldsymbol{\gamma}\}$ is restricted to be nonnegative to ensure the estimated VaR quantile and ES do not cross. Additionally, the use of autoregressive expression can essentially smooth the magnitude of exceedances beyond the corresponding VaR threshold.  

\textbf{CAViaR-ES Parameters Estimation}

\citeauthor{taylor2017forecasting} extended the CAViaR models to the CAViaR-ES framework, which allows the joint estimation of VaR and ES. The parameters are estimated by minimising a new scoring function based on the AL density. The new scoring function, AL log score, is also considered as an evaluation function to test the model performance; the details of this score function will be presented in Section \ref{backtesting} later. According to \citet{taylor2017forecasting}, the AL log score is defined as:
\begin{equation*}
    \mathcal{S} (\mathbf{VaR}_{t}(\alpha),\mathbf{ES}_{t}(\alpha),r_{t}) = -\ln\Big(\frac{\alpha-1}{\mathbf{ES}_{t}(\alpha)}\Big) - \frac{(r_t-\mathbf{VaR}_{t}(\alpha)(\alpha-\mathbf{I}(r_t\leq\mathbf{VaR}_{t}(\alpha))}{\alpha\mathbf{ES}_{t}(\alpha)},
\end{equation*}
where the expressions of $\mathbf{VaR}_t(\alpha)$ and $\mathbf{ES}_{t}(\alpha)$ can be any combination of the CAViaR models and the additional ES components from \citeauthor{taylor2017forecasting}'s joint framework.
\subsubsection*{Conditional Autoregressive Expectiles}
\citet{taylor2008estimating} proposed the conditional autoregressive expectiles (CARE) framework, which incorporates expectile theory into the estimation of VaR and ES.

\textbf{Expectile Theory}

The concept of expectiles was originally introduced by \citet{aigner1976estimation} to construct an estimator that assigns different weights to positive and negative residuals in the classic regression model estimation process. In their work, for any univariate continuous random variable $Y$, the corresponding $\tau$-level expectile $\mu_{\tau}$ is defined as:
\begin{equation}
    \mu_{\tau} = \underset{\mu_\tau}{\text{argmin}}~E\Big[|\tau - \mathbf{I}[Y<\mu_{\tau}]|(Y - \mu_{\tau})^2\Big]\label{expectile},
\end{equation}
where the $\tau$-level expectile $\mu_{\tau}$ can be estimated by minimising the expectation function \eqref{expectile} with $\tau\in(0,1)$. Note that $\mu_{0.5} = E(Y)$ because the weights are equally assigned to values greater and less than the corresponding expectile $\mu_{\tau}$.

Therefore, given the sample returns $r_1,\cdots,r_n$ on $Y$, with a fixed $\tau$, the constant $\tau$-level expectile on $Y$, $\mu_{\tau}$, can be estimated by minimising the asymmetric sum of squares function, denoted as:
\begin{equation}
    \sum^n_{t=1}\Big[|\tau - \mathbf{I}[r_t<\mu_{\tau}]|(r_t-\mu_{\tau})^2\Big]\label{ALS}
\end{equation}

\textbf{CARE Modelling}

\citeauthor{taylor2008estimating} took the CAViaR framework from \citet{engle2004caviar} and replaced the dynamic quantile terms with dynamic expectile terms and thus introduced three models, CARE-SAV, CARE-AS and CARE-IG, where the model expressions are defined by the following specifications:

\textbf{CARE-SAV}:
\begin{flalign*}
    ~~~~~~~~~~\mu_t = \beta_1 + \beta_2 + \beta_3|r_{t-1}|,&&
\end{flalign*}
where the parameter set $\{\boldsymbol{\beta}\}$ is unrestricted.

\textbf{CARE-AS}:
\begin{flalign*}
    ~~~~~~~~~~\mu_t = \beta_1 + \beta_2\mu_{t-1} + \Big(\beta_3\mathbf{I}(r_{t-1}>0) + \beta_4\mathbf{I}(r_{t-1}<0)\Big)|r_{t-1}|,&&
\end{flalign*}
where the conditional expectile responds to positive and negative shocks asymmetrically, which is similar to the CAViaR-AS model.

\textbf{CARE-IG}:
\begin{flalign*}
~~~~~~~~~~\mu_t = -\sqrt{\beta_1+\beta_2\mu_{t-1}^2 + \beta_3r_{t-1}^2},&&
\end{flalign*}
where the parameter set $\{\boldsymbol{\beta}\}$ is restricted to be positive to ensure positivity under the square root. 

\textbf{CARE Parameters Estimation}

The parameters of CARE models are estimated by minimising the asymmetric sum of squares function, as shown in expression \eqref{ALS}. The estimation process of the parameter set $\{\boldsymbol{\beta}\}$ can be expressed as:
\begin{equation*}
    \{\boldsymbol{\beta}\} = \underset{\{\boldsymbol{\beta}\}}{\text{argmin}}~\sum^n_{t=1}\Big[|\tau - \mathbf{I}[r_t-\mu_{\tau}]|(r_t-\mu_{\tau})^2\Big]
\end{equation*}

\textbf{CARE VaR and ES Estimators}

For the VaR, \citeauthor{taylor2008estimating} used the $\tau$-level expectile as the estimator of the corresponding VaR quantile at the same level. This one-to-one mapping from expectiles to quantiles is supported by the theoretical works of \citet{jones1994expectiles}, \citet{abdous1995relating}, etc.

To connect the expectiles to the corresponding ES, \citeauthor{taylor2008estimating} showed that the ES estimator is a simple function of the expectile based on the theoretical work from \citet{newey1987asymmetric}. The one-to-one connection function that links the expectile and ES is defined as:
\begin{equation*}
    \mathbf{ES}_{t}(\alpha) = \Big(1 + \frac{\tau}{(1-2\tau)\alpha_{\tau}}\Big)\mu_{\tau},
\end{equation*}
where the expectile $\mu_{\tau}$ is equivalent to the corresponding $\alpha$-level quantile of the given return series.

\section{Forecast Evaluation}\label{backtesting}
The violation ratio, UC test, CC test, DQ test and quantile loss function are employed to test the forecast accuracy of the VaR estimation. These tests are now standard. For the ES estimation, the quantile loss function is replaced by the AL log score function. In addition, the model confidence set that incorporates the AL log score function is applied to compare the joint VaR and ES forecast performances.
\subsection{Violation Rate/Ratio}
The violation rate is an informal indicator for comparing the VaR quantile and ES forecast performances by testing whether the empirical violation rate equals the theoretical expectation. We define the corresponding VaR and ES violation rates as Expressions \eqref{VRate} and \eqref{ESRate}, respectively.
\begin{equation}
    \textbf{VRate} = \frac{\sum^{T+m}_{t=T+1}\mathbf{I}(r_t<\mathbf{VaR}_{t}(\alpha))}{m}\label{VRate}
\end{equation}
\begin{equation}
    \textbf{ESRate} = \frac{\sum^{T+m}_{t=T+1}\mathbf{I}(r_t<\mathbf{ES}_{t}(\alpha))}{m}\label{ESRate},
\end{equation}
where $m$ is the forecasting period, and $T$ is the estimation sample size, in-sample size or initial window size.

The use of the violation rate is slightly different between VaR and ES because of their mathematical properties. For the $\alpha$-level VaR, which can be treated like the $\alpha$-level quantile, the best model should be able to produce the VaR forecast with an $\alpha$ violation rate. For example, the $95\%$ confidence level VaR should have a violation ratio of $5\%$. A model with a higher or lower violation rate should be rejected since it is not consistent with the theoretical expectation, even if the model performs well in terms of the other evaluation criteria.

The concept of violation ratio will be employed to visualise the violation rate and make it easy to compare. The violation ratio is defined as:
\begin{equation*}
\begin{split}
&\textbf{VRatio} = \frac{\textbf{VRate}}{\alpha}\\
&\textbf{ESRatio} = \frac{\textbf{ESRate}}{\delta_{\alpha}},
\end{split}
\end{equation*}
where $\alpha$ is the quantile level, and  $\delta_{\alpha}$, which is slightly different between each model and depends on the distribution assumption, denotes the corresponding nominal level of the ES forecasts.

For the ES forecasts, the denominator of the violation ratio is not the corresponding significance level $\alpha$ because the ES forecasts cannot be treated as $\alpha$-level quantile forecasts. Nevertheless, \citet{chen2012bayesian} and \citet{gerlach2014bayesian} provided recommendations for testing ES forecasts by treating them as quantile forecasts at specific quantile levels. They further proved that such levels appear to be robust to the underlying conditional distributions. Table \ref{tab:ES Nominal levels} summarises the corresponding denominator $\delta_{\alpha}$ that should be used to calculate the violation ratio for parametric models with different distribution assumptions \citep{gerlach2014bayesian}. The single number in each bracket is the DOF of the corresponding Student's t or skewed Student's t-distribution. Furthermore, for the nonparametric and semiparametric models, \citet{gerlach2014bayesian} proposed that the $1\%$ and $5\%$ ES values will occur close to $0.36\%$ and $1.8\%$, respectively, for daily return data. For example, see \citet{gerlach2016forecasting}. Therefore, the target violation ratio of each VaR and ES forecast is $1$, with the corresponding denominator used in the calculation. Additionally, a violation ratio $< 1$ indicates that the risk and loss forecasts are higher than the true values (conservative); alternatively, when a violation ratio $> 1$ indicates that risk and loss forecasts are lower than the true values (anti-conservative). Financial institutions must allocate sufficient capital and funds to cover future potential losses; thus, overestimating the risk level is preferred to underestimating it, that is, a violation ratio $= 0.95$ is preferred to a violation ratio $= 1.05$. For example, see \citet{gerlach2011bayesian}.
\begin{table}[t]
  \centering
  \caption{Nominal Levels for ES Forecasts}
    \begin{tabular}{ccccccccc}
    \toprule
          & \multicolumn{8}{c}{Nominal Levels $\delta_{\alpha}$} \\
\cmidrule{2-9}    $\alpha$ & N(0,1) & AL    & t*(10) & t*(6) & t*(4) & Sk-t*(6) & Sk-t*(4) & TW \\
    \midrule
    0.01  & 0.0038 & 0.0037 & 0.0036 & 0.0034 & 0.0032 & 0.0034 & 0.0032 & 0.0035-0.0037 \\
    0.05  & 0.0196 & 0.0184 & 0.0184 & 0.0175 & 0.0164 & 0.0175 & 0.0164 & 0.0184-0.0187 \\
    \bottomrule
    \end{tabular}%
 \label{tab:ES Nominal levels}%

\end{table}%

\subsection{UC Test}
The UC test from \citet{kupiec1995techniques} is a formal statistical procedure to assess the accuracy of VaR quantile forecasts by testing whether the violation rates are equal to the theoretical expectation. The ES forecasts are tested using the corresponding quantile level, as shown in Table \ref{tab:ES Nominal levels}. Mathematically, the null and alternative hypothesis are defined as:
\begin{align*}
    &H_0: \frac{N}{m} = \alpha\\
    &H_1: \frac{N}{m} \neq \alpha ,
\end{align*}
where $N$ is the number of violations during the forecasting period $m$. The corresponding likelihood test statistic is defined as:
\begin{align}
    t_{uc}=2\ln\left[\left(1-\frac{N}{m}\right)^{T-m}\left(\frac{N}{m}\right)^N\right]-2\ln\left[(\alpha)^{m-N}(1-\alpha)^N\right]\sim\chi^2(1)\label{UC Test}
\end{align}

The UC test follows the Chi-squared distribution with DOF $= 1$, that is, the corresponding critical values are 3.84 and 6.63 at the 5\% and 1\% significance levels, respectively. The UC test can reject VaR or ES forecasts with too many or too few violations. 
\subsection{CC Test}
The CC test is another hypothesis test proposed by \citet{christoffersen1998evaluating} to jointly investigate the following:
\begin{itemize}
    \item[$-$] Whether the number of violations is statistically consistent with the theoretically expected value (unconditional coverage property).
    \item[$-$] Whether the violations are independently distributed, e.g., no violation clustering.
    \item[$-$] Whether the violation clustering is within allowable limits (conditional coverage property).
\end{itemize}
The corresponding likelihood test statistic is defined as:
\begin{align}
    t_{cc} = -2\ln[(\alpha)^{m-N}(1-\alpha)^N]+2\ln[(1-\pi_{01})^{\pi_{00}}\pi_{01}^{n_{01}}(1-\pi_{11})^{n_{10}}\pi_{11}^{n_{11}}]\sim\chi^2(2)\label{CC Test},
\end{align}
where:
\begin{itemize}
    \item[$-$] $n_{00}$ denotes the number of cases in which no failure is followed by no failure.
    \item[$-$] $n_{10}$ denotes the number of cases in which failure is followed by no failure.    
    \item[$-$] $n_{01}$ denotes the number of cases in which no failure is followed by failure.
    \item[$-$] $n_{11}$ denotes the number of cases in which failure is followed by failure.
    \item[$-$] $\pi_{01}$ denotes the probability of having a failure at time $t$ given that no failure occurred at time $t-1$.  
    \item[$-$] $\pi_{10}$ denotes the probability of no failure at time $t$ given that a failure occurred at time $t-1$.
    \item[$-$] $\pi_{11}$ denotes the probability of having a failure at time $t$ given that a failure occurred at time $t-1$.
    \item[$-$] $\pi_{00}$ denotes the probability of having no failure at time $t$ given that a failure occurred at time $t-1$.
\end{itemize}

The CC test asymptotically follows a Chi-squared distribution with DOF $=2$, that is, the corresponding critical values are 5.99 and 9.21 at the 5\% and 1\% significance levels, respectively.
\subsection{DQ Test}
Similar to the UC and CC tests, the DQ test proposed by \citet{engle2004caviar} is a joint test for the coverage and independence properties of the VaR and ES forecasts. The DQ test is more powerful than the UC and CC tests \citep{berkowitz2011evaluating}. Both in-sample and out-of-sample tests were introduced by \citet{engle2004caviar}, but we focus on only the out-of-sample test because the DQ test is employed to assess the VaR and ES forecasts with rolling windows. To begin, at the given significance level $\alpha$, a series of `hits' are calculated. The variable \textbf{H} is defined as:
\begin{equation*}
    H_t = \mathbf{I_t}(\mathbf{VaR}_{t}(\alpha)<r_t) - \alpha,
\end{equation*}
where $t\in m$ is the out-of-sample testing period, which must belong to the forecasting window $m$. 
Statistically, the expectation of hits, $E(\textbf{H})$, is zero. Therefore, the null hypothesis of the DQ test is that the series of hits \textbf{H} is composed of i.i.d. variables with the hit rate equal to the significance level $\alpha$.  

Furthermore, the DQ test constructs a regression as follows:
\begin{equation}
    H_t = \beta_0+\sum^K_{k=1}\beta_kH_{t-k}+\epsilon_t\label{DQ},
\end{equation}
where k is the lag number. With the regression \eqref{DQ}, the DQ test jointly assesses the null hypothesis, which is defined as:
\begin{align}
H_0: \beta_0=\beta_1=\dots=\beta_k=0,
\end{align}
where $\beta_0 = 0$ indicates $E(\textbf{H})=0$, and $\beta_1=\dots=\beta_k=0$ indicates the independence property. The lag value k was set to $1$ or $4$, as recommended by \citet{engle2004caviar}, since \citet{chen2012bayesian} illustrated that the test outcome is not sensitive to the choice of lag value $k$.

The corresponding test statistic is defined as:
\begin{equation*}
    DQ = \frac{\mathbf{H}'\mathbf{W}(\mathbf{W}'\mathbf{W})^{-1}\mathbf{W}'\mathbf{H}}{\alpha(1-\alpha)},
\end{equation*}
where for $k=1$, $W_t'= \Big(1,H_{t-1},\mathbf{VaR}_t(\alpha)\Big)$, and $k = 4$, $W_t' = \Big(H_{t-1},\dots,H_{t-4},\mathbf{VaR}_t(\alpha)\Big)$. Asymptotically, the DQ test follows a Chi-squared distribution with DOF = k+2, that is, the corresponding critical values are 7.82 and 11.34 at the 5\% and 1\% significance levels, respectively, for a lag value equal to 1. The corresponding critical values are 12.59 and 16.81 at the 5\% and 1\% significance levels, respectively, for a lag value equal to 4.  

\subsection{Quantile Loss Function}
The CC test, UC test and DQ test are all statistical tests that do not consider differences in magnitude; that is, the tests have less power to distinguish between VaR forecasts with very similar dynamics and violation ratios. The quantile loss function, proposed by \citet{koenker1978regression}, can be used as a criterion function to assess VaR forecasts by accounting for the violations and corresponding magnitude; the most accurate VaR quantile forecast should minimise the quantile loss function. The quantile loss function is defined as:
\begin{equation}
    \text{QLF} = \sum^{T+m}_{t=T+1}\Big(r_t-\mathbf{VaR}_t(\alpha)\Big)\Big(\alpha-\mathbf{I}(r_t<\mathbf{VaR}_t(\alpha))\Big)\label{QLF}
\end{equation}

The quantile loss function expressed as Equation \eqref{QLF} is a piecewise function that can be used to assess VaR quantile forecasts by treating violation and non-violation scenarios in different ways. For example, for a 95\% confidence level VaR forecast, if a violation occurs, 95\% of the violation magnitude is penalised; by contrast, 5\% of the non-violation magnitude is penalised in non-violation cases. Therefore, the VaR quantile forecasts with the lowest loss function value are preferred. However, for ES forecasts, the best forecasts may not minimise the quantile loss function or other loss functions, a condition that is called not elicitable \citep{fissler2016higher}. 

\subsection{AL Log Score}
The term \textit{elicitable} is used to describe whether the correct forecast of a certain measure can uniquely minimise the expectation of at least one loss or scoring function \citep{fissler2016higher}. \citet{gneiting2011making} showed that the ES measure is not elicitable. Nevertheless, \citet{fissler2016higher} illustrated that although ES is not elicitable, the VaR and ES measures are jointly elicitable; for example, the mean and variance are jointly elicitable, even through the variance is not elicitable itself. They presented the form of the strictly consistent scoring functions used for jointly testing the VaR and ES forecasts:

\begin{equation}
\begin{split}
    \mathcal{S}(\mathbf{VaR}_t(\alpha),\mathbf{ES}_{t}(\alpha), r_t) &= \Big(\mathbf{I}(r_t\leq\mathbf{VaR}_t(\alpha))-\alpha\Big)G_1\Big(\mathbf{VaR}_{t}(\alpha)\Big) - \mathbf{I}(r_t\leq\mathbf{VaR}_t(\alpha))G_1(r_t)\\ &+ G_2\Big(\mathbf{ES}_{t}(\alpha)\Big)\Big(\mathbf{ES}_t(\alpha)-\mathbf{VaR}_t(\alpha)+\mathbf{I}(r_t\leq\mathbf{VaR}_{t}(\alpha))\\ & \times (\mathbf{VaR}_t(\alpha) - r_t)/\alpha\Big) - \zeta_2\Big(\mathbf{ES}_{t}(\alpha)\Big) + a(r_t),\label{Joint}
\end{split}
\end{equation}
where $G_1, G_2$ and $\zeta_2$ are functions that must satisfy certain conditions, namely, $G_2 = \zeta_2'$, $G_1$ is increasing,  and $\zeta_2$ is increasing and convex. \citet{taylor2017forecasting} proposed the AL log score function based on Expression \eqref{Joint}, as well as the AL log-likelihood function, which is defined as:
\begin{equation}
    \text{AL} =\frac{1}{m}\sum^{T+m}_{t=T+1}\Big[ -\ln\Big(\frac{\alpha-1}{\mathbf{ES}_t(\alpha)}\Big) - \frac{(r_t-\mathbf{VaR}_t(\alpha))(\alpha-\mathbf{I}(r_t\leq\mathbf{VaR}_t(\alpha)))}{\alpha\mathbf{ES}_t(\alpha)}\Big]\label{AL}
\end{equation}

The AL log score function is strictly consistent and can be used as a loss function to jointly assess the accuracy of VaR and ES forecasts \citep{gneiting2007strictly}. Models that can minimise the AL log score function are preferred.

\subsection{Model Confidence Set}
The model confidence set (MCS) proposed by \cite{hansen2011model} is further employed to compare the ES forecast performances between each applied method and identify the set of the best models. The MCS summarises the relative performance of an entire set of models by selecting the models that are statistically superior at a given confidence level. 

According to \cite{hansen2011model}, the starting point of the MCS is a finite set of forecasting models, denoted as $\mathcal{M}_0$. Next, the MCS aims to identify the best model set $\mathcal{M}^{*}$ through a sequence of significance tests defined as:
\begin{equation*}
    \mathcal{M^{*}} = \{i\in\mathcal{M}_0 : u_{i,j}\leq 0 \quad \text{for all}\quad j \in \mathcal{M}_0\}
\end{equation*}
where $u_{i,j} = \mathbf{E}(d_{ij,t})$ is the expected loss differential between each model pair and $d_{ij,t} = L_{i,t} - L_{j,t}$ is the model loss differential. We set the loss function $L$ as the AL log score during the ES evaluation process.

Therefore, when given an initial set of forecasting models $\mathcal{M}_0$, the MCS algorithm searches for the set of models that cannot be rejected statistically at the given confidence level. The corresponding null hypothesis is defined as:
\begin{equation*}
    H_{0,\mathcal{M}}: u_{ij} = 0, \quad\text{for all}\quad i, j \in \mathcal{M}.
\end{equation*}

The t-statistic associated with the null hypothesis of the MCS is defined as:
\begin{equation*}
    t_{ij} = \frac{\bar{d}_{ij}}{\sqrt{\hat{var}(\bar{d}_{ij})}},
\end{equation*}
\begin{equation*}
    T_{\mathcal{M}} = \max_{i,j\in\mathcal{M}}|t_{ij}|
\end{equation*}
where $\bar{d}_{ij} = n^{-1}\sum^n_{t=1}d_{ij,t}$ measures the relative loss between model i and model j. Further, the MCS employs a stationary bootstrap to address the issue that the asymptotic distribution of the test statistic is non-standard.

Simply, the MCS detects the best set of models based on the following algorithm:
\begin{itemize}
    \item[$-$] Step 1. Enter the model set $\mathcal{M} = \mathcal{M}_0 $, which contains all the forecast models.
    \item[$-$] Step 2. Test the null hypothesis $H_{0,\mathcal{M}}$ at the given significance level.
    \item[$-$] Step 3. If the null hypothesis $H_{0,\mathcal{M}}$ is rejected, the MCS eliminates an object or a model from $\mathcal{M}$ and repeats Step 2 until the null hypothesis is accepted at the given significance level. In the final result, there is no significance difference between each model included in MCS $\mathcal{M}$. 
\end{itemize}

\section{Proposed Joint Combination Framework}\label{proposed} 
In this section, we propose a new methodology that allows the joint VaR and ES combinations from a group of models. The idea was inspired and motivated by \citet{taylor2017forecasting}, who developed a VaR and ES joint forecast framework based on the proposed AL score function. Similarly, our new methodology combines the VaR and ES forecasts and obtains the optimal weights by minimising the AL score function. 

If a number of VaR and ES forecasts are available at time t, the combined VaR and ES forecasts are defined as:
\begin{equation}
    \mathbf{VaR}_t(\alpha) = \sum^n_{i=1}\beta_t^i\mathbf{VaR}_t^{i}(\alpha)\label{combine VaR}
\end{equation}
where $n$ is the number of VaR forecast models available at time t, and $\beta_t^i$ is the weight assigned to model $i$ at time t. Expression \eqref{combine VaR} returns the combined VaR forecast at time t. 
\begin{equation}
    \mathbf{ES}_t(\alpha) = \sum^h_{j=1}\gamma_t^j\mathbf{ES}_t^{j}(\alpha)\label{combine ES}
\end{equation}
where $h$ is the number of ES forecast models available at time t, and $\gamma_t^j$ is the weight assigned to model $j$ at time t. Expression \eqref{combine ES} returns the combined ES forecast at time t.

Clearly, two sets of weights must be estimated under the proposed combination framework. The corresponding weight estimators are defined as:
\begin{equation}
\begin{split}
    \{\boldsymbol{\beta}, \boldsymbol{\gamma}\} &= \underset{\boldsymbol{\beta}, \boldsymbol{\gamma}}{\text{argmin}}~E\Big[ -\ln\Big(\frac{\alpha-1}{\mathbf{ES}_t(\alpha)}\Big) - \frac{(r_t-\mathbf{VaR}_t(\alpha))(\alpha-\mathbf{I}(r_t\leq\mathbf{VaR}_t(\alpha)))}{\alpha\mathbf{ES}_t(\alpha)}\Big]\\ &= \underset{\boldsymbol{\beta}, \boldsymbol{\gamma}}{\text{argmin}}~\frac{1}{m}\sum^{T+m}_{t=T+1}\Big[ -\ln\Big(\frac{\alpha-1}{\mathbf{ES}_t(\alpha)}\Big) - \frac{(r_t-\mathbf{VaR}_t(\alpha))(\alpha-\mathbf{I}(r_t\leq\mathbf{VaR}_t(\alpha)))}{\alpha\mathbf{ES}_t(\alpha)}\Big]
\end{split}
\end{equation}
where $\mathbf{VaR}_t(\alpha)$ and $\mathbf{ES}_t(\alpha)$ are the combined VaR and ES forecasts, which are defined as Expressions \eqref{combine VaR} and \eqref{combine ES}, respectively.

Additionally, the following constraints are set to increase the convexity because in the forecast combination literature, the convexity restriction on the combined weights often helps to improve the forecast performance. For example, see \citet{timmermann2006forecast}. The convex weights are nonnegative and sum to one, which is denoted as:
\begin{equation}
\begin{split}
    &\sum^n_{i=1}\beta^i_t = 1,\quad \sum^h_{j=1}\gamma_t^{j} = 1\\
    &\beta^i_t\geq0,\quad \gamma^j_t\geq0
\end{split}
\end{equation}
Furthermore, one additional inequality constraint is enforced to avoid the crossing between the combined VaR and ES. We avoid the crossing by employing a straightforward function, as shown in Expression \eqref{crossing}.
\begin{equation}
\sum^h_{j=1}\gamma_t^j\mathbf{ES}_t^{j}(\alpha)\leq\sum^n_{i=1}\beta_t^i\mathbf{VaR}_t^{i}(\alpha)\label{crossing}
\end{equation}

Therefore, the proposed method combines multiple VaR and ES forecasts by constructing a convex optimisation problem, and the weights are estimated by solving the convex optimisation problem.

\chapter{Empirical Study}\label{empirical}
\section{Data Description and Analysis}
Five hourly cryptocurrency market prices are analysed: Bitcoin (BTC); Bitcoin Cash (BCH); Ethereum Classic (ETC); Ethereum (ETH) and Litecoin (LTC). Hourly price data from 27 October 2017 19:00 to 03 May 2018 09:00 are obtained from Binance, one of the largest cryptocurrency trading platforms. One reason we use data from this period is that the data before this period may not reflect the recent characteristics and risks of the cryptocurrency market.

\begin{figure}[H]
\begin{center}
\caption{Hourly Prices from 27/10/2017 19:00 to 28/10/2017 19:00}
\includegraphics[width= \textwidth,height=0.6\textheight]{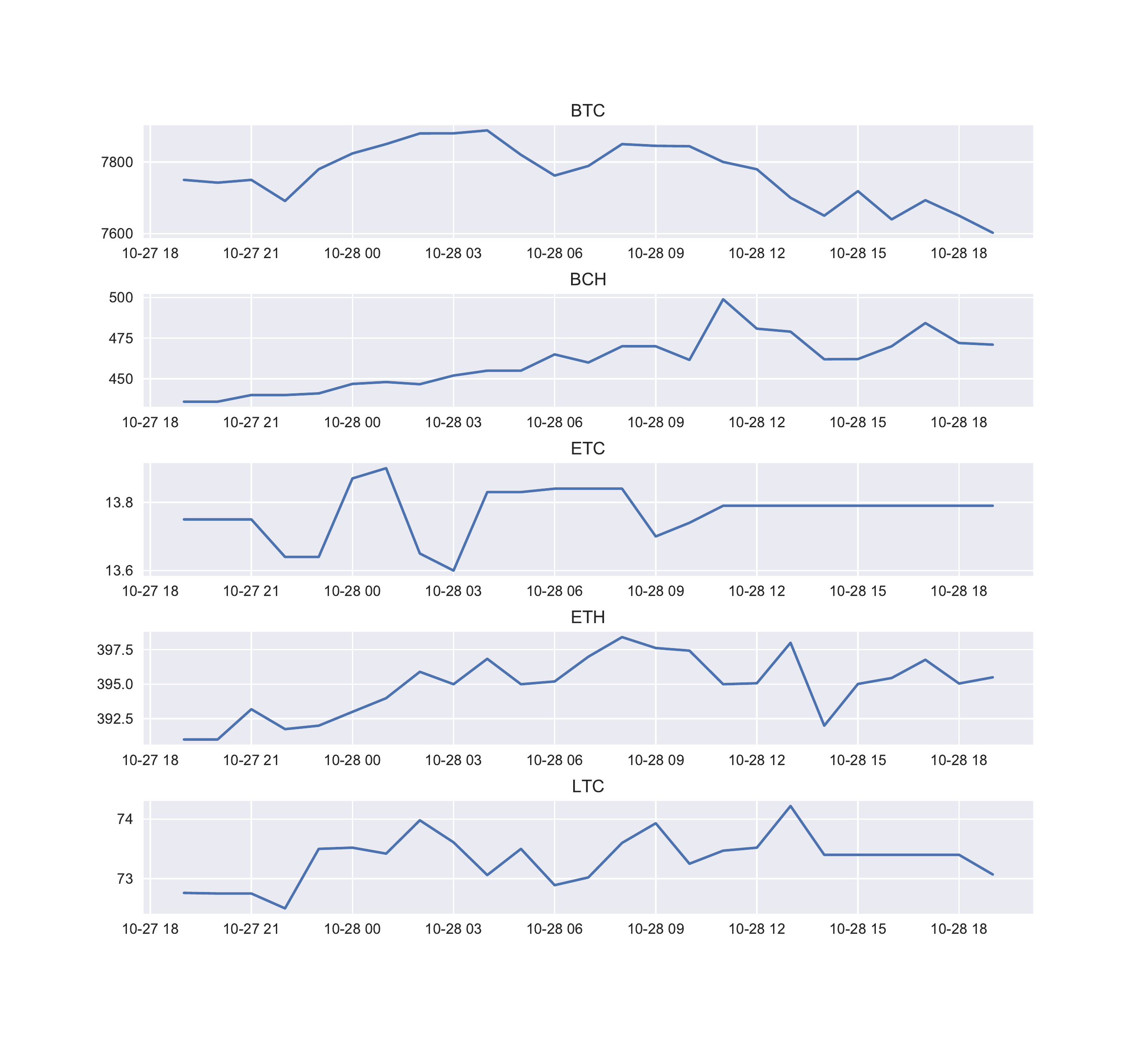}
\label{hourly}
\end{center}
\end{figure}

Furthermore, we use hourly data to analyse the risk characteristics of the given cryptocurrencies since the high-frequency data contains more information than do daily data. Moreover, the cryptocurrency market may present higher intraday volatility than traditional financial stock markets. Figure \ref{hourly} plots the intraday prices for the five cryptocurrencies from 27 October 2017 19:00 to 28 October 2017 19:00 (the first 25 observations of each data set). Fluctuating price movements are clearly observable, and the volatility is substantial.  

With the hourly price data, the percentage log return series are defined and generated as:
\begin{equation*}
    y_t = \Big(\ln(P_t) - \ln(P_{t-1})\Big) \times 100
\end{equation*}
where $P_t$ is the (closing) price in hour t. 
Table \ref{Descriptive} summarises the descriptive statistics of the log return series for the five specified cryptocurrencies. Each series has the same amount of sample data, that is, $T + m = 4502$, where T is the estimation sample and m is the forecasting period, which were defined in Section \ref{methodology}. The values of $T$ and $m$ are specified in the next section, \textit{Forecasting Study}. The last two indicators, skewness and kurtosis, reflect the distribution behaviour. The log return series of all the cryptocurrencies exhibit positive skewness (but skewness values of less than 1), which indicates that the tail on the right side of the distribution is slightly longer than that on the left side. Moreover, the five cryptocurrencies also exhibit high kurtosis values, which indicates that the corresponding distributions tend to have heavy tails or many outliers. For example, the kurtosis value of ETH is 17.010, indicating that the distribution of the ETH log return series has heavy tails compared to those other cryptocurrencies and the normal distribution (with a kurtosis value $\approx 3$). Moreover, the plots of the autocorrelation function (ACF) and partial autocorrelation function (PACF) for the squared returns show that the squared return (defined as $((r_t/r_{t-1})-1)^2$) series of all five cryptocurrencies exhibit significant autocorrelation (see Appendix \ref{ACF and PACF}). This result is consistent with our expectation, as autocorrelation is a characteristic of financial time series \citep{10.2307/2350752}. Therefore, cryptocurrencies exhibit the characteristics that similar to those of traditional financial time series, except cryptocurrencies exhibit positive skewness while most traditional financial assets, such as stock returns, exhibit negative skewness.

\begin{table}[H]
  \centering
  \caption{Descriptive Statistics Summary Table}
    \begin{tabular}{lccccc}
    \toprule
          & BTC   & BCH   & ETC   & ETH   & LTC \\
    \midrule
    count    & 4502.000 & 4502.000 & 4502.000 & 4502.000 & 4502.000 \\
    mean        & 0.010 & 0.033 & 0.017 & 0.020 & 0.023 \\
    std          & 1.518 & 2.630 & 2.472 & 1.715 & 2.085 \\
    min       & -14.550 & -25.412 & -21.504 & -19.060 & -19.547 \\
    25\%  & -0.595 & -1.127 & -1.063 & -0.627 & -0.849 \\
    50\%  & 0.000 & 0.000 & 0.000 & 0.030 & 0.000 \\
    75\%  & 0.638 & 1.065 & 1.103 & 0.729 & 0.877 \\
    max         & 14.377 & 20.909 & 20.320 & 20.041 & 21.853 \\
    skewness & 0.610 & 0.068 & 0.361 & 0.548 & 0.484 \\
    kurtosis & 13.650 & 11.109 & 8.831 & 17.010 & 14.014 \\
    \bottomrule
    \bottomrule
    \end{tabular}%
  \label{Descriptive}%
\end{table}%
A unit root test called the augmented Dickey-Fuller (ADF) test, proposed by \citet{dickey1981likelihood}, are employed to assess the stationary property of the given log return series. The ADF test is relatively robust to sample size compared to other unit root tests \citep{fedorova2016selection}. The null hypothesis of the ADF test is that the given time series has a unit root, that is, the time series exhibits some time-dependent structure and is not stationary. Thus, a stationary time series will reject the ADF test. The critical values of the ADF test depend on the sample size and trend, and the critical values are all $> -5$. We applied the ADF test to the five log return series; the test results are summarised in Table \ref{ADF}. The null hypothesis was rejected in each time series, which indicates the hourly log return series of the five cryptocurrencies are stationary. 

\begin{table}[H]
  \centering
  \caption{Augmented Dickey-Fuller Test Results Summary Table}
    \begin{tabular}{cccccc}
    \toprule
          & BTC   & BCH   & ETC   & ETH   & LTC \\
    \midrule
    ADF Statistic & -12.6124 & -12.1315 & -13.0000 & -12.9536 & -12.9536 \\
    ADF p-value & 0.0000 & 0.0000 & 0.0000 & 0.0000 & 0.0000 \\
    Test Result & Reject & Reject & Reject & Reject & Reject \\
    \bottomrule
    \bottomrule
    \end{tabular}%
  \label{ADF}%
\end{table}%

\section{Forecasting Study}
\begin{figure}[H]
\begin{center}
\caption{Log Return Plots}
\includegraphics[width=\textwidth,height=0.5\textheight]{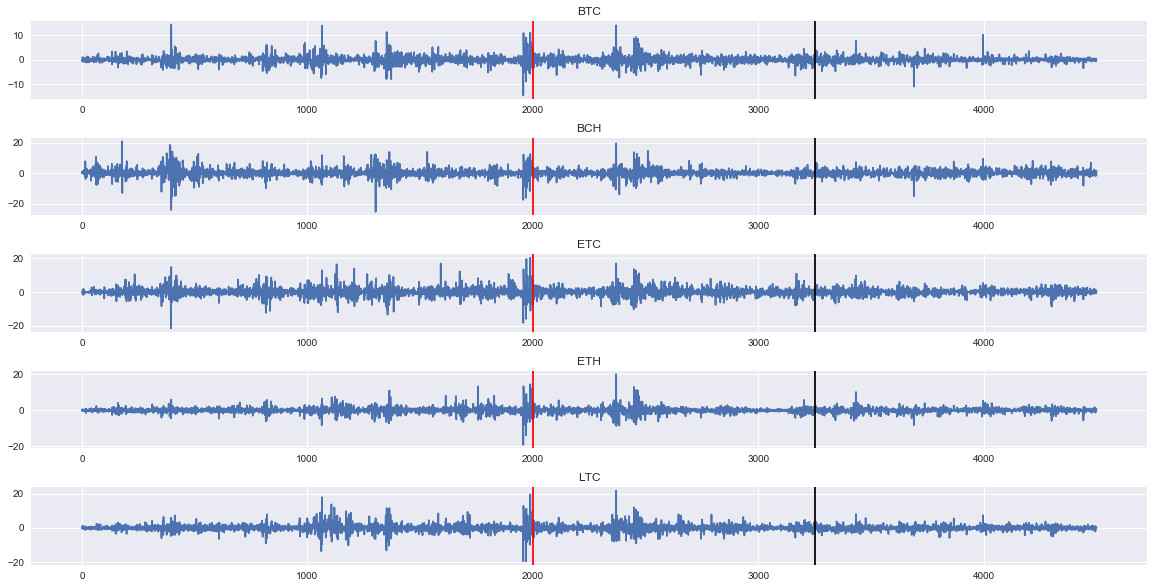}
\label{LogReturn}
\end{center}
\textit{Notes:} The red vertical line splits each data set into an estimation sample and forecast sample for the individual models. The black vertical line splits the data set on the right side of the red line into an estimation sample and forecast sample for the forecast combination methods.
\end{figure}

The one-step-ahead VaR and ES forecasts, incorporating a rolling window with initial window size $T = 2000$, are generated for each hour in the forecasting sample with size $m = 2502$ in each series, using a range of competing models, including the nonparametric, semiparametric and parametric models specified in Section \ref{methodology}. Figure \ref{LogReturn} plots the log return series for the five cryptocurrencies, where the red vertical line in each series splits the data sets into the first estimation period (or window) and the forecast period.
We employed the HS method, denoted as HS-168, as the nonparametric method. The estimation sample size used for the HS method is 168, the last seven trading days (7*24 = 168 hours). This sample size is different from those of the other methods, which use an estimation sample size of 2000. 

Taylor's CAViaR-ES and CARE frameworks are used as the semiparametric methods. We employed the CAViaR-ES-SAV-AR and CAViaR-ES-AS-EXP models; that is, the VaR forecasts are generated by the CAViaR-SAV and CAViaR-AS models, and the AR and AS formulas are employed to connect the ES forecasts with the corresponding VaR forecasts. For the CARE frameworks, we employed the CARE-SAV and CARE-AS models. All the models were specified in Section \ref{methodology}.

The EWMA, GARCH, GJRGARCH and EGARCH models are applied as the parametric methods to generate the one-step-ahead VaR and ES forecasts for the five cryptocurrencies. Furthermore, as these parametric methods require a distribution assumption for the residuals, we employed the standard normal distribution, standardised Student's t-distribution and standard skewed Student's t-distribution. Those models are denoted as EWMA, GARCH-T, GARCH-SKT, GJRGARCH-T, GJRGARCH-SKT, EGARCH-N, EGARCH-T and EGARCH-SKT.

A total of 2502 VaR forecasts and 2502 ES forecasts are generated (for HS-168, we simply take the last 2502 forecasts for consistency). We employed two standard forecast combination approaches as benchmarks: the simple average and the median forecast methods. Furthermore, the proposed joint combination method (denoted as proposed method) is applied in each data set by incorporating the rolling windows approach with a window size $n = 1251$ (50\% of 2502 forecasts), resulting in $1251$ combined VaR forecasts and $1251$ combined ES forecasts produced jointly. 

To compare the performance of each method, we applied the evaluation procedures specified in Section \ref{backtesting} to assess the final 1200 VaR forecasts and 1200 ES forecasts of each method. We tested the last 1200 forecasts instead of the last 1251 forecasts to obtain integer violation numbers that are easy to compare. For example, for the forecasts at 95\% and 99\% confidence levels, the expected violation numbers are 60 ($1200\times 5\%$) and 12 ($1200 \times 1\%$), respectively.

Moreover, the GARCH family models with the normal distribution and Student's t-distribution specifications were implemented in Python by using the ARCH package developed by Kevin Sheppard. The skewed Student's t-distribution and the other specified models, including the proposed joint combination framework, were all developed by the author because corresponding Python packages are not available. Additionally, for the evaluation procedures, all the tests were developed by the author and implemented in Python. Those python codes will be merged to the ARCH package in the future as one contribution of this thesis.

\subsection{Forecasting VaR}
\subsubsection*{VaR Forecasts at the 95\% Confidence Level}

\begin{table}[H]
\begin{center}
\caption{Forecast Comparison of the VaR Violation Ratio at the 95\% Confidence Level}
\resizebox{\linewidth}{!}{%
\begin{tabular}{lcccccccc}
    \toprule
    Model & \multicolumn{1}{l}{BTC} & \multicolumn{1}{l}{BCH} & \multicolumn{1}{l}{ETC} & \multicolumn{1}{l}{ETH} & \multicolumn{1}{l}{LTC} & \multicolumn{1}{l}{Mean} & \multicolumn{1}{l}{Median} & \multicolumn{1}{l}{RMSE} \\
    \midrule
    EWMA  & \textbf{0.250} & 0.700 & \textbf{0.400} & \textbf{0.517} & \textbf{0.317} & \textbf{0.437} & \textbf{0.400} & \textbf{0.563} \\
    HS-168 & \textcolor[rgb]{ 1,  0,  0}{0.883} & 0.850 & 1.150 & 1.200 & 1.250 & 1.067 & 1.150 & 0.173 \\
    GARCH-T & 0.833 & 1.017 & 0.867 & \framebox{1.000} & 0.867 & 0.917 & 0.867 & 0.090 \\
    GARCH-SKT & 0.833 & 1.017 & 0.883 & 0.900 & 0.883 & 0.903 & 0.883 & 0.103 \\
    GJRGARCH-T & 0.867 & 1.100 & \textcolor[rgb]{ 1,  0,  0}{0.900} & 0.950 & \textcolor[rgb]{ 1,  0,  0}{0.900} & \framebox{0.943} & \textcolor[rgb]{ 1,  0,  0}{0.900} & 0.097 \\
    GJRGARCH-SKT & 0.850 & 1.050 & 0.883 & 0.883 & 0.883 & 0.910 & 0.883 & 0.110 \\
    EGARCH-N & 0.667 & 0.800 & 0.800 & 0.833 & 0.700 & 0.760 & 0.800 & 0.240 \\
    EGARCH-T & \textcolor[rgb]{ 1,  0,  0}{0.883} & 1.017 & 0.883 & \textcolor[rgb]{ 1,  0,  0}{0.983} & \textcolor[rgb]{ 1,  0,  0}{0.900} & \textcolor[rgb]{ 1,  0,  0}{0.933} & \textcolor[rgb]{ 1,  0,  0}{0.900} & \textcolor[rgb]{ 1,  0,  0}{0.073} \\
    EGARCH-SKT & 0.867 & \framebox{1.000} & \framebox{0.933} & 0.867 & \textcolor[rgb]{ 1,  0,  0}{0.900} & 0.913 & \textcolor[rgb]{ 1,  0,  0}{0.900} & 0.087 \\
    CARE-SAV & \framebox{0.900} & 1.783 & 1.217 & 1.217 & 1.167 & 1.257 & 1.217 & 0.297 \\
    CARE-AS & 0.767 & \textbf{1.833} & 1.217 & 1.167 & 1.133 & 1.223 & 1.167 & 0.317 \\
    CAViaR-ES-SAV-AR & 0.333 & 0.933 & 0.450 & 0.700 & 0.517 & 0.587 & 0.517 & 0.413 \\
    CAViaR-ES-AS-EXP & 0.417 & 1.033 & 0.500 & 0.633 & 0.600 & 0.637 & 0.600 & 0.377 \\
          &       &       &       &       &       &       &       &  \\
    Simple Average & 0.683 & \framebox{1.000} & 0.783 & 0.817 & 0.750 & 0.807 & 0.783 & 0.193 \\
    Median Forecast & 0.783 & \textcolor[rgb]{ 1,  0,  0}{0.983} & 0.850 & 0.900 & 0.833 & 0.870 & 0.850 & 0.130 \\
          &       &       &       &       &       &       &       &  \\
    Proposed Joint Method & \textcolor[rgb]{ 1,  0,  0}{0.883} & \framebox{1.000} & \textcolor[rgb]{ 1,  0,  0}{0.900} & 1.017 & \framebox{0.917} & \framebox{0.943} & \framebox{0.917} & \framebox{0.063} \\
\bottomrule
\bottomrule
\end{tabular}}
\label{VR95}
\end{center}
\footnotesize{\textit{Notes:} Boxes indicate the best model for each currency, red shading indicates the 2nd best model for each currency, bold indicates the worst model for each currency. `RMSE' stands for the square root of the average squared difference between the violation ratio for each currency and the target value of 1.}
\end{table}

Tabel \ref{VRate} summarises the violation ratios for the 95\% VaR forecasts, as well as the mean and median of the ratios across the five cryptocurrency markets, for each model. As illustrated in Section \ref{backtesting}, the target value of the VaR violation ratio is 1. For the BTC market, all the models produce violation ratios less than 1, that is, all 16 methods overestimate the risk exhibited by the BTC market. The violation ratios from the parametric models, except for EGARCH-N (0.667), are similar. This result is not surprising because EGARCH-N is the only GARCH family model that incorporates the standard normal distribution, which may not be consistent with the true distribution of the BTC market (see Table \ref{Descriptive} and Figure \ref{Distribution Plots} in Appendix \ref{AppendixA}). Furthermore, the CARE-SAV model (0.900) produces the best forecast in terms of the violation ratio while the EWMA method (0.250) produces the worst forecast, as it also assumes that the underlying financial asset follows a standard normal distribution. HS-168, EGARCH-T and the proposed method produce the second best forecasts with a violation ratio of 0.883. Additionally, the CAViaR-ES joint models tend to highly overestimate the risk exhibited by the market. Recall that HS-168 produces the 2nd best forecast with a violation ratio of 0.883. One possible reason is that it uses information from the past 168 hours (1 week), which can provide a better representation of the current risk level of the BTC market than can the past 2000 hours, which are used by the other methods. 

For the BCH market, the violation ratios produced by each model are much better (closer to the target value of 1) than they are in the BTC market, except for the CARE-SAV and CARE-AS. The CARE-SAV and CARE-AS models produce VaR forecasts that are highly anti-conservative, with violation ratios equal to 1.783 and 1.833, respectively. Both models tend to underestimate the risk exhibited by the market. The CARE-AS model has the worst performance, with a violation ratio of 1.833. EGARCH-N also tends to overestimate the risk exhibited by the market, with a violation ratio of 0.800. Moreover, EGARCH-SKT, simple average and the proposed method produce the best VaR forecasts, with violation ratios exactly equal to the target ratio of 1. The median forecast method produces the second best forecast, with a violation ratio of 0.983.

For the ETC market, every model, except HS-168, CARE-SAV and CARE-AS, is conservative; that is, they tend to overestimate the risk exhibited by the market with a violation ratio less than 1. EGARCH-SKT generates the best VaR forecast, with a corresponding violation ratio of 0.933, whereas the EWMA method performs the worst, with a violation ratio of 0.517. The proposed method produces the second best VaR forecast, with a violation ratio of 0.900. 

For the ETH market, GARCH-T produces the best VaR forecast, with a violation ratio equal to the target value of 1. EGARCH-T (0.983) produces the second best VaR forecast, and the EWMA method (0.517) produces the worst forecast and tends to overestimate the risk level. Although the proposed method produces the third best forecast, with a violation ratio of 1.017, only a slight difference exists between the proposed method and EGARCH-T (violation ratio of 0.983). Moreover, GARCH-T, GJRGARCH-T and EGARCH-T all produce relatively favourable forecasts, which suggests that Student's t-distribution fits the ETH market well and is a good representation of the true distribution of the market. 

For the LTC market, the proposed method produces the best VaR forecast, with a violation ratio of 0.917. The GJRGARCH-T, EGARCH-T and EGARCH-SKT models produce the second best VaR forecasts, with violation ratios all equal to 0.900. Furthermore, all the applied methods, except HS-168, CARE-SAV and CARE-AS, tend to overestimate the risk exhibited by the market.

Therefore, for the 95\% VaR forecasts, EWMA, EGARCH-N, CAViaR-ES-SAV-AR and CAViaR-ES-AS-EXP models are all clearly inaccurate at the 95\% confidence level in terms of the violation ratio, with a target value of 1, across all five cryptocurrency markets. The mean, median and RMSE show the same results. For example, the EWMA method has the highest RMSE (0.563), that is, the average distance between the violation ratios and the target value of 1 is the largest, indicating the high inaccuracy of the VaR forecasts of the EWMA method. By contrast, the proposed method has the lowest RMSE (0.063), which indicates the relatively high accuracy of the VaR forecast. EGARCH-T has the second lowest RMSE (0.073), as EGARCH-T is quite stable across the five markets. Generally, the proposed method produces relatively stable and accurate VaR forecasts at the 95\% confidence level: the proposed method performs the best in two markets, the second best in two markets and the third best in one market. Furthermore, the proposed method always performs the best in terms of the mean, median and RMSE.

However, a model with a violation ratio close to the target value of 1 is a necessary but not sufficient condition to guarantee the forecast accuracy, as it only ensures that the forecasts are consistent with the theoretical expectation of the violation ratio. 
Therefore, the results of the application of four standard tests and one loss function are summarised in Table \ref{test95} and Table \ref{loss95} to further assess the forecast performances.

\begin{figure}[H]
\begin{center}
\caption{95\% VaR Forecasts for the BTC}
\includegraphics[width=\textwidth,height=0.3\textheight]{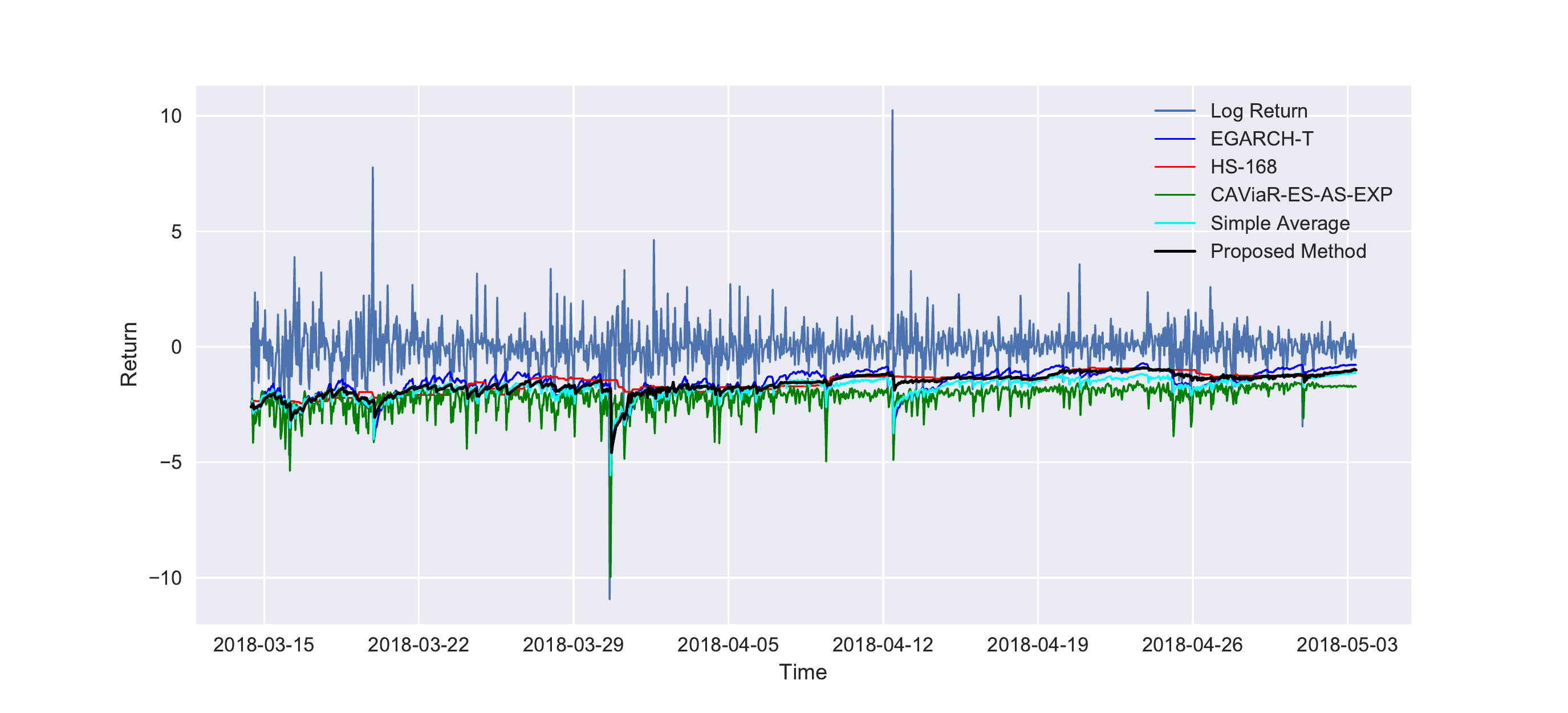}
\label{BTC_VaR_95}
\end{center}
\end{figure}

Regarding the four standard statistical tests conducted at the 5\% significance level, the VaR forecasts produced by the EWMA method were rejected the most times by each test, across the five markets. This result is consistent with expectations, as those forecasts perform the worst in terms of the violation ratio. Moreover, the CAViaR-ES joint models also perform poorly as the forecasts were rejected more than 3 times by each test and were rejected more times in each market. For example, the CAViaR-ES-SAV-AR model is rejected 4, 4, 4 and 3 times by the UC, CC, DQ1 and DQ4 tests, respectively. In general, more than one-half of the models perform well in the UC and CC tests, whereas fewer models perform well in the DQ1 and DQ4 tests because the DQ test is more powerful than the UC and CC tests. Both the proposed method and the GJRGARCH-T model perform well: they pass all the tests across all five markets. The second best models are the GJRGARCH-SKT model and the median forecast method, which are rejected only 1 time by the DQ test.

\begin{table}[H]
\begin{center}
\caption{Counts of Model Rejections by the Statistical Tests for the 95\% VaR Forecasts}
\resizebox{0.4\textheight}{!}{%
\begin{tabular}{lccccc}
\toprule
    Model & \multicolumn{1}{c}{UC} & \multicolumn{1}{c}{CC} & \multicolumn{1}{c}{DQ1} & \multicolumn{1}{c}{DQ4} & \multicolumn{1}{c}{Total} \\
    \midrule
    EWMA  & \textbf{5} & \textbf{5} & 4     & \textbf{4} & \textbf{5} \\
    HS-168 & \framebox{0}     & \framebox{0}     & 2     & \textcolor[rgb]{ 1,  0,  0}{1} & 2 \\
    GARCH-T & \framebox{0}     & \framebox{0}     & \textcolor[rgb]{ 1,  0,  0}{1} & \textcolor[rgb]{ 1,  0,  0}{1} & 2 \\
    GARCH-SKT & \framebox{0}     & \framebox{0}     & \textcolor[rgb]{ 1,  0,  0}{1} & \textcolor[rgb]{ 1,  0,  0}{1} & 2 \\
    GJRGARCH-T & \framebox{0}     & \framebox{0}     & \framebox{0}     & \framebox{0}     &\framebox{0} \\
    GJRGARCH-SKT & \framebox{0}     & \framebox{0}     & \textcolor[rgb]{ 1,  0,  0}{1}     & \framebox{0}     & \textcolor[rgb]{ 1,  0,  0}{1} \\
    EGARCH-N & \textcolor[rgb]{ 1,  0,  0}{1} & \framebox{0}     & 3     & \textcolor[rgb]{ 1,  0,  0}{1} & 3 \\
    EGARCH-T & \framebox{0}     & \framebox{0}     & 2     & 2     & 3 \\
    EGARCH-SKT & \framebox{0}     & \framebox{0}     & \textcolor[rgb]{ 1,  0,  0}{1} & \textcolor[rgb]{ 1,  0,  0}{1} & 2 \\
    CARE-SAV & \framebox{0}     & \textcolor[rgb]{ 1,  0,  0}{1} & 2     & 2     & 2 \\
    CARE-AS & \framebox{0}     & \textcolor[rgb]{ 1,  0,  0}{1} & 2     & 2     & 2 \\
    CAViaR-ES-SAV-AR & 4     & 4     & 4     & 3     & \textbf{5} \\
    CAViaR-ES-AS-EXP & 4     & 4     & \textbf{5} & 3     & \textbf{5} \\
          &       &       &       &       &  \\
    Simple Average & 2     & \textcolor[rgb]{ 1,  0,  0}{1} & \textcolor[rgb]{ 1,  0,  0}{1} & \framebox{0}     & 2 \\
    Median Forecast & \framebox{0}     & \framebox{0}     & \framebox{0}     & \textcolor[rgb]{ 1,  0,  0}{1} & \textcolor[rgb]{ 1,  0,  0}{1} \\
          &       &       &       &       &  \\
     Proposed Joint Method & \framebox{0}     & \framebox{0}     & \framebox{0}     &\framebox{0}     & \framebox{0} \\
\bottomrule
\bottomrule
\end{tabular}}
\label{test95}
\end{center}
\footnotesize{\textit{Notes:} Boxes indicate the best model, red shading indicates the 2nd best model, and bold indicates the worst model. `Total' indicates the number of markets in which the model was rejected by at least one test. 
} 
\end{table}

The quantile loss function is employed to assess the accuracy of the VaR forecasts by considering the magnitude of the difference between the VaR forecasts and the corresponding return series. Table \ref{loss95} summarises the quantile loss function values of the VaR forecasts from each model across all five markets. For the BTC market, the EWMA method produces the worst VaR forecasts with the highest loss function value of 161.148, and the CAViaR-ES joint models perform the second worst, with corresponding loss function values of 155.95 and 153.328. This result is consistent with the violation ratios and statistical tests results, where the EWMA method is rejected either 4 or 5 times by the four formal tests across the five markets. Additionally, Figure \ref{BTC_VaR_95} plots several VaR forecasts of the test period (the last 1000 observations of log returns) and clearly explains why the EWMA method performs worse under those evaluation procedures. The VaR forecasts produced by the EWMA method (red line) are far from the return series, which results in a low violation ratio and high loss function value. The risk level exhibited in the BTC market is highly overestimated by the EWMA method and the CAViaR-ES joint models. However, the CAViaR-ES joint models are the only models that attempt to capture extreme returns. Moreover, all the other individual models, except EGARCH-N, produce VaR forecasts with similar loss function values of approximately 138.00 to 140.00. The proposed method performs the best, with the lowest quantile loss function value of 137.542, followed by HS-168 and CARE-AS, which have the 2nd and 3rd best performance, respectively.

For the BCH market, the individual model performances are similar to one another in terms of the loss function values. The CARE-AS model performs the worst, with a loss function value of 275.337. Similarly, the other three semiparametric models all have loss function values greater than 265. The three forecast combination approaches perform well: the proposed method has the lowest loss function value of 250.108, followed by the simple average (253.653) and median forecast (255.024).

For the ETC market, the proposed method has the lowest loss function value of 227.437, followed by GARCH-SKT (227.486) and GARCH-T (227.666). The CAViaR-ES joint models and the EWMA method perform considerably worse than the other approaches, as indicated by the higher loss function values. The simple average and median forecast approaches are both ranked among the top performers.

For the ETH market, the semiparametric and parametric models produce VaR forecasts with similar loss function values, and the EWMA method performs the worst with a loss function value of 201.509. The three forecast combination methods perform well: the simple average method has the lowest loss function value of 178.625, followed by the median forecast (180.118) and the proposed method (180.207).

For the LTC market, similar to the performances in the ETC market, the proposed method performs the best, with a loss function value of 179.087, followed by GARCH-SKT and GARCH-T, with loss function values of 179.539 and 179.647, respectively. The EWMA method has the highest loss function value of 210.186, and the CAViaR-ES joint models perform as poorly as they do in other market.

\begin{table}[H]
\begin{center}
\caption{Quantile Loss Function for the 95\% VaR Forecasts}
\resizebox{0.7\linewidth}{!}{%
\begin{tabular}{lccccc}
    \toprule
    Model & \multicolumn{1}{l}{BTC} & \multicolumn{1}{l}{BCH} & \multicolumn{1}{l}{ETC} & \multicolumn{1}{l}{ETH} & \multicolumn{1}{l}{LTC} \\
    \midrule
    EWMA  & \textbf{161.148} & 267.943 & \textbf{255.850} & \textbf{201.509} & \textbf{210.186} \\
    HS-168 & \textcolor[rgb]{ 1,  0,  0}{138.006} & 263.544 & 242.394 & 194.412 & 189.966 \\
    GARCH-T & 140.942 & 257.341 & \textcolor[rgb]{ 0,  .439,  .753}{227.666} & 181.259 & \textcolor[rgb]{ 0,  .439,  .753}{179.647} \\
    GARCH-SKT & 141.258 & 257.177 & \textcolor[rgb]{ 1,  0,  0}{227.486} & 182.149 & \textcolor[rgb]{ 1,  0,  0}{179.539} \\
    GJRGARCH-T & 139.431 & 261.688 & 229.253 & 180.455 & 179.719 \\
    GJRGARCH-SKT & 139.926 & 261.797 & 229.437 & 181.737 & 179.830 \\
    EGARCH-N & 143.987 & 258.109 & 231.661 & 181.897 & 184.690 \\
    EGARCH-T & 138.752 & 256.870 & 230.621 & 180.817 & 181.008 \\
    EGARCH-SKT & 139.062 & 256.579 & 230.405 & 181.777 & 180.858 \\
    CARE-SAV & 138.916 & 272.917 & 232.933 & 183.957 & 185.268 \\
    CARE-AS & \textcolor[rgb]{ 0,  .439,  .753}{138.136} & \textbf{275.337} & 232.261 & 187.989 & 187.206 \\
    CAViaR-ES-SAV-AR & 155.995 & 266.213 & 253.110 & 185.514 & 198.067 \\
    CAViaR-ES-AS-EXP & 153.328 & 268.959 & 249.299 & 188.684 & 197.811 \\
          &       &       &       &       &  \\
    Simple Average & 140.156 & \textcolor[rgb]{ 1,  0,  0}{253.653} & 228.460 & \framebox{178.625} & 181.886 \\
    Median Forecast & 138.842 & \textcolor[rgb]{ 0,  .439,  .753}{255.024} & 227.780 & \textcolor[rgb]{ 1,  0,  0}{180.118} & 180.375 \\
          & \multicolumn{5}{c}{} \\
    Proposed Method & \framebox{137.542} & \framebox{250.108} & \framebox{227.437} & \textcolor[rgb]{ 0,  .439,  .753}{180.207} & \framebox{179.087} \\
\bottomrule
\bottomrule
\end{tabular}}
\label{loss95}
\end{center}
\footnotesize{\textit{Notes:} Boxes indicate the best model in each market, red shading indicates the 2nd best model and blue shading indicates the 3rd best model. Bold indicates the worst model.
}
\end{table}

In summary, for the VaR forecasts at the 95\% confidence level, the EWMA method always performs the worst, except in the BCH market. The reason for the poor performance is straightforward: the EWMA method assumes that the return series of the underlying currencies follow a standard normal distribution, which is inconsistent with reality. Additionally, it estimates the volatility with the empirical data instead of forecasting the true volatility. Therefore, the method is hard capture the tailed behaviour of the underlying currencies. Similarly, the HS-168 method is not competitive. This result is consistent with expectations, as HS-168 uses the empirical distribution instead of estimating the true distribution of the given return series and relies strongly on the choice of window size. For example, HS-168 performs well in the BTC market but performs poorly in the other four markets, which indicates that the window size of 186 hours is representative of the current market risk level of the BTC market but not for the other four markets. Moreover, for the performances of the four semiparametric models are not as competitive as those of the GARCH family models (parametric models), possibly because the distribution assumption in the GARCH family models fits the true distributions of the given currencies well, resulting in low forecast errors and quantile loss function values. For example, the models that incorporate Student's-t and skewed Student's t-distributions always perform better than EGARCH-N, which assumes the residual series follow a normal distribution. The semiparametric methods may benefit from the lack of distribution restriction in some cases but are highly subject to the bias-variance tradeoff and dependent on the fitness of measurement equation. For example, Figure \ref{BTC_VaR_95} plots the 95\% confidence level VaR forecasts for one semiparametric model (CAViaR-ES-AS-EXP) and one nonparametric model (HS-168). Clearly, both HS-168 and CAViaR-ES-AS-EXP generate series of VaR forecasts with high forecast errors (far away from the return series) and, consequently, high loss function values. Additionally, although CAViaR-ES-AS-EXP tends to capture the dynamics of the return series, the measurement equation may not be the true equation of the time series process and produces worse forecast performances.

\begin{figure}[H]
\begin{center}
\caption{Individual Weights of the Combined VaR Forecasts at the 95\% Confidence Level for the BTC Market}
\includegraphics[width=\textwidth,height=0.3\textheight]{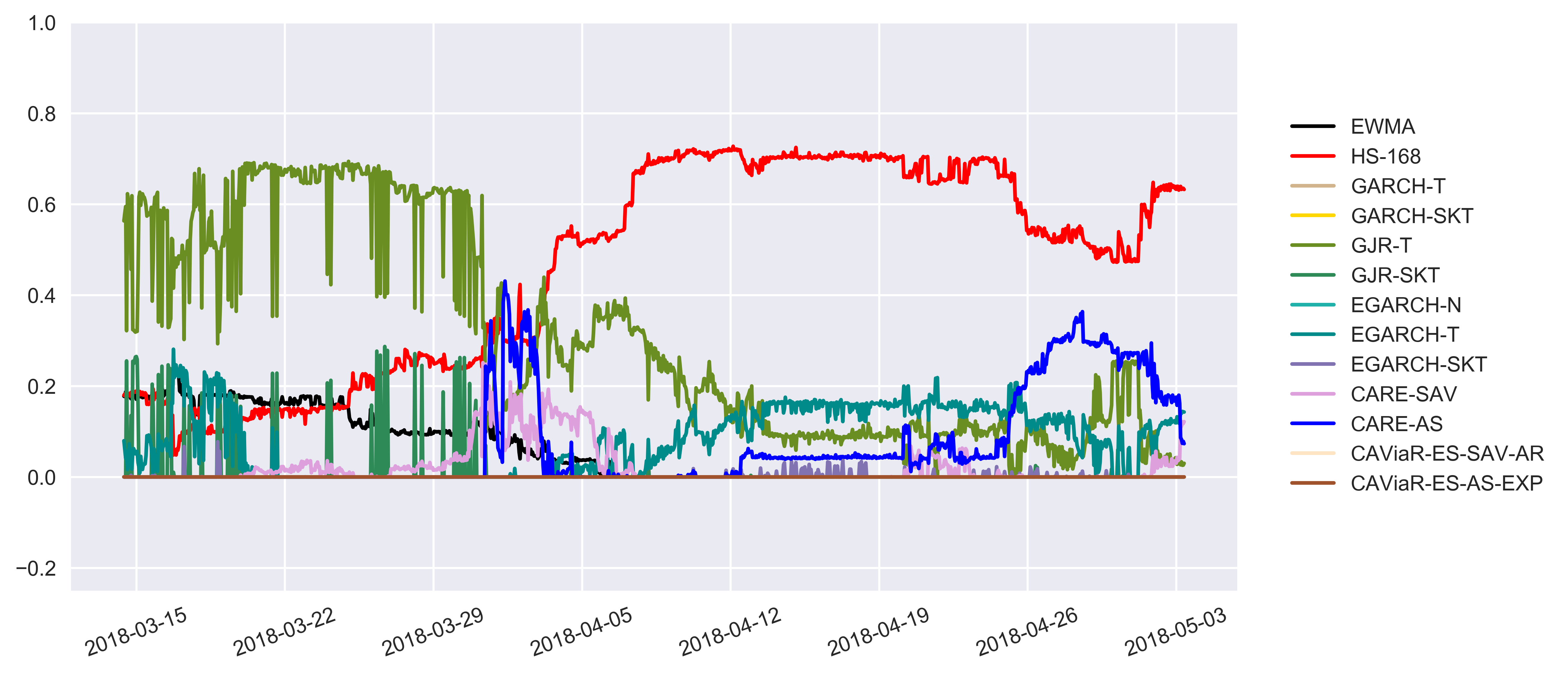}
\label{BTC_VaR_95_Weight}
\end{center}
\end{figure}

Furthermore, the forecast performances of the applied forecast combination approaches are relatively favourable and stable across each market. The simple average and median forecast approaches are stable and robust to the different currency markets, although they may not perform the best. This result is consistent with \cite{mcneil2005quantitative}, who suggest that the forecast combination approaches, especially the median forecast method, performed better during the global financial crisis. In contrast to the simple average and median forecast, which combine the individual forecasts by either taking a simple average or taking the 50th percentile after reordering the individual forecasts at each point, the proposed method combines the individual forecasts by assigning the optimal weight to each individual forecast, where the weights are estimated by optimising the tick loss function (AL log score). Figure \ref{BTC_VaR_95_Weight} plots the weights assigned to individual forecasts to produce the combined forecast under the proposed method in the BTC market. Clearly, the weights assigned to individual models are not constant over time. The EWMA method (black line) sometimes has a weight of 20\%, even though it performs the worst in the BTC market. These weights help to smooth the combined VaR series (see the black line in Figure \ref{BTC_VaR_95}): both the dynamic variance and forecast error of the proposed method decrease, leading to a lower loss function value. HS-168 (red line) and CARE-AS (blue line) have large weights across time as they are the best and second best individual models, with the lowest and second lowest loss function values, respectively. The high weights assigned to those two models guarantee the forecast accuracy of the combined VaR forecasts and reduce the loss function value. The weight plots and VaR series plots of the other four markets have been placed in the Appendix to save space (see Appendix \ref{appendixB plot}).

\subsubsection*{VaR Forecasts at the 99\% Confidence Level}

\begin{table}[H]
\begin{center}
\caption{Forecast Comparison of the VaR Violation Ratios at the 99\% Confidence Level}
\resizebox{\linewidth}{!}{
\begin{tabular}{lcccccccc}
    \toprule
    Model & \multicolumn{1}{l}{BTC} & \multicolumn{1}{l}{BCH} & \multicolumn{1}{l}{ETC} & \multicolumn{1}{l}{ETH} & \multicolumn{1}{l}{LTC} & \multicolumn{1}{l}{Mean} & \multicolumn{1}{l}{Median} & \multicolumn{1}{l}{RMSE} \\
    \midrule
    EWMA  & 0.333 & 0.833 & 0.583 & 0.583 & 0.500 & 0.567 & 0.583 & 0.433 \\
    HS-168 & 0.750 & 0.900 & 1.417 & \textbf{2.583} & 1.500 & 1.430 & 1.417 & 0.570 \\
    GARCH-T & \framebox{1.000} & 0.833 & 0.667 & 1.750 & \textcolor[rgb]{ 1,  0,  0}{1.083} & 1.067 & \framebox{1.000} & 0.267 \\
    GARCH-SKT & \framebox{1.000} & 0.750 & 0.750 & 1.583 & 1.167 & 1.050 & \framebox{1.000} & 0.250 \\
    GJRGARCH-T & 1.167 & 0.833 & 0.833 & 1.917 & \framebox{1.000} & 1.150 & \framebox{1.000} & 0.283 \\
    GJRGARCH-SKT & 1.167 & 0.750 & 0.833 & 1.250 & \framebox{1.000} & \framebox{1.000} & \framebox{1.000} & 0.167 \\
    EGARCH-N & \textbf{1.750} & 1.250 & 1.500 & 2.083 & 1.417 & 1.600 & 1.500 & 0.600 \\
    EGARCH-T & 1.167 & 0.667 & \framebox{0.917} & 1.250 & \textcolor[rgb]{ 1,  0,  0}{1.083} & 1.017 & \textcolor[rgb]{ 1,  0,  0}{1.083} & 0.183 \\
    EGARCH-SKT & 1.167 & 0.667 & \textcolor[rgb]{ 1,  0,  0}{1.083} & \framebox{1.000} & \textcolor[rgb]{ 1,  0,  0}{1.083} & \framebox{1.000} & \textcolor[rgb]{ 1,  0,  0}{1.083} & \textcolor[rgb]{ 1,  0,  0}{0.133} \\
    CARE-SAV & 0.417 & 2.667 & 1.583 & 1.833 & 1.500 & 1.600 & 1.583 & 0.833 \\
    CARE-AS & 0.333 & \textbf{2.833} & \textbf{1.917} & 1.667 & \textbf{1.583} & \textbf{1.667} & \textbf{1.667} & \textbf{0.933} \\
    CAViaR-ES-SAV-AR & 0.250 & 0.583 & 0.333 & 0.500 & 0.750 & 0.483 & 0.500 & 0.517 \\
    CAViaR-ES-AS-EXP & 0.250 & \textcolor[rgb]{ 1,  0,  0}{0.917} & 0.417 & 0.417 & 0.667 & 0.533 & 0.417 & 0.467 \\
          &       &       &       &       &       &       &       &  \\
    Simple Average & 0.583 & \framebox{1.000} & 0.750 & \framebox{1.000} & \textcolor[rgb]{ 1,  0,  0}{1.083} & 0.883 & \framebox{1.000} & 0.150 \\
    Median Forecast & \framebox{1.000} & 0.833 & 0.750 & \textcolor[rgb]{ 1,  0,  0}{1.167} & \textcolor[rgb]{ 1,  0,  0}{1.083} & 0.967 & \framebox{1.000} & \textcolor[rgb]{ 1,  0,  0}{0.133} \\
          &       &       &       &       &       &       &       &  \\
    Proposed Method & \textcolor[rgb]{ 1,  0,  0}{0.900} & 1.083 & \textcolor[rgb]{ 1,  0,  0}{1.083} & \framebox{1.000} & \framebox{1.000} & \textcolor[rgb]{ 1,  0,  0}{1.013} & \framebox{1.000} & \framebox{0.053} \\

\bottomrule
\bottomrule
\end{tabular}}
\label{VRatio99}%
\end{center}
\footnotesize{\textit{Notes:} Boxes indicate the best model for each currency, red shading indicates the 2nd best model for each currency, and bold indicates the worst model for each currency. `RMSE' stands for the square root of the average squared difference between the violation ratio for each currency and the target value of 1.
}
\end{table}%
Table \ref{VRatio99} summarises the violation ratios for the 99\% VaR forecasts across the five cryptocurrency markets with a target ratio of 1. For the BTC market, the GARCH-T, GARCH-SKT and median forecast models perform the best as they produce VaR forecasts with violation ratios exactly equal to 1. The proposed method is the second performing model, with a violation ratio of 0.9; that is, it slightly overestimates the risk exhibited by the BTC market. The other parametric models tend to underestimate the risk exhibited by the BTC market as they all produce VaR forecasts with violation ratios greater than 1. Similar to the performance at the 95\% confidence level, EGARCH-N, which assumes that the underlying asset's residuals follow a standard normal distribution, is the worst performing method: it produces a VaR forecast with a violation ratio of 1.75 which is highly anti-conservative. The four semiparametric models perform as poorly as they do at the 95\% confidence level: they are highly conservative (all less than 0.42) and tend to overestimate the risk level.

For the BCH market, simple average is the best method, with a violation ratio exactly equal to 1. EGARCH-T is the second best method, with a violation ratio of 0.917, and the proposed method is the third best method, with a violation ratio of 1.083. The worst method is the CARE-AS model, with a violation ratio of 1.917. All the parametric and nonparametric models, except EGARCH-N, tend to overestimate the risk level exhibited by the BCH market as they all produce VaR forecasts with violation ratios less than one.

For the ETC market, EGARCH-T is the best model, with a violation ratio of 0.917. EGARCH-SKT and the proposed method are the second best models, with violation ratios of 1.083. By contract, the CARE-AS model is the worst model, with a violation ratio equals to 1.917, which is highly anti-conservative with 91.7\% more VaR violations than expected. Another CARE family model, CARE-SAV, also tends to underestimate the risk level. The CARE-SAV model is also highly anti-conservative, with 58.3\% more VaR violations than expected. By contrast, CAViaR-ES-SAV-AS and CAViaR-ES-AS-EXP are highly conservative, with violation ratios of 0.333 and 0.417, respectively. Moreover, the two formal forecast combination approaches produce VaR forecast series with violation ratios equal to 0.75.

For the ETH market, EGARCH-SKT, simple average and the proposed method are the best approaches, with corresponding violation ratios all equal to the target ratio of 1. Median forecast is the second best method with a violation ratio of 1.167. By contrast, the worst model is HS-168, which is highly anti-conservative, with 158.3\% more violations than expected. Moreover, the parametric models tend to underestimate the risk level exhibited by the ETH market as they are all anti-conservative.

For the LTC market, GJRGARCH-T, GJRGARCH-SKT and the proposed method are the best methods; they produce VaR forecast series with violation ratios exactly equal to 1. The GARCH-T, EGARCH-T, EGARCH-SKT, simple average and median forecast approaches are the second best methods, as they all produce VaR forecasts with identical violation ratios of 1.083. The two CARE family models perform the worst and are highly anti-conservative with violation ratios are $\geq 1.500$. The parametric models all tend to underestimate the risk level exhibited by the LTC market, and the VaR forecast series produced by the parametric models all have violation ratios greater than 1.

Therefore, the proposed method is the most favoured model in terms of the violation ratio: the proposed method performs the best in the ETH and ETC markets, the second best in the BTC and ETC markets and the third best in the BCH market. Although the proposed method is not the best in terms of the violation ratio in some markets, the corresponding VaR forecasts produced by the proposed method have stable forecast performances across the five cryptocurrency markets. Furthermore, the proposed method performs well in terms of the mean, median and RMSE, which also indicates the stability of the proposed method.

\begin{table}[H]
\begin{center}
\caption{Counts of Model Rejections by the Statistical Tests for the 99\% VaR Forecasts}
\resizebox{0.4\textheight}{!}{%
\begin{tabular}{lccccc}
\toprule
    Model & \multicolumn{1}{c}{UC} & \multicolumn{1}{c}{CC} & \multicolumn{1}{c}{DQ1} & \multicolumn{1}{c}{DQ4} & \multicolumn{1}{c}{Total} \\
    \midrule
    EWMA  & \textbf{5}     & \textbf{5}     & 4     & \textbf{4}     & \textbf{5} \\
    HS-168 & \framebox{0}     & \framebox{0}     & 2     & \textcolor[rgb]{ 1,  0,  0}{1} & 2 \\
    GARCH-T & \framebox{0}     & \framebox{0}     & \textcolor[rgb]{ 1,  0,  0}{1} & \textcolor[rgb]{ 1,  0,  0}{1} & 2 \\
    GARCH-SKT & \framebox{0}     & \framebox{0}     & \textcolor[rgb]{ 1,  0,  0}{1} & \textcolor[rgb]{ 1,  0,  0}{1} & 2 \\
    GJRGARCH-T & \framebox{0}     & \framebox{0}     & \framebox{0}     & \framebox{0}     & \framebox{0} \\
    GJRGARCH-SKT & \framebox{0}     & \framebox{0}     & \textcolor[rgb]{ 1,  0,  0}{1}     & \framebox{0}     & \textcolor[rgb]{ 1,  0,  0}{1} \\
    EGARCH-N & \textcolor[rgb]{ 1,  0,  0}{1} & \framebox{0}     & 3     & \textcolor[rgb]{ 1,  0,  0}{1} & 3 \\
    EGARCH-T & \framebox{0}     & \framebox{0}     & 2     & 2     & 3 \\
    EGARCH-SKT & \framebox{0}     & \framebox{0}     & \textcolor[rgb]{ 1,  0,  0}{1} & \textcolor[rgb]{ 1,  0,  0}{1} & 2 \\
    CARE-SAV & \framebox{0}     & \textcolor[rgb]{ 1,  0,  0}{1} & 2     & 2     & 2 \\
    CARE-AS & \framebox{0}     & \textcolor[rgb]{ 1,  0,  0}{1} & 2     & 2     & 2 \\
    CAViaR-ES-SAV-AR & 4     & 4     & 4     & 3     & 5 \\
    CAViaR-ES-AS-EXP & 4     & 4     & \textbf{5}     & 3     & 5 \\
          &       &       &       &       &  \\
    Simple Average & 2     & \textcolor[rgb]{ 1,  0,  0}{1} & \textcolor[rgb]{ 1,  0,  0}{1} & \framebox{0}     & 2 \\
    Median Forecast & \framebox{0}     & \framebox{0}     & \framebox{0}     & \textcolor[rgb]{ 1,  0,  0}{1} & \textcolor[rgb]{ 1,  0,  0}{1} \\
          &       &       &       &       &  \\
    Proposed Method & \framebox{0}     & \framebox{0}     & \framebox{0}     & \framebox{0}     & \framebox{0} \\
    \bottomrule
    \bottomrule
\end{tabular}}%
\label{99VaRtest}%
\end{center}
\footnotesize{\textit{Notes:} Boxes indicate the best model, red shading indicates the 2nd best model, and bold indicates the worst model. `Total' indicates the number of markets in which the model was rejected by at least one test.} 
\end{table}%

Table \ref{99VaRtest} summarises the results of four applied statistical tests. Clearly, GJRGARCH-T and the proposed method are the best methods for obtaining the 99\% VaR forecasts, as they were not rejected by any of the four tests across the five markets. The CAViaR-ES joint models and the EWMA method perform the worst, with the largest number of rejections. 

\begin{table}[H]
  \begin{center}
  \caption{Quantile Loss Function for the 99\% VaR Forecasts}
\resizebox{0.7\linewidth}{!}{%
    \begin{tabular}{lccccc}
    \toprule
    Model & \multicolumn{1}{l}{BTC} & \multicolumn{1}{l}{BCH} & \multicolumn{1}{l}{ETC} & \multicolumn{1}{l}{ETH} & \multicolumn{1}{l}{LTC} \\
    \midrule
    EWMA  & 49.755 & \textcolor[rgb]{ 1,  0,  0}{80.686} & 71.638 & 59.248 & 58.112 \\
    HS-168 & 47.742 & 82.142 & \textbf{76.140} & \textbf{71.354} & 56.788 \\
    GARCH-T & 49.188 & 83.505 & 67.966 & 60.980 & \textcolor[rgb]{ 1,  0,  0}{55.797} \\
    GARCH-SKT & 49.268 & 83.745 & 67.519 & 60.413 & \textcolor[rgb]{ 1,  0,  0}{55.797} \\
    GJRGARCH-T & 48.112 & 81.818 & 67.961 & 59.974 & \textcolor[rgb]{ 0,  .439,  .753}{55.882} \\
    GJRGARCH-SKT & 48.115 & 82.346 & 67.938 & 59.375 & 55.971 \\
    EGARCH-N & 48.466 & 81.018 & 69.081 & 61.231 & 57.109 \\
    EGARCH-T & 48.531 & 83.537 & 70.335 & 59.601 & 56.142 \\
    EGARCH-SKT & 48.561 & 83.937 & 70.059 & 60.011 & 56.126 \\
    CARE-SAV & 47.843 & 88.142 & 70.689 & 59.009 & 62.088 \\
    CARE-AS & \textcolor[rgb]{ 1,  0,  0}{47.070} & \textbf{92.964} & 72.342 & 59.092 & 62.039 \\
    CAViaR-ES-SAV-AR & \textbf{52.876} & 88.274 & 77.476 & 60.943 & 63.859 \\
    CAViaR-ES-AS-EXP & 50.062 & 86.517 & 72.312 & 59.368 & \textbf{64.213} \\
          &       &       &       &       &  \\
    Simple Average & \textcolor[rgb]{ 0,  .439,  .753}{47.232} & \textcolor[rgb]{ 0,  .439,  .753}{80.845} & \textcolor[rgb]{ 0,  .439,  .753}{67.346} & \framebox{57.037} & 56.404 \\
    Median Forecast & 47.577 & 81.315 & \framebox{66.977} & \textcolor[rgb]{ 0,  .439,  .753}{57.990} & 55.978 \\
          & \multicolumn{5}{c}{} \\
    Proposed Method & \framebox{46.959} & \framebox{80.447} & \textcolor[rgb]{ 1,  0,  0}{67.102} & \textcolor[rgb]{ 1,  0,  0}{57.316} & \framebox{55.556} \\
    \bottomrule
    \end{tabular}}%
  \label{99QLF}%
\end{center}
\footnotesize{\textit{Notes:} Boxes indicate the best model in each market, red shading indicates the 2nd best model, and the blue shading indicates the 3rd best model. Bold indicates the worst model.
}
\end{table}%
We also applied the quantile loss function to further assess the model performances across markets. Table \ref{99QLF} summarises the loss function values of the models across the five cryptocurrency markets. The model with the lowest loss function value is preferred. The three most preferred and the least preferred models in each market are highlighted. For the BTC market, the proposed method is the best method, with a quantile loss function of 46.959. The CARE-AS model is the second best method, with a loss function of 47.070. By contrast, CAViaR-ES-AS-EXP performs the worst, as it has the highest loss function value of 52.876. Additionally, the three forecast combination approaches are competitive, with lower loss function values. 

For the BCH market, the proposed method produces the best VaR forecasts, with the lowest quantile loss function value of 80.447. The EWMA method is the second best method, with a loss function value of 80.686; however, the corresponding violation ratio is 0.833, which is different from the target ratio of 1 and inconsistent with the theoretical expectation. The CARE-AS model performs the worst, with the highest loss function value of 92.964. In general, the three forecast combination approaches perform better than the individual models, with lower loss function values. 

For the ETC market, the three forecast combination approaches perform well and rank as the top 3 methods. Median forecast has the lowest loss function value of 66.977, followed by the proposed method and simple average, with loss function values of 67.102 and 67.346, respectively. By contrast, HS-168 performs the worst, with the highest loss function value of 76.140. Additionally, the parametric models generally perform better than the nonparametric models (including the semiparametric models). 

For the ETH market, the three forecast combination approaches also perform significantly better than the individual models. The simple average method has the lowest loss function value of 57.037, followed by the proposed method and median forecast, with the loss function values of 57.316 and 57.990, respectively. The loss function values of all the other methods, except HS-168, are approximately 60.000. HS-168 has the highest loss function value of 71.354, which is substantially higher than those of the other methods.  

For the LTC market, the proposed method (55.556) performs the best in terms of the quantile loss function. The GARCH family models, which incorporate Student's t and skewed Student's t-distributions, perform better than the other methods, with corresponding loss function values of approximately 56.000. The four semiparametric models perform the worst, with loss function values all greater than 62.000. Finally, CAViaR-ES-AS-EXP has the highest loss function value of 64.213. 

\begin{figure}[H]
\begin{center}
\caption{99\% VaR Forecasts for the BTC Market}
\includegraphics[width=\textwidth,height=0.3\textheight]{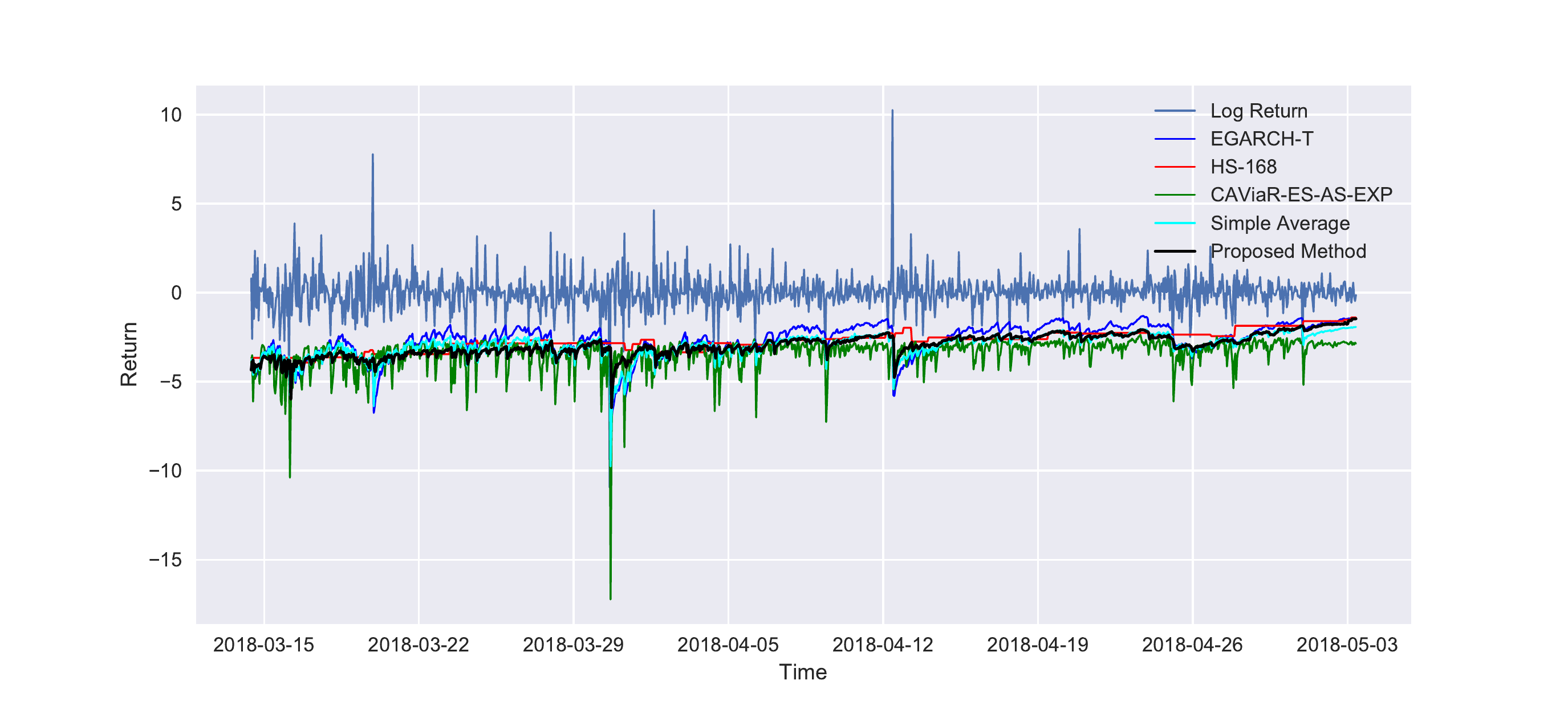}
\label{BTC_VaR_99}
\end{center}
\end{figure}

In summary, the proposed method can generate accurate forecasts for the 99\% VaR quantile. Clearly, the three forecast combination approaches have considerable advantages over the individual models. For example, the proposed method can generate stable and relatively accurate 99\% VaR forecasts across the five cryptocurrency markets in terms of the violation ratio, statistical tests and quantile loss function. The VaR series produced by the proposed method across the five markets all have corresponding violation ratios near the target ratio of 1, and they are not rejected by any of the four applied statistical tests. Moreover, in terms of the quantile loss function, the proposed method has the lowest loss function values in three markets (BTC, BCH and LTC) and the second lowest loss function values in two markets (ETC and ETH). Additionally, Figure \ref{BTC_VaR_99} plots the 99\% VaR forecasts for several methods and the proposed method in the BTC market. The advantages of the proposed method over the other methods can be observed in this plot. For example, the VaR series produced by the proposed method (black line) is smoother than that produced by CAViaR-ES-AS-EXP (green line) and has lower forecast error (closer to the return series). Furthermore, GJRGARCH-SKT provides relatively competitive VaR forecasts compared to the individual models, with a better violation ratio, test results and loss function values. 

\begin{figure}[H]
\begin{center}
\caption{Individual Weights of the Combined VaR Forecasts at the 99\% Confidence Level for the BTC Market}
\includegraphics[width=\textwidth,height=0.3\textheight]{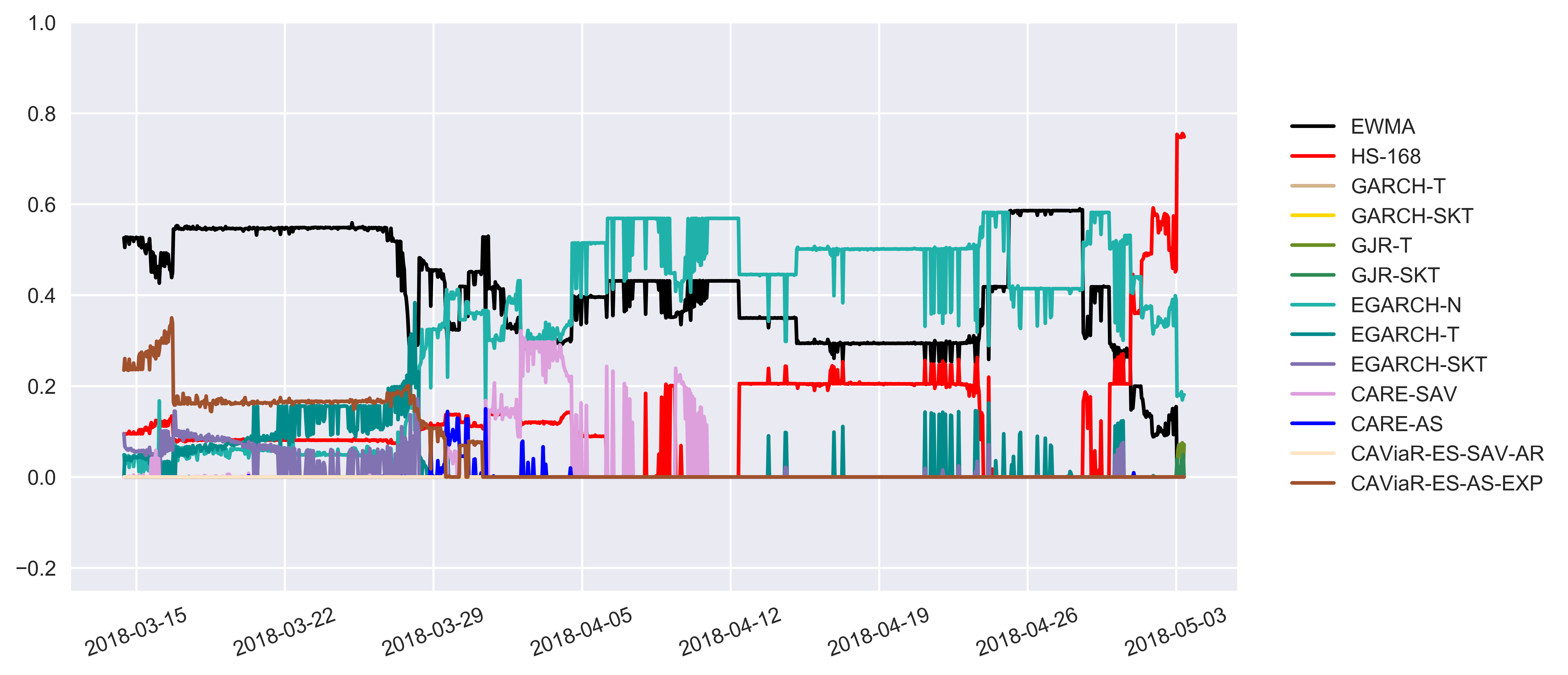}
\label{BTC_VaR_99_Weight}
\end{center}
\end{figure}

Figure \ref{BTC_VaR_99_Weight} plots the optimal weights assigned to the individual models over time for the BTC market. Clearly, the weights are vary considerably at each forecast point. The proposed method generates more accurate VaR series by assigning optimal weights to different individual forecasts over time.
For example, HS-168 (red line) and EWMA (black line) always take some weights to smooth the combined VaR series. Additionally, different weights are assigned to the other models to improve the forecast accuracy of the combined VaR series.  

\subsection{Forecasting ES}
The same set of models is applied to generate the one-step-ahead 95\% and 99\% ES forecasts for the five given cryptocurrency markets. However, the procedures that used to assess the ES forecasts are slightly different. We still employ the violation ratio and four statistical tests to assess the performances of the parametric models by treating them as VaR forecasts with the corresponding target violation rates (Table \ref{targetERate}), whereas for the other models, the best way to assess the ES forecasts is via the joint VaR-ES loss function (AL log score) and the MCS. 
\subsubsection*{ES Forecasts at 95\% Confidence Level}
\begin{table}[H]
  \centering
  \caption{Summary of the Target ESRate for the Parametric Models}
  
    \begin{tabular}{lccc}
\cmidrule{1-3}    Model & 95\%  & 99\%  &  \\
\cmidrule{1-3}    EWMA  & 1.96\% & 0.38\% &  \\
    GARCH-T & 1.64\% & 0.32\% &  \\
    GARCH-SKT & 1.64\% & 0.32\% &  \\
    GJRGARCH-T & 1.64\% & 0.32\% &  \\
    GJRGARCH-SKT & 1.64\% & 0.32\% &  \\
    EGARCH-N & 1.96\% & 0.38\% &  \\
    EGARCH-T & 1.64\% & 0.32\% &  \\
    EGARCH-SKT & 1.64\% & 0.32\% &  \\
    \bottomrule
    \bottomrule
    \end{tabular}%
  \label{targetERate}%
\end{table}%
Recall that Section \ref{backtesting} illustrated that the ES forecasts with daily return data can be treated as VaR forecasts to be assessed. However, for hourly data, this technique may not work. Nevertheless, we can still treat the parametric models as VaR forecasts by fitting the corresponding distribution into each estimation window and estimating the corresponding distribution parameters. Table \ref{targetERate} summarises the target violation rates used for each method. We fit the Student's t and skewed Student's t-distributions into the last 1200 estimation windows (with window size = 2000) to compute the corresponding DOF for each market. The results are summarised in Table \ref{t DOF} and Table \ref{skt DOF} in Appendix \ref{AppendixA}. The mean DOF across each window for each market is approximately 4 for both.

\begin{table}[H]
  \begin{center}
  \caption{Forecast Comparison of the ES Violation Ratios at the 95\% Confidence Level}
    \resizebox{\linewidth}{!}{
    \begin{tabular}{lcccccccc}
    \toprule
    Model & \multicolumn{1}{l}{BTC} & \multicolumn{1}{l}{BCH} & \multicolumn{1}{l}{ETC} & \multicolumn{1}{l}{ETH} & \multicolumn{1}{l}{LTC} & \multicolumn{1}{l}{Mean} & \multicolumn{1}{l}{Median} & \multicolumn{1}{l}{RMSE} \\
    \midrule
    EWMA  & \textbf{0.2551} & \framebox{1.0204} & \textbf{0.5952} & \textbf{0.5952} & \textbf{0.3827} & \textbf{0.5697} & \textbf{0.5952} & \textbf{0.4384} \\
    GARCH-T & 0.8638 & 0.5589 & 0.6606 & 1.5244 & 0.8638 & 0.8943 & 0.8638 & 0.3154 \\
    GARCH-SKT & 0.7622 & 0.5589 & 0.6606 & 1.3720 & 0.8638 & 0.8435 & 0.7622 & 0.3053 \\
    GJRGARCH-T & \framebox{0.9654} & 0.7622 & \textcolor[rgb]{ 1,  0,  0}{0.7114} & 1.4736 & \framebox{1.0163} & \framebox{0.9858} & \framebox{0.9654} & \textcolor[rgb]{ 1,  0,  0}{0.2102} \\
    GJRGARCH-SKT & 0.8638 & 0.7114 & \textcolor[rgb]{ 1,  0,  0}{0.7114} & 1.3211 & 0.9146 & \textcolor[rgb]{ 1,  0,  0}{0.9045} & 0.8638 & 0.2240 \\
    EGARCH-N & \textcolor[rgb]{ 1,  0,  0}{1.1054} & \textcolor[rgb]{ 1,  0,  0}{0.9779} & \framebox{1.1054} & 1.3180 & \textcolor[rgb]{ 1,  0,  0}{0.9779} & 1.0969 & 1.1054 & \framebox{0.1146} \\
    EGARCH-T & 1.1179 & \textbf{0.5081} & 0.6098 & \textcolor[rgb]{ 1,  0,  0}{1.2703} & 0.9146 & 0.8841 & \textcolor[rgb]{ 1,  0,  0}{0.9146} & 0.2711 \\
    EGARCH-SKT & 1.1179 & \textbf{0.5081} & 0.6606 & \framebox{1.1687} & 0.8638 & 0.8638 & 0.8638 & 0.2508 \\
    \bottomrule
    \bottomrule
    \end{tabular}}%
  \label{ES95Ratio}%
 \end{center}
\footnotesize{\textit{Notes:} Boxes indicate the best model for each currency, red shading indicates the 2nd best model for each currency, and bold indicates the worst model for each currency. `RMSE' stands for the square root of the average squared difference between the violation ratio for each currency and the target value of 1.}
\end{table}%

Table \ref{ES95Ratio} summarises the violation ratios for each 95\% ES forecast from the parametric models across all five cryptocurrency markets.  Clearly, no single model can adapt to all five markets well. In general, GJRGARCH-T performs well as it has the best violation ratios in the BTC and LTC markets. Moreover, GJRGARCH-T is also favourable in terms of the mean and median violation ratios across the five markets. By contrast, the EWMA method always produces the worst ES forecasts in terms of the violation ratio. 

\begin{table}[H]
  \begin{center}
      
    \caption{Counts of Model Rejections by the Statistical Tests for the 95\% ES Forecasts}
    \resizebox{0.5\linewidth}{!}{%
    \begin{tabular}{lcccccc}
    \toprule
    Model & \multicolumn{1}{c}{UC} & \multicolumn{1}{c}{CC} & \multicolumn{1}{c}{DQ1} & \multicolumn{1}{c}{DQ4} & \multicolumn{1}{c}{Total} \\
    \midrule
    EWMA  & \textbf{4} & \textbf{2} & \textbf{3} & \textbf{4} & \textbf{5} \\
    GARCH-T & 2     & \framebox{0}     & \textcolor[rgb]{ 1,  0,  0}{1}     & \framebox{0}     & 2 \\
    GARCH-SKT & \textcolor[rgb]{ 1,  0,  0}{1}     & \framebox{0}     & \textcolor[rgb]{ 1,  0,  0}{1}     & \framebox{0}     & 2 \\
    GJRGARCH-T & \textcolor[rgb]{ 1,  0,  0}{1}     & \framebox{0}     & \textcolor[rgb]{ 1,  0,  0}{1}     & \framebox{0}     & \textcolor[rgb]{ 1,  0,  0}{1} \\
    GJRGARCH-SKT & \framebox{0}     & \framebox{0}     & \framebox{0}     & \framebox{0}     & \framebox{0} \\
    EGARCH-N & \framebox{0}     & \framebox{0}     & \textcolor[rgb]{ 1,  0,  0}{1}     & \framebox{0}     & \textcolor[rgb]{ 1,  0,  0}{1} \\
    EGARCH-T & \textcolor[rgb]{ 1,  0,  0}{1}     & \textbf{2} & \framebox{0}     & \framebox{0}     & 2 \\
    EGARCH-SKT & \textcolor[rgb]{ 1,  0,  0}{1}     & \textbf{2} & \textcolor[rgb]{ 1,  0,  0}{1}     & \framebox{0}     & 3 \\
    \bottomrule
    \bottomrule
    \end{tabular}}
\label{95EScounts}%
\end{center}
\footnotesize{\textit{Notes:} Boxes indicate the best model, red shading indicates the 2nd best model, and bold indicates the worst model. `Total' indicates the number of markets in which the model was rejected by at least one test.} 
\end{table}%

Four statistical tests are applied with their target violation rates (not ES level) as the input. The test results are summarised in Table \ref{95EScounts}. Clearly, the GJRGARCH-SKT model has the best performances in the statistical tests as it is not rejected by any test in any market. By contrast, the EWMA method has the worst performances and is rejected by at least one test in each market. 

\begin{table}[H]
\begin{center}
  \caption{AL Log Score for the 95\% ES Forecasts}
  \resizebox{0.7\linewidth}{!}{%

    \begin{tabular}{lccccc}
    \toprule
    Model & \multicolumn{1}{l}{BTC} & \multicolumn{1}{l}{BCH} & \multicolumn{1}{l}{ETC} & \multicolumn{1}{l}{ETH} & \multicolumn{1}{l}{LTC} \\
    \midrule
    EWMA  & \textbf{2.0389} & 2.5503 & \textbf{2.5077} & \textbf{2.2602} & \textbf{2.3105} \\
    HS-168 & 1.9369 & 2.5477 & 2.4802 & 2.2272 & 2.2574 \\
    GARCH-T & 1.9045 & 2.5177 & 2.3723 & 2.1667 & 2.1492 \\
    GARCH-SKT & 1.9029 & 2.5093 & \framebox{2.3707}& 2.1650 & \textcolor[rgb]{ 1,  0,  0}{2.1487} \\
    GJRGARCH-T & \textcolor[rgb]{ 1,  0,  0}{1.8871} & 2.5229 & 2.3751 & 2.1530 & \textcolor[rgb]{ 0,  .439,  .753}{2.1490} \\
    GJRGARCH-SKT & \textcolor[rgb]{ 1,  0,  0}{1.8871} & 2.5173 & 2.3768 & 2.1532 & 2.1504 \\
    EGARCH-N & 1.9287 & \textcolor[rgb]{ 1,  0,  0}{2.4977} & 2.3860 & 2.1732 & 2.1718 \\
    EGARCH-T & 1.8923 & 2.5248 & 2.3988 & 2.1539 & 2.1580 \\
    EGARCH-SKT & 1.8904 & 2.5127 & 2.3959 & 2.1561 & 2.1574 \\
    CARE-SAV & 1.9472 & \textbf{2.5997} & 2.4338 & 2.2018 & 2.2294 \\
    CARE-AS & 1.9585 & 2.5953 & 2.4303 & 2.2484 & 2.2444 \\
    CAViaR-ES-SAV-AR & 2.0203 & 2.5539 & 2.5110 & 2.1732 & 2.2640 \\
    CAViaR-ES-AS-EXP & 1.9920 & 2.5605 & 2.4890 & 2.1852 & 2.2593 \\
          &       &       &       &       &  \\
    Simple Average & 1.9043 & \framebox{2.4929} & 2.3915 & \framebox{2.1342} & 2.1725 \\
    Median Forecast & \textcolor[rgb]{ 0,  .439,  .753}{1.8877} & \textcolor[rgb]{ 1,  0,  0}{2.4977} & \textcolor[rgb]{ 0,  .439,  .753}{2.3718} & \textcolor[rgb]{ 1,  0,  0}{2.1459} & 2.1539 \\
          & \multicolumn{5}{c}{} \\
    Proposed Method & \framebox{1.8791} & \textcolor[rgb]{ 0,  .439,  .753}{2.4999} & \textcolor[rgb]{ 1,  0,  0}{2.3717} & \textcolor[rgb]{ 0,  .439,  .753}{2.1472} & \framebox{2.1486} \\
    \bottomrule
    \bottomrule
    \end{tabular}}
  \label{AL95}
\end{center}
\footnotesize{\textit{Notes:} Boxed indicates the most favoured model in each market, red shading indicates the 2nd favoured model and the blue shading indicates the 3rd favoured model. While bold indicates the least favoured model.}
\end{table}%

For the hourly data, we can treat the ES forecasts as VaR quantile forecasts according to Table \ref{targetERate} only if they are produced by parametric models, that is, we need to know the distribution assumptions. The AL log score function and MCS are employed to further assess the forecast performances and compare each model. Table \ref{AL95} summarises the AL log scores for each model, and Table \ref{ESMCS95} summarises the 75\% and 90\% MCS results. 

In terms of the AL log score, the model with the lowest score is preferred. The proposed method has the lowest score (1.8791) in the BTC market and is thus the most favoured model. GJRGARCH-T and GJRGARCH-SKT are the second most favoured method, with scores of 1.8871. Median forecast is the third most favoured method, with a score of 1.8877. Similar to finding for the VaR forecast performance, the EWMA method has the highest AL score and is therefore the least favoured. The semiparametric models are not as competitive as the parametric models in general. However, the three forecast combination approaches are competitive. 

For the BCH market, the three forecast combination methods are competitive and rank as the top three methods with similar scores. The semiparametric models perform as poorly as observed before, and CARE-SAV has the highest score of 2.5997. 

For the ETC market, GARCH-SKT has the lowest score (2.3707) and is therefore the most favoured model, followed by the proposed method (2.3717) and median forecast (2.3718), which are the second and third best models, respectively. Additionally, GARCH-SKT, GJRGARCH-T and GJRGARCH-SKT also provide competitive ES forecasts, with AL scores (approximately 2.375) similar to those of the top three models. 

For the ETH market, the top three models are simple average (2.1342), median forecast (2.1459) and the proposed method (2.1472). In general, the parametric models perform better than the nonparametric models (including the semiparametric models). 

For the LTC market, the proposed method performs the best, with a score of 2.1486, followed by GARCH-SKT (2.1487) and GJRGARCH-T (2.1490), which are the second and third best models, respectively. By contrast, the EWMA method provides the worst ES forecasts, with the highest score of 2.3105. In general, similar to the findings for most of the other markets, the parametric models perform much better than the nonparametric (including semiparametric) models.

\begin{table}[H]
\begin{center}
  \caption{Summary Results of the MCS for the 95\% ES Forecasts}
  \resizebox{0.8\linewidth}{!}{%

    \begin{tabular}{lccccccc}
    \toprule
    Model & \multicolumn{1}{l}{BTC} & \multicolumn{1}{l}{BCH} & \multicolumn{1}{l}{ETC} & \multicolumn{1}{l}{ETH} & \multicolumn{1}{l}{LTC} & \multicolumn{1}{l}{Mean} & \multicolumn{1}{l}{Total} \\
    \midrule
    EWMA  & 0.0000 & \textcolor[rgb]{ 1,  0,  0}{0.1530} & 0.0010 & 0.0340 & 0.0000 & 0.0376 & 0 \\
    HS-168 & 0.0090 & \textcolor[rgb]{ 1,  0,  0}{0.1530} & 0.0000 & \textcolor[rgb]{ 1,  0,  0}{0.2470} & 0.0000 & 0.0818 & 0 \\
    GARCH-T & \textcolor[rgb]{ 0,  .439,  .753}{0.6200} & 0.0070 & \textcolor[rgb]{ 0,  .439,  .753}{0.6860} & \textcolor[rgb]{ 0,  .439,  .753}{0.5500} & \textcolor[rgb]{ 0,  .439,  .753}{0.8820} & 0.5490 & 4 \\
    GARCH-SKT & \textcolor[rgb]{ 0,  .439,  .753}{0.7260} & \textcolor[rgb]{ 1,  0,  0}{0.1060} & \textcolor[rgb]{ 0,  .439,  .753}{1.0000} & \textcolor[rgb]{ 0,  .439,  .753}{0.2720} & \textcolor[rgb]{ 0,  .439,  .753}{0.9990} & 0.6206 & 4 \\
    GJRGARCH-T & \textcolor[rgb]{ 0,  .439,  .753}{1.0000} & 0.0150 & \textcolor[rgb]{ 0,  .439,  .753}{0.7630} & \textcolor[rgb]{ 0,  .439,  .753}{0.7620} & \textcolor[rgb]{ 0,  .439,  .753}{0.9990} & 0.7078 & 4 \\
    GJRGARCH-SKT & \textcolor[rgb]{ 0,  .439,  .753}{0.9990} & 0.0870 & \textcolor[rgb]{ 0,  .439,  .753}{0.5490} & \textcolor[rgb]{ 0,  .439,  .753}{0.7620} & \textcolor[rgb]{ 0,  .439,  .753}{0.3520} & 0.5498 & 4 \\
    EGARCH-N & 0.0570 & \textcolor[rgb]{ 0,  .439,  .753}{0.8390} & \textcolor[rgb]{ 0,  .439,  .753}{0.6680} & \textcolor[rgb]{ 0,  .439,  .753}{0.2720} & \textcolor[rgb]{ 1,  0,  0}{0.1090} & 0.3890 & 3 \\
    EGARCH-T & \textcolor[rgb]{ 0,  .439,  .753}{0.9150} & 0.0000 & \textcolor[rgb]{ 1,  0,  0}{0.1070} & \textcolor[rgb]{ 0,  .439,  .753}{0.7620} & \textcolor[rgb]{ 0,  .439,  .753}{0.7550} & 0.5078 & 3 \\
    EGARCH-SKT & \textcolor[rgb]{ 0,  .439,  .753}{0.9970} & \textcolor[rgb]{ 1,  0,  0}{0.1530} & \textcolor[rgb]{ 1,  0,  0}{0.2020} & \textcolor[rgb]{ 0,  .439,  .753}{0.3740} & \textcolor[rgb]{ 0,  .439,  .753}{0.7810} & 0.5014 & 3 \\
    CARE-SAV & \textcolor[rgb]{ 1,  0,  0}{0.1260} & 0.0150 & 0.0110 & 0.0890 & 0.0000 & 0.0482 & 0 \\
    CARE-AS & 0.0240 & 0.0300 & 0.0110 & 0.0050 & 0.0000 & 0.0140 & 0 \\
    CAViaR-ES-SAV-AR & 0.0000 & 0.0660 & 0.0000 & \textcolor[rgb]{ 1,  0,  0}{0.1350} & 0.0000 & 0.0402 & 0 \\
    CAViaR-ES-AS-EXP & 0.0000 & 0.0870 & 0.0010 & 0.0840 & 0.0000 & 0.0344 & 0 \\
          &       &       &       &       &       &       &  \\
    Simple Average & \textcolor[rgb]{ 0,  .439,  .753}{0.7260} & \textcolor[rgb]{ 0,  .439,  .753}{1.0000} & \textcolor[rgb]{ 0,  .439,  .753}{0.5490} & \textcolor[rgb]{ 0,  .439,  .753}{1.0000} & \textcolor[rgb]{ 1,  0,  0}{0.1090} & 0.6768 & 4 \\
    Median Forecast & \textcolor[rgb]{ 0,  .439,  .753}{0.9990} & \textcolor[rgb]{ 0,  .439,  .753}{0.5880} & \textcolor[rgb]{ 0,  .439,  .753}{0.8750} & \textcolor[rgb]{ 0,  .439,  .753}{0.7620} & \textcolor[rgb]{ 0,  .439,  .753}{0.8420} & 0.8132 & \framebox{5} \\
          & \multicolumn{5}{c}{\textcolor[rgb]{ 0,  .439,  .753}{}} &       &  \\
    Proposed Method & \textcolor[rgb]{ 0,  .439,  .753}{0.9990} & \textcolor[rgb]{ 0,  .439,  .753}{0.8390} & \textcolor[rgb]{ 0,  .439,  .753}{0.6930} & \textcolor[rgb]{ 0,  .439,  .753}{0.5710} & \textcolor[rgb]{ 0,  .439,  .753}{1.0000} & \framebox{0.8204} & \framebox{5} \\
    \bottomrule
    \bottomrule

    \end{tabular}}
  \label{ESMCS95}%
\end{center}
\footnotesize{\textit{Notes:} Models with blue shading are included by the 75\% MCS; models with either red shading or blue shading are included by the 90\% MCS. `Total' indicates the number of markets in which the corresponding model is included by the 75\% MCS. Boxes indicate the best model.}
\end{table}%

With the MCS, we can further assess the ES forecast performances of the models. Table \ref{ESMCS95} summarises the test results of the MCS for the 95\% ES forecasts. We implement both the 75\% and 90\% confidence levels for the MCS. Blue shading in Table \ref{ESMCS95} highlights the models included by the 75\% MCS while red shading and blue shading jointly highlight the models included by the 90\% MCS. We also compute the mean p-value for each model across the five markets. A higher average p-value indicates greater preferability of the corresponding model. Moreover, in the last column, we highlight the number of times the model was included by the 75\% MCS, which is more strict than the 90\% MCS. 

Clearly, the proposed method and median forecast are the only two models included by the 75\% MCS across all five markets; the proposed method has a higher average p-value (0.8204). GARCH-T, GARCH-SKT, GJRGARCH-T and GJRGARCH-SKT are included by the 75\% MCS in four markets. Moreover, the three EGARCH family models are included by the 75\% MCS in three markets. The other six models are not included by the 75\% MCS in any market. Furthermore, GARCH-SKT, EGARCH-SKT, simple average, median forecast and the proposed method are included by the 90\% MCS in all five markets. By contrast, CARE-AS and CAViaR-ES-AS-EXP are not included by the 90\% MCS in any market. CARE-AS and CAViaR-ES-SAV-AR are included by the 90\% MCS only once, in the BTC market and the ETH market, respectively.
\begin{figure}[H]
\begin{center}
\caption{95\% ES Forecasts for the BTC Market}
\includegraphics[width=\textwidth,height=0.3\textheight]{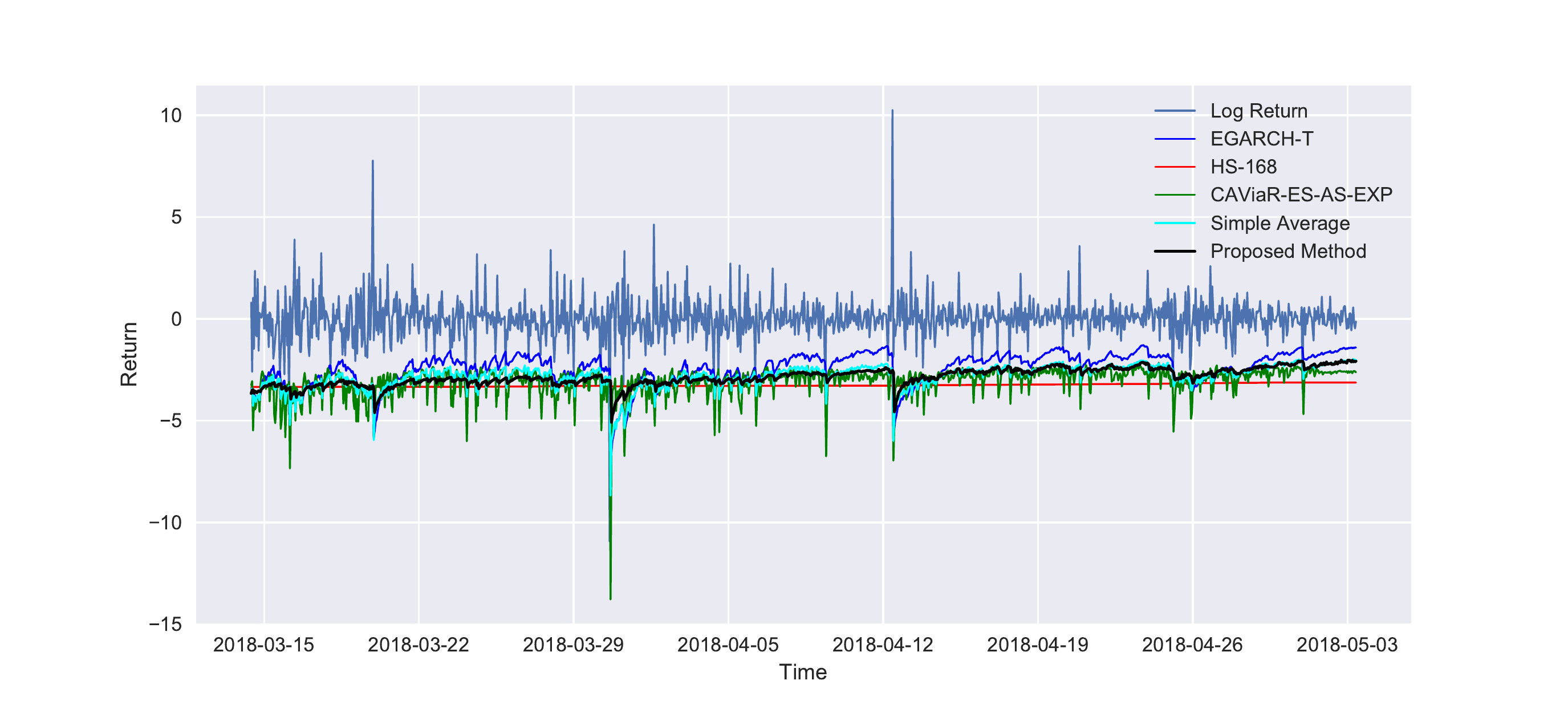}
\label{BTC_ES_95}
\end{center}
\end{figure}

\begin{figure}[H]
\begin{center}
\caption{Individual Weights of the Combined ES Forecasts at the 95\% Confidence Level for the BTC Market}
\includegraphics[width=\textwidth,height=0.3\textheight]{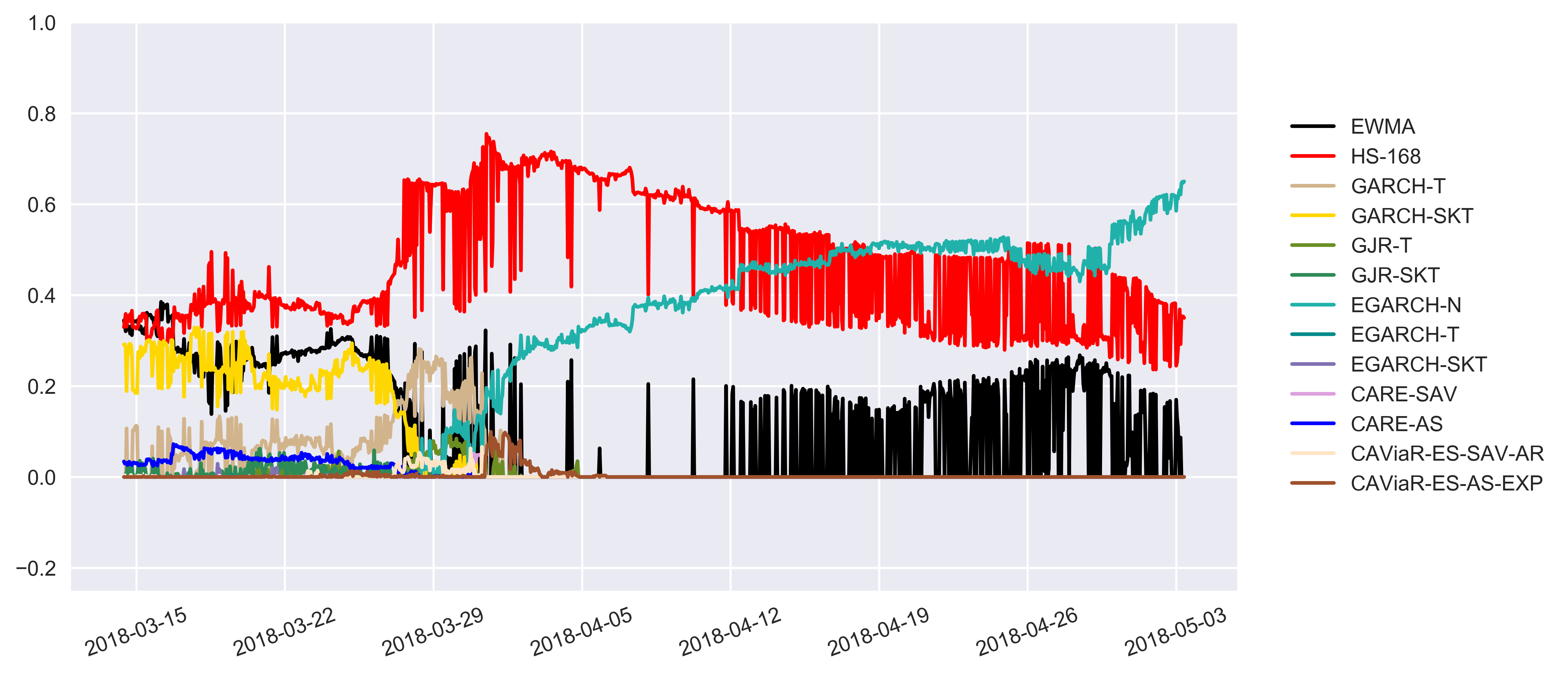}
\label{BTC_ES_95_weights}
\end{center}
\end{figure}

In summary, in the case of the 95\% ES forecasts, the proposed method is the best approach in terms of the AL log score and MCS. Similar to findings for the VaR forecasts, the proposed method can generate relatively accurate and stable ES forecasts by assigning the optimal weights to individual forecasts and obtaining the benefits of combination. Figure \ref{BTC_ES_95_weights} presents the weights assigned to each individual forecast to generate the combined ES forecast at each time point for the BTC market. Clearly, different optimal weights are assigned to each individual forecast over time to generate the combined ES series. Additionally, the advantage of the ES series from the proposed method over the ES series from the individual forecasts can be roughly observed in Figure \ref{BTC_ES_95}: EGARCH-T generates the ES series that is close to the return series, which results in a higher violation ratio (1.1179); thus, EGARCH-T tends to underestimate the risk level exhibited by the BTC market. CAViaR-ES-AS-EXP tends to capture the dynamics of the return series, whereas the measurement equation may not be consistent with the true dynamics of the time series process, which results in an increased forecast errors of the corresponding ES series. Both forecast combination approaches, the simple average method and the proposed method, obtain advantages from averaging individual forecasts. The difference between the methods is that simple average is an equal-weighted approach, whereas the proposed method is an optimal-weighted approach. The performance of simple average is subject to the individual model specifications, as it combines all the individual models. By contrast, the optimal-weighted approach may assign zero weights to some individual models during the combination process. Therefore, the combined forecasts from these two approaches perform differently. The median forecast method also produces accurate and stable forecasts across the five markets, which is consistent with \citet{mcneil2005quantitative}. Moreover, similar to their VaR quantile forecast performances, the GARCH family models incorporating Student's t and skewed Student's t-distributions are accurate compared to the other individual models.   

\subsubsection*{ES Forecasts at the 99\% Confidence Level}
We also conduct a 99\% ES forecasts study with the same model set and data set. The same evaluation procedures are employed to assess the forecast accuracy.
\begin{table}[H]
\begin{center}
  \caption{Forecast Comparison of the ES Violation Ratios at the 99\% Confidence Level}
\resizebox{0.98\linewidth}{!}{%
\begin{tabular}{lcccccccc}
    \toprule
    Model & \multicolumn{1}{l}{BTC} & \multicolumn{1}{l}{BCH} & \multicolumn{1}{l}{ETC} & \multicolumn{1}{l}{ETH} & \multicolumn{1}{l}{LTC} & \multicolumn{1}{l}{Mean } & \multicolumn{1}{l}{Median} & \multicolumn{1}{l}{RMSE} \\
    \midrule
    EWMA  & \framebox{0.6579} & \framebox{0.8772} & \framebox{0.8772} & \textcolor[rgb]{ 1,  0,  0}{1.0965} & \framebox{0.6579} & \framebox{0.8333} & \framebox{0.8772} & \framebox{0.2053} \\
    GARCH-T & \textcolor[rgb]{ 1,  0,  0}{0.5208} & 0.2604 & 0.2604 & 1.5625 & \textcolor[rgb]{ 1,  0,  0}{0.2604} & \textcolor[rgb]{ 1,  0,  0}{0.5729} & \textcolor[rgb]{ 1,  0,  0}{0.2604} & \textcolor[rgb]{ 1,  0,  0}{0.6521} \\
    GARCH-SKT & 1.8229 & 0.5208 & \textcolor[rgb]{ 1,  0,  0}{0.7813} & 2.3438 & 2.3438 & 1.5625 & 1.8229 & 0.8417 \\
    GJRGARCH-T & \textcolor[rgb]{ 1,  0,  0}{0.5208} & 0.2604 & 0.2604 & 1.5625 & \textcolor[rgb]{ 1,  0,  0}{0.2604} & \textcolor[rgb]{ 1,  0,  0}{0.5729} & \textcolor[rgb]{ 1,  0,  0}{0.2604} & \textcolor[rgb]{ 1,  0,  0}{0.6521} \\
    GJRGARCH-SKT & 1.8229 & 0.5208 & \textcolor[rgb]{ 1,  0,  0}{0.7813} & 2.0833 & 2.0833 & 1.4583 & 1.8229 & 0.7375 \\
    EGARCH-N & 2.1930 & \textbf{1.9737} & \textbf{2.6316} & \textbf{3.0702} & \textbf{2.6316} & \textbf{2.5000} & \textbf{2.6316} & \textbf{1.5000} \\
    EGARCH-T & \textcolor[rgb]{ 1,  0,  0}{0.5208} & 0.2604 & 0.0000 & \framebox{1.0417} & 0.0000 & 0.3646 & \textcolor[rgb]{ 1,  0,  0}{0.2604} & \textcolor[rgb]{ 1,  0,  0}{0.6521} \\
    EGARCH-SKT & \textbf{2.3438} & \textcolor[rgb]{ 1,  0,  0}{0.7813} & \textcolor[rgb]{ 1,  0,  0}{0.7813} & 2.0833 & 2.0833 & 1.6146 & 2.0833 & 0.7896 \\
    \bottomrule
    \bottomrule
    \end{tabular}}%
  \label{ES99VR}%
\end{center}
\footnotesize{\textit{Notes:} Boxed indicates the favoured model for each currency, red shading indicates the 2nd favoured model for each currency, bold indicates the least favoured model for each currency. `RMSE' stands for the square root of the average squared difference between the five violation ratios and their target value of 1.}
\end{table}%

Table \ref{ES99VR} summarises the corresponding ES violation ratios for each model across the five markets. The 99\% violation ratios are computed by dividing the violation rate by the target ES violation rate, which is defined in Table \ref{targetERate}. In terms of the ES violation ratio, the EWMA method is best approach in the BTC market, with a violation ratio of 0.6579. However, EWMA is still highly conservative and tends to overestimate the risk level exhibited by the BTC market. The three parametric models, GARCH-T, GJRGARCH-T and EGARCH-T, which incorporate Student's t-distribution, generate the second best ES forecasts, with violation ratios of 0.5208. By contrast, EGARCH-SKT is highly anti-conservative, as it has 134\% more violations than expected and it highly underestimates the risk level. 

For the BCH market, the EWMA method produces the best ES forecasts with respect to the violation ratio. EGARCH-SKT has the second best violation ratio of 0.7813. By contrast, EGARCH-N has the worst performance, with 97\% more violations than expected. Moreover, EGARCH-N is the only model that underestimates the risk level exhibited by the BCH market. 

For the ETC market, the EWMA method has the best performance in terms of the violation ratio. GARCH-SKT, GJRGARCH-SKT and EGARCH-SKT produce the second best ES forecasts, with violation ratios equal to 0.7813. Unsurprisingly, EGARCH-N, which incorporates the standard normal distribution, has the worst performance and is highly anti-conservative, with 163\% more violations than expected. 

For the ETH market, EGARCH-T produces the best ES forecasts, with a violation ratio of 1.0417, which is close to the expected ratio. The EWMA method has the second best performance, with a violation ratio of 1.0965. By contrast, EGARCH-N is highly anti-conservative, with 207\% more violations than expected, and highly underestimates the risk level. Additonally, all the models tend to underestimate the risk level exhibited by the ETH market: they all have corresponding violation ratios greater than 1. 

For the LTC market, the best model is still EWMA, and the worst model is EGARCH-N. However, all the methods tend to highly overestimate or underestimate the risk level exhibited by the market, as they are either highly conservative or highly anti-conservative. For example, the second best models are GARCH-T and GJRGARCH-T, both of which have corresponding violation ratios of 0.2604, which is highly conservative and overestimates the risk level. Moreover, in terms of the mean, median and RMSE, the EWMA method is the best approach, whereas EGARCH-N is the worst.

\begin{table}[H]
\begin{center}
  \caption{Counts of Model Rejections by Statistical Tests for the 99\% ES Forecasts}
  \resizebox{0.4\textheight}{!}{%
    \begin{tabular}{lccccc}
    \toprule
    Model & \multicolumn{1}{c}{UC} & \multicolumn{1}{c}{CC} & \multicolumn{1}{c}{DQ1} & \multicolumn{1}{c}{DQ4} & \multicolumn{1}{c}{Total} \\
    \midrule
    EWMA  & \framebox{0}     & \framebox{0}     & \framebox{0}     & \textcolor[rgb]{ 1,  0,  0}{1} & \textcolor[rgb]{ 1,  0,  0}{1} \\
    GARCH-T & \framebox{0}     &\framebox{0}     & \framebox{0}     & \framebox{0}     & \framebox{0} \\
    GARCH-SKT & 2     & 2     & \textcolor[rgb]{ 1,  0,  0}{2} & \textcolor[rgb]{ 1,  0,  0}{1} & 3 \\
    GJRGARCH-T & \framebox{0}     & \framebox{0}     & \framebox{0}     & \framebox{0}     & \framebox{0} \\
    GJRGARCH-SKT & 0     & \textcolor[rgb]{ 1,  0,  0}{1} & \textcolor[rgb]{ 1,  0,  0}{2} & \textcolor[rgb]{ 1,  0,  0}{1} & 3 \\
    EGARCH-N & \textbf{4} & \textbf{4} & \textbf{5} & \textbf{3} & \textbf{5} \\
    EGARCH-T & 2     & \framebox{0}     & \framebox{0}     & \framebox{0}     & 2 \\
    EGARCH-SKT & \textcolor[rgb]{ 1,  0,  0}{1} & 2     & 3     & \framebox{0}     & 4 \\
    \bottomrule
    \bottomrule
    \end{tabular}}
  \label{ES99test}%
  \end{center}
\footnotesize{\textit{Notes:} Boxed indicates the most favoured model, red shading indicates the 2nd favoured model and bold indicates the least favoured model. `Total' indicates the number of markets in which the model was rejected by at least one test.}
\end{table}%

We also employ the four statistical tests to assess the forecast performances of the parametric models. GARCH-T and GJRGARCH-T achieve the best test results and are not rejected by any test across the five markets. The EWMA method is the second best approach. By contrast, EGARCH-N has the worst performance and is rejected more than three times by each test and at least one time in each market. These results are consistent with the results of the violation ratio.

\begin{table}[H]
\begin{center}
  \caption{AL Log Score for the 99\% ES Forecasts}
  \resizebox{0.7\linewidth}{!}{%

    \begin{tabular}{lccccc}
    \toprule
    Model & \multicolumn{1}{l}{BTC} & \multicolumn{1}{l}{BCH} & \multicolumn{1}{l}{ETC} & \multicolumn{1}{l}{ETH} & \multicolumn{1}{l}{LTC} \\
    \midrule
    EWMA  & 2.4208 & 2.9390 & 2.8024 & 2.6087 & 2.5855 \\
    HS-168 & \framebox{2.3509} & 2.9404 & \textbf{2.8647} & \textbf{2.8502} & 2.5962 \\
    GARCH-T & 2.4365 & 2.9886 & 2.7872 & 2.6745 & 2.5783 \\
    GARCH-SKT & 2.4487 & 2.9106 & \framebox{2.7211} & 2.7040 & \textcolor[rgb]{ 1,  0,  0}{2.5483} \\
    GJRGARCH-T & 2.3921 & 2.9566 & 2.7751 & 2.6465 & 2.5787 \\
    GJRGARCH-SKT & 2.4046 & \framebox{2.8862} & \textcolor[rgb]{ 0,  .439,  .753}{2.7303} & 2.6650 & 2.5597 \\
    EGARCH-N & \textbf{2.5065} & 2.9306 & 2.7940 & 2.8394 & 2.6125 \\
    EGARCH-T & 2.4275 & 3.0059 & 2.8285 & 2.6400 & 2.5809 \\
    EGARCH-SKT & 2.4559 & 2.9155 & 2.7611 & 2.6685 & \textcolor[rgb]{ 0,  .439,  .753}{2.5539} \\
    CARE-SAV & 2.4809 & 3.0045 & 2.8292 & 2.6496 & 2.6963 \\
    CARE-AS & 2.4597 & \textbf{3.0764} & 2.8302 & 2.6484 & \textbf{2.7102} \\
    CAViaR-ES-SAV-AR & 2.5017 & 2.9924 & 2.8622 & 2.6615 & 2.6903 \\
    CAViaR-ES-AS-EXP & 2.4376 & 2.9813 & 2.7952 & 2.6052 & 2.6883 \\
          &       &       &       &       &  \\
    Simple Average & \textcolor[rgb]{ 1,  0,  0}{2.3783} & \textcolor[rgb]{ 0,  .439,  .753}{2.8980} & 2.7423 & \framebox{2.5755} & 2.5652 \\
    Median Forecast & 2.3868 & 2.8997 & 2.7319 & \textcolor[rgb]{ 1,  0,  0}{2.5897} & 2.5579 \\
          & \multicolumn{5}{c}{} \\
    Proposed Method & \textcolor[rgb]{ 0,  .439,  .753}{2.3791} & \textcolor[rgb]{ 1,  0,  0}{2.8964} & \textcolor[rgb]{ 1,  0,  0}{2.7213} & \textcolor[rgb]{ 0,  .439,  .753}{2.5927} & \framebox{2.5411} \\
    \bottomrule
    \bottomrule
    \end{tabular}}
  \label{ES99AL}%
\end{center}
\footnotesize{\textit{Notes:} Boxed indicates the most favoured model in each market, red shading indicates the 2nd favoured model and the blue shading indicates the 3rd favoured model. While bold indicates the least favoured model.}
\end{table}%

The AL log score and the MCS are powerful evaluation methods to assess ES forecast performance. We applied these two procedures to assess the 99\% ES forecast performances of the models across each market. Table \ref{ES99AL} summarises the AL log score results. HS-168 is the best model for the BTC market, as it has the lowest score of 2.3509. The simple average method and the proposed method are the second and third best models in terms of the AL log score. By contrast, EGARCH-N has the worst performance, as it has the highest score of 2.5065. Generally, the parametric models perform better than the semiparametric models. Furthermore, the three forecast combination approaches perform better than the individual models in the BTC market. 

For the BCH market, GJRGARCH-SKT performs the best and has the lowest score of 2.8862, followed by the proposed method and simple average, which perform the second and third best, respectively. On the other hand, CARE-AS has the worst performance, with a score of 3.0764. For the parametric models, the models that incorporate the skewed Student's t-distribution produce competitive ES forecasts, which indicates that the skewed Student's t-distribution fits the BCH market well. Additionally, the three forecast combination approaches perform well, second only to GJRGARCH-SKT, which is the best model with respect to the AL log score. 

For the ETC market, GARCH-SKT is the best model in terms of the AL log score, followed by the proposed method and GJRGARCH-SKT, which are the second and third best models, respectively. By contrast, HS-168 has the highest score of 2.8647 and is therefore the least favoured method. Similar to their performances in the BCH market, the parametric models that incorporate Student's t-distribution are highly competitive. Additionally, the three forecast combination approaches maintain competitive performance, as in other markets.

For the ETH market, the three forecast combination approaches are the most competitive models. Simple average is the best model, with the lowest score of 2.5755, followed by median forecast and the proposed method. For the individual models, HS-168 is the worst model, with the highest score of 2.8502, and EGARCH-N is the second worst model, with a score of 2.8394. The other models produce ES forecasts with similar scores ranging from 2.60 to 2.70.

For the LTC market, the proposed method is the best model, and GARCH-SKT and EGARCH-SKT are the second and third best models, respectively. All the other parametric models, except EGARCH-N, have similar scores. On the other hand, CARE-AS has the highest score of 2.7102, and the other three semiparametric models have poor performance with respect to the AL log score. Additionally, the two general forecast combination approaches perform better than most individual models.

\begin{table}[H]
\begin{center}
  \caption{Summary Results of the MCS for the 99\% ES Forecasts}
  \resizebox{0.8\linewidth}{!}{%

    \begin{tabular}{lccccccc}
    \toprule
    Model & \multicolumn{1}{l}{BTC} & \multicolumn{1}{l}{BCH} & \multicolumn{1}{l}{ETC} & \multicolumn{1}{l}{ETH} & \multicolumn{1}{l}{LTC} & \multicolumn{1}{l}{Mean} & \multicolumn{1}{l}{Total} \\
    \midrule
    EWMA  & \textcolor[rgb]{ 1,  0,  0}{0.1790} & \textcolor[rgb]{ 0,  .439,  .753}{0.8950} & \textcolor[rgb]{ 0,  .439,  .753}{0.5510} & \textcolor[rgb]{ 0,  .439,  .753}{0.9260} & \textcolor[rgb]{ 0,  .439,  .753}{0.9350} & 0.6972 & 4 \\
    HS-168 & \textcolor[rgb]{ 0,  .439,  .753}{1.0000} & \textcolor[rgb]{ 0,  .439,  .753}{0.5750} & \textcolor[rgb]{ 1,  0,  0}{0.2310} & \textcolor[rgb]{ 1,  0,  0}{0.1590} & \textcolor[rgb]{ 0,  .439,  .753}{0.7220} & 0.5374 & 3 \\
    GARCH-T & 0.0300 & 0.0470 & 0.0110 & 0.0510 & \textcolor[rgb]{ 0,  .439,  .753}{0.7150} & 0.1708 & 1 \\
    GARCH-SKT & \textcolor[rgb]{ 0,  .439,  .753}{0.6620} & \textcolor[rgb]{ 0,  .439,  .753}{0.5340} & \textcolor[rgb]{ 0,  .439,  .753}{1.0000} & \textcolor[rgb]{ 0,  .439,  .753}{0.3120} & \textcolor[rgb]{ 0,  .439,  .753}{0.9350} & 0.6886 & \framebox{5} \\
    GJRGARCH-T & \textcolor[rgb]{ 0,  .439,  .753}{0.6620} & \textcolor[rgb]{ 1,  0,  0}{0.1890} & 0.0370 & \textcolor[rgb]{ 0,  .439,  .753}{0.5180} & \textcolor[rgb]{ 0,  .439,  .753}{0.6360} & 0.4084 & 3 \\
    GJRGARCH-SKT & \textcolor[rgb]{ 0,  .439,  .753}{0.6620} & \textcolor[rgb]{ 0,  .439,  .753}{1.0000} & \textcolor[rgb]{ 0,  .439,  .753}{0.7210} & \textcolor[rgb]{ 0,  .439,  .753}{0.5220} & \textcolor[rgb]{ 0,  .439,  .753}{0.9350} & 0.7680 & \framebox{5} \\
    EGARCH-N & \textcolor[rgb]{ 0,  .439,  .753}{0.6620} & \textcolor[rgb]{ 0,  .439,  .753}{0.8950} & \textcolor[rgb]{ 0,  .439,  .753}{0.6210} & \textcolor[rgb]{ 1,  0,  0}{0.2290} & \textcolor[rgb]{ 0,  .439,  .753}{0.7220} & 0.6258 & 4 \\
    EGARCH-T & \textcolor[rgb]{ 0,  .439,  .753}{0.3390} & 0.0790 & 0.0040 & 0.0030 & \textcolor[rgb]{ 0,  .439,  .753}{0.7220} & 0.2294 & 2 \\
    EGARCH-SKT & \textcolor[rgb]{ 0,  .439,  .753}{0.6620} & \textcolor[rgb]{ 0,  .439,  .753}{0.5750} & \textcolor[rgb]{ 0,  .439,  .753}{0.2560} & \textcolor[rgb]{ 0,  .439,  .753}{0.2970} & \textcolor[rgb]{ 0,  .439,  .753}{0.9350} & 0.5450 & \framebox{5} \\
    CARE-SAV & 0.0670 & \textcolor[rgb]{ 0,  .439,  .753}{0.5340} & 0.0850 & \textcolor[rgb]{ 0,  .439,  .753}{0.5220} & 0.0020 & 0.2420 & 2 \\
    CARE-AS & \textcolor[rgb]{ 1,  0,  0}{0.2040} & \textcolor[rgb]{ 0,  .439,  .753}{0.2590} & \textcolor[rgb]{ 1,  0,  0}{0.1190} & \textcolor[rgb]{ 0,  .439,  .753}{0.5220} & 0.0020 & 0.2212 & 2 \\
    CAViaR-ES-SAV-AR & 0.0000 & \textcolor[rgb]{ 0,  .439,  .753}{0.3260} & 0.0040 & \textcolor[rgb]{ 1,  0,  0}{0.1360} & 0.0080 & 0.0948 & 1 \\
    CAViaR-ES-AS-EXP & 0.0260 & \textcolor[rgb]{ 0,  .439,  .753}{0.6400} & 0.0280 & \textcolor[rgb]{ 0,  .439,  .753}{0.9260} & 0.0330 & 0.3306 & 2 \\
          &       &       &       &       &       &       &  \\
    Simple Average & \textcolor[rgb]{ 0,  .439,  .753}{0.5080} & \textcolor[rgb]{ 0,  .439,  .753}{0.8950} & \textcolor[rgb]{ 0,  .439,  .753}{0.6950} & \textcolor[rgb]{ 0,  .439,  .753}{1.0000} & \textcolor[rgb]{ 0,  .439,  .753}{0.7710} & 0.7738 & \framebox{5} \\
    Median Forecast & \textcolor[rgb]{ 0,  .439,  .753}{0.6620} & \textcolor[rgb]{ 0,  .439,  .753}{0.8950} & \textcolor[rgb]{ 0,  .439,  .753}{0.7210} & \textcolor[rgb]{ 0,  .439,  .753}{0.9260} & \textcolor[rgb]{ 0,  .439,  .753}{0.9350} & \framebox{0.8278} & \framebox{5} \\
          & \multicolumn{5}{c}{}                  &       &  \\
    Proposed Method & \textcolor[rgb]{ 0,  .439,  .753}{0.4870} & \textcolor[rgb]{ 0,  .439,  .753}{0.8950} & \textcolor[rgb]{ 0,  .439,  .753}{0.7210} & \textcolor[rgb]{ 0,  .439,  .753}{0.9260} & \textcolor[rgb]{ 0,  .439,  .753}{1.0000} & 0.8058 & \framebox{5} \\
    \bottomrule
    \bottomrule
    \end{tabular}}%
  \label{ES99MCS}%
\end{center}
\footnotesize{\textit{Notes:} Models with blue shading are included by the 75\% MCS, models with either red shading or blue shading are included by the 90\% MCS. `Total' indicates the number of markets that the corresponding model is included by the 75\% MCS. Boxing indicates the most favoured model.}

\end{table}%

Table \ref{ES99MCS} summarises the MCS results at both the 75\% and 90\% confidence levels, which can help us to further assess the ES forecast performances. The p-values with blue shading indicate models included by the 75\% MCS. The p-values with either red shading or blue shading indicate models included by the 90\% MCS. Clearly, the three parametric models that incorporate Student's t-distribution and the three forecast combination approaches produce competitive 99\% ES forecasts in terms of the 75\% MCS. The forecasts of these methods are included by the 75\% MCS in all five markets. GARCH-T and CAViaR-ES-SAV-AR are the least competitive models and are included by the 75\% MCS in only one market. Moreover, the median forecast method has the highest average p-value of 0.8278, and the proposed method has the second highest average p-value of 0.8058. 

\begin{figure}[H]
\begin{center}
\caption{99\% ES Forecasts for the BTC Market}
\includegraphics[width=\textwidth,height=0.3\textheight]{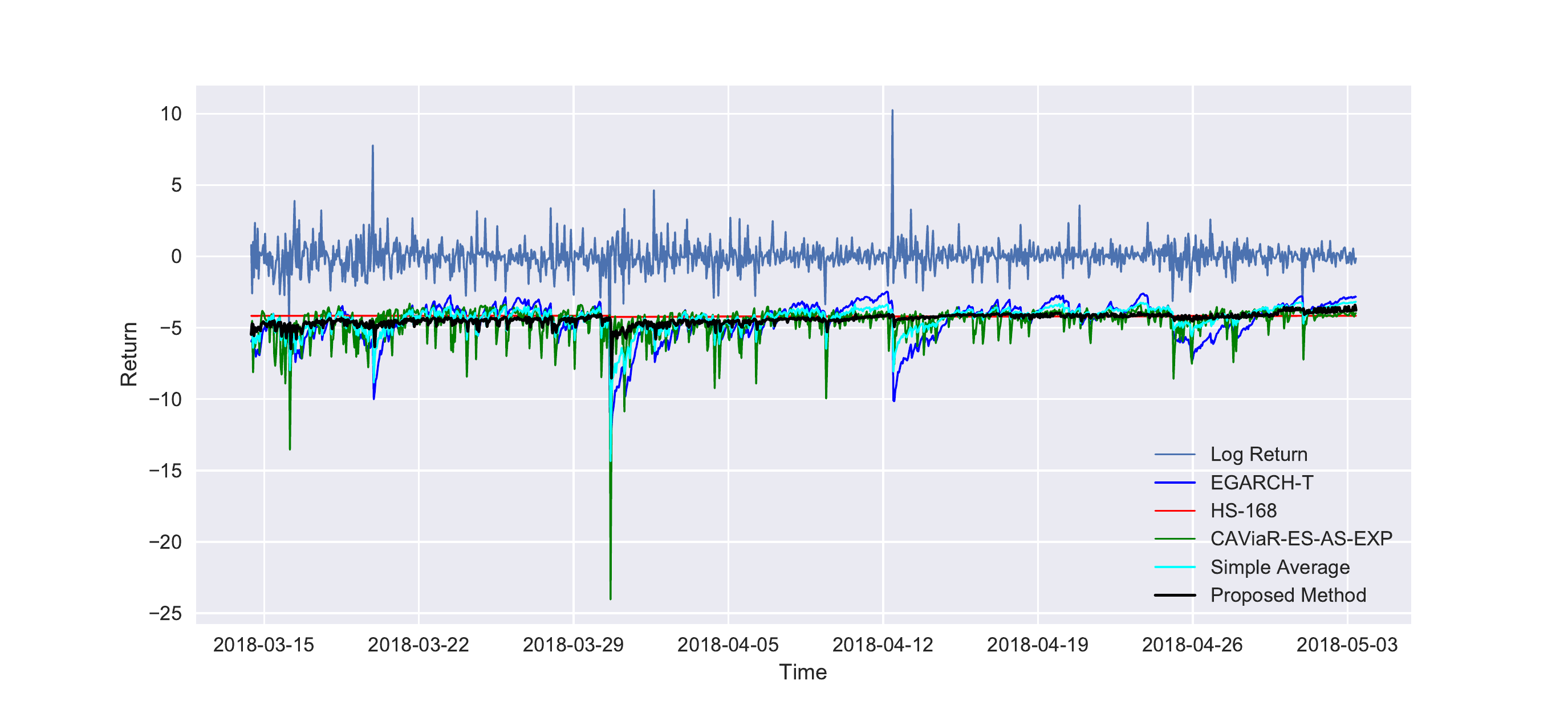}
\label{BTC_99_plots}
\end{center}
\end{figure}

\begin{figure}[H]
\begin{center}
\caption{Individual Weights of the Combined ES Forecasts at the 99\% Confidence Level for the BTC}
\includegraphics[width=\textwidth,height=0.3\textheight]{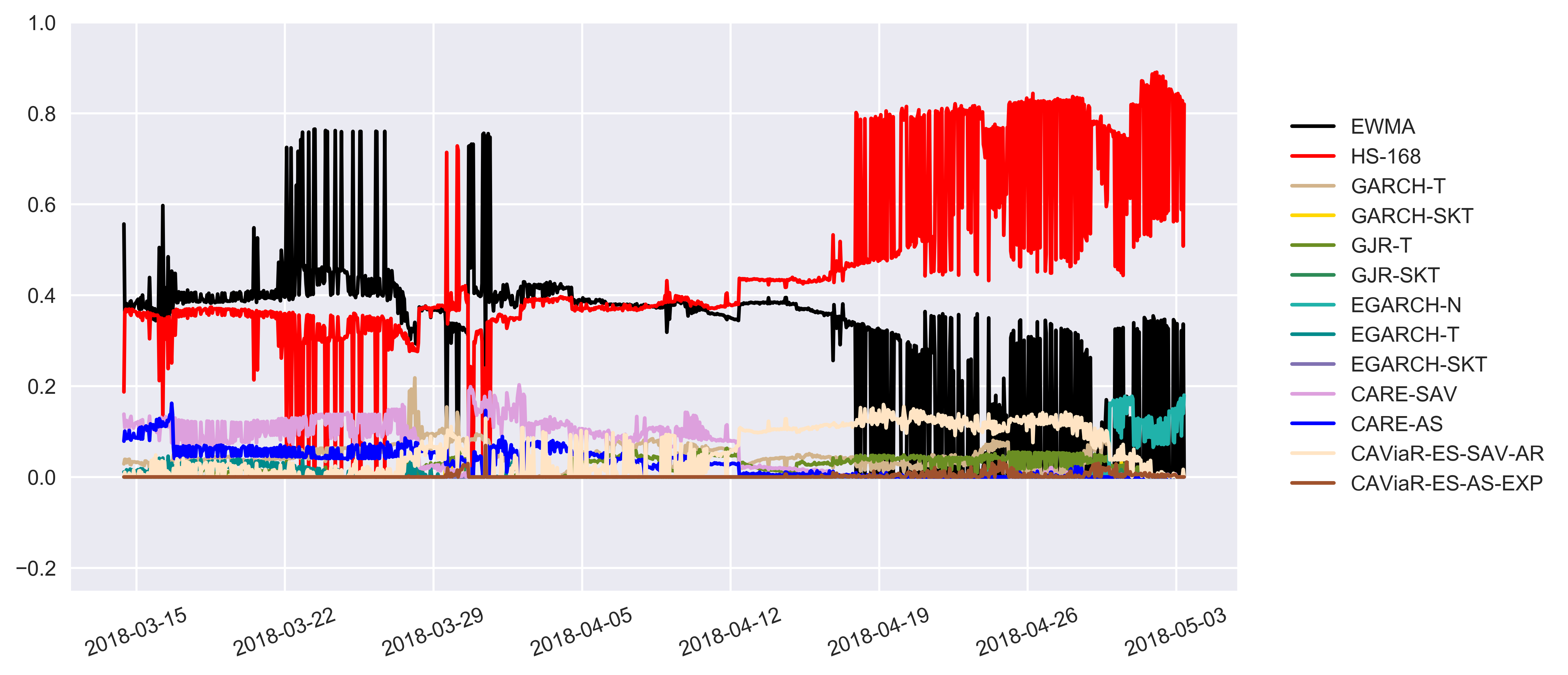}
\label{BTC_ES_99_weights}
\end{center}
\end{figure}

In summary, the 99\% ES forecasting study shows that the proposed method and the two other forecast combination approaches are competitive and stable in terms of the AL log score and the MCS. Specifically, the performances of the parametric models are highly dependent on the distribution assumption. For example, the parametric models that incorporate the skewed Student's t-distribution perform better than the others in most cases. This result may be due to the fact that the skewed Student's t-distribution fits the underlying log return residuals better than do the normal distribution and Student's t-distribution as the skewed Student's t-distribution may be more skewed than Student's t-distribution and more fat-tailed than the standard normal distribution. Recall that in Table \ref{Descriptive}, the five log return series exhibit different levels of positive-skewed and leptokurtic behaviour. In general, the semiparametric models perform as poorly as they do in the VaR forecasts, which may be because the measurement equations employed by those models cannot fit the underlying time series well, as observed in Figure \ref{BTC_99_plots}. The ES series produced by CAViaR-ES-AS-EXP is roughly lower than other plotted series even though CAViaR-ES-AS-EXP attempts to capture the time series dynamics, the measurement equation may not fit the underlying process well which results in higher forecast errors. 
The performance of HS-168 is not stable and is highly dependent on the estimation window size. For example, HS-168 is highly competitive in the BTC market but performs the worst in the ETC and ETH markets. Moreover, the three forecast combination approaches present competitive forecast accuracy and stability and perform better than most individual models across the five markets. Figure \ref{BTC_ES_99_weights} plots the weights assigned to each individual model by the proposed method. Clearly, the proposed method gains advantages by combining the individual forecasts at each time point to produce the ES forecasts. Figure \ref{BTC_99_plots} shows that the ES series produced by the proposed method is smoother than the other series (except HS-168) and has lower forecast error (closer to the return series), possibly because the proposed method attempts to generate more accurate forecasts by assigning optimal weights to the individual forecasts. Additionally, the forecasting power of the simple average and median forecast methods has been demonstrated in several empirical studies. However, their performances are subject to the individual model specifications. Hence, the proposed method is the best approach for the 99\% ES forecasts in terms of the AL log score and MCS.

\chapter{Simulation Study}\label{Simulation}
In this chapter, we report results from a simulation study at the 95\% confidence level which illustrates the general framework of the first proposed methodology, called Quantile-ES regression. The main objective is to describe how does the proposed regression franework can combine ES forecasts by treating them as VaR quantile forecasts.

Our first proposed methodology is a parametric method and based on the previous work of \cite{gerlach2014bayesian} which is presented in Table \ref{tab:ES Nominal levels}; this work allows us to treat ES series as certain level of VaR series that they are fall in and combine them by employing the existing quantile regression framework such as quantile regression \citep{halbleib2012improving}, quantile LASSO regression \citep{li20081}, etc.

The model chosen for the return series $\{{r_t}\}$ in the simulation is a EGARCH (1,1) model with the specific error distribution is the Laplace distribution, $L (0,1)$. Samples with size $n = 3500$ are simulated, specified as:
\begin{equation*}
\begin{split}
    &\ln({\sigma_{t+1}^2}) = 0.025 + 0.25\frac{|\epsilon_{t}|+0.035\epsilon_{t}}{\sigma_{t}}+0.065\ln(\sigma^2_{t})\\
    &r_{t+1}=0.025\mu_{t+1}+\epsilon_{t+1}\\
    &\epsilon_{t+1} = 0.85\sigma_{t+1}z_{t+1}\\
    &z_t\sim L (0,1),
\end{split}
\end{equation*}
where the notations used are consistent with Chapter \ref{methodology}.

With the simulated data set, we simply applied five parametric models, denoted as: EWMA, GARCH-N, GARCH-T, GJRGARCH-N and GJRGARCH-T. Additionally, four semiparametric models are employed, denoted as: CARE-AS, CARE-SAV, CAViaR-ES-SAV-AR and CAViaR-ES-AS-EXP. Those models were all defined in Chapter \ref{methodology}. 2500 one-step-ahead VaR and 2500 one-step-ahead ES forecasts are generated by employing the rolling window approach with fixed window size of 1000.

We further employ the quantile LASSO regression proposed by \cite{li20081} to combine each individual forecasts by treating them as predictor variables and the corresponding response variable is the underlying return series. We keep the notations consistent with the original paper, the quantile LASSO regression is defined as:
\begin{equation*}
    \underset{\beta_0, \beta}{\text{min}}\sum^n_{i=1}\rho_\tau(y_i - \beta_0 - \boldsymbol{\beta}^T\boldsymbol{x_i}) + \lambda ||\boldsymbol{\beta}||_1
\end{equation*}
where $y_i \in \mathcal{R}$ is the response variable, $\boldsymbol{x_i} = (x_{i1}, x_{i2}, \dots, x_{ip})$ are the predictors, $\beta_0$ is the intercept, $\boldsymbol{\beta}$ is the coefficients set of the predictor variables and $\lambda$ is the shrinkage parameter. Different from the LASSO regression, the quantile LASSO regression requires an additional parameter, $\tau$, which is the specified VaR quantile level. For the ES forecasts, $\tau$ will be the specified VaR quantile level that they are fall in. Additionally, function $\rho_\tau (*)$ is the quantile loss function specified in Section \ref{QLF}.

Therefore, with the 2500 VaR forecasts from 9 models, we apply the quantile LASSO regression (denoted as QLASSO) to combine them for comparison purpose. For the 2500 ES forecasts from 9 models, we also apply the quantile LASSO regression to combine them by treating them as VaR forecasts with different distribution assumptions. we employ the standard normal distribution with nominal level of 0.0196, standardised Student's-t distribution (DOF = 4) with nominal level of 0.0164, standardised Student's-t distribution (DOF = 6) with nominal level of 0.0175 and AL distribution with nominal level of 0.0184. They are denoted as QLASSO-N. QLASSO-T(4), QLASSO-T(6) and QLASSO-AL. 

We apply the violation rate, violation ratio, DQ4 and QLF to evaluate the forecast performance of each applied method, the details of those evaluation procedures are specified in Section \ref{backtesting}. For the VaR forecasts, we input the corresponding quantile level (5\%) to evaluate them, whereas for the ES forecasts, we input the corresponding nominal quantile level that they are fall in to evaluate them. Table \ref{simulationstudyresult} summarises the test results of those procedures.

\begin{figure}[H]
\begin{center}
\caption{Distribution of Simulated Return Process}
\includegraphics{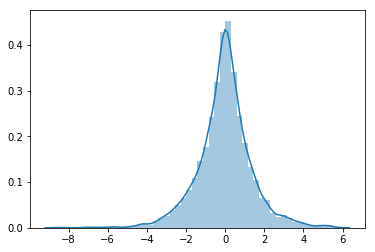}
\label{distributionofsimulatin}
\end{center}
\end{figure}

\begin{table}[H]
  \centering
  \caption{Backtesting Results Summary for the Simulation Study}
  \resizebox{\linewidth}{!}{%
    \begin{tabular}{c|ccccc|ccccc}
    \toprule
          & \multicolumn{5}{c|}{VaR}              & \multicolumn{5}{c}{ES} \\
    \midrule
    Model & VRate    & VRatio & DQ4   & QLF   & QLF Rank & ESRate    & ESRatio & DQ4   & QLF   & QLF Rank \\
    \midrule
    EWMA  & 0.065 & 1.300 & ACCEPT & 152.858 & 7     & 0.033 & 1.684 & REJECT & 74.114 & 13 \\
    GARCH-N & 0.056 & 1.120 & ACCEPT & 151.242 & 2     & 0.027 & 1.378 & ACCEPT & 72.121 & 12 \\
    GARCH-T & 0.068 & 1.360 & REJECT & 152.467 & 5     & 0.006 & 0.366 & ACCEPT & 68.970 & 8 \\
    GJRGARCH-N & 0.055 & 1.100 & ACCEPT & 151.159 & 1     & 0.028 & 1.429 & ACCEPT & 72.117 & 11 \\
    GJRGARCH-T & 0.067 & 1.340 & REJECT & 153.439 & 8     & 0.005 & 0.305 & ACCEPT & 68.220 & 6 \\
    CARE-AS & 0.115 & 2.300 & REJECT & 171.161 & 9     & 0.012 & 0.652 & ACCEPT & 69.024 & 9 \\
    CARE-SAV & 0.113 & 2.260 & REJECT & 171.550 & 10    & 0.012 & 0.652 & ACCEPT & 68.404 & 7 \\
    CAViaR-ES-SAV-AR & 0.053 & 1.060 & ACCEPT & 152.428 & 4     & 0.019 & 1.033 & ACCEPT & 67.426 & 5 \\
    CAViaR-ES-AS-EXP & 0.056 & 1.120 & ACCEPT & 152.586 & 6     & 0.018 & 0.978 & ACCEPT & 66.195 & 3 \\
    \midrule
    QLASSO & 0.053 & 1.060 & ACCEPT & 151.398 & 3     & \multicolumn{5}{c}{} \\
    \midrule
    QLASSO-N & \multicolumn{5}{c|}{\multirow{4}[2]{*}{}} & 0.017 & 0.867 & ACCEPT & 70.539 & 10 \\
    QLASSO-T(4) & \multicolumn{5}{c|}{}                 & 0.018 & 1.098 & ACCEPT & 60.937 & 1 \\
    QLASSO-T(6) & \multicolumn{5}{c|}{}                 & 0.018 & 1.029 & ACCEPT & 64.092 & 2 \\
    QLASSO-AL & \multicolumn{5}{c|}{}                 & 0.018 & 0.978 & ACCEPT & 66.892 & 4 \\
    \bottomrule
    \bottomrule
    \end{tabular}}%
  \label{simulationstudyresult}%
\end{table}%
For the VaR forecasts, the GARCH-N and GJRGARCH-N models have the best performances in terms of the applied test procedures. This is not surprise, as the simulated return series is close to normal distribution without much tailed behaviours (see Figure \ref{distributionofsimulatin}). The quantile LASSO regression also performs well as it has the violation ratio closes to the expected value of 1, accepted by the DQ4 test and the 3rd lowest loss function value. The two CARE models and two CAViaR-ES models perform differently as the CARE model estimates the expectile as the quantile while the CAViaR-ES model estimates the VaR straightly.

For the ES forecasts, the performances are different. The QLASSO-T(4) and QLASSO-T(6) perform the best as they have violation ratios close to target value of 1, lower loss function values and been accepted by the DQ4 test. Clearly, the performances of the proposed parametric quantile-ES regression are highly dependent on the distribution assumptions, the QLASSO-N performs worse which indicates the normal distribution may not be able to capture the true ES of the simulated return distribution, whereas the Student's-t distribution with DOF of 4 may fit the underlying distribution well.

The same idea can be applied to other quantile regression frameworks to combine ES forecasts. However, this methodology is still in progress and need to be improved and the forecast performance cannot be guaranteed as it does not been applied in a empirical study with real world time series data. Additionally, we will further improve this Simulation Study by simulating more time series data with different distribution and repeating thousand of times.

\chapter{Conclusion}\label{conclusion}
In this thesis, a new semiparametric joint VaR and ES combination framework is proposed for combining VaR and ES forecasts. Through the joint combination, improvements over a range of competitive individual models, including the well-developed GARCH family models, CAViaR-ES models and CARE models, in terms of both out-of-sample VaR and ES forecasting are observable. Specifically, the proposed joint combination model has strong competitive advantages over the other applied methods, as it generated accurate VaR and ES forecasts in the empirical study of 5 well-known cryptocurrencies' return series. In terms of several VaR and ES backtesting procedures, the proposed method always produced more favourable forecasts at both the 95\% and 99\% confidence levels. For example, regarding the VaR forecasting study, the proposed method generally produced VaR series that were more consistent with the statistical theoretical expectation (violation ratios closer to 1). Also, they were less likely to be rejected by the four statistical tests than the other tested models. Additionally, the quantile loss function also favours the proposed method; that is, the VaR series produced by the proposed method were closer to the log return series and were thus more accurate. On the other hand, the ES series that were the joint products of the proposed method were also favoured by the AL log score and MCS. To the best of the author's knowledge, the joint VaR-ES loss function and MCS are the best ways to assess ES forecast performances. Moreover, the proposed method regularly generates less extreme tail risk forecast series, which can allow financial institutions to preserve smaller amounts of regulatory capital for extreme market events. Another reason that the proposed joint combination framework should be preferred is the dynamic of financial time series is changing, a model with favourable performance within certain period cannot guarantee the performance of next period. Thus, a method such as the proposed joint combination framework that can generate accurate forecasts under different conditions and are robust to different markets should be considered by the financial institutions. 

Another parametric ES combination regression framework named as Quantile-ES regression has been presented via a simulation study which can be regarded as an extension of the existing quantile regression frameworks. By treating ES forecasts as VaR quantile forecasts, combined ES series can be produced by employing the quantile regression framework. However, this method still need to be further improved with several aspects due to the time limitation, the author will implement this in the future study instead of in this thesis.

In addition to the proposed joint VaR and ES combination framework and Quantile-ES regression framework, this thesis implements a risk management analysis for the emerging cyptocurrency market by applying the existing well-established VaR and ES forecast methods as well as the proposed method. The cryptocurrency market is well known as having higher risk (as well as higher returns) than the traditional financial stock market; therefore, the need for risk management in the cryptocurrency market is increased. This thesis further highlights that the use of currently well-established risk management procedures, such as tail risk measures, is necessary and efficient even though the maturity of the cryptocurrency market is debatable. We found that the performance of parametric models is highly dependent on the distribution assumptions. Nonparametric models, especially semiparametric models, perform worse in the cryptocurrency market than in the traditional financial stock market, potentially because the measurement equations employed by those models may not fit the cryptocurrency return series well. 

In conclusion, the proposed joint combination method should be considered by financial institutions for forecasting tail risk, as the proposed method can generate accurate and stable VaR and ES series jointly. VaR and ES can jointly provide more comprehensive risk quantification, which allows financial institutions to allocate capital efficiently under the Basel \RomanNumeralCaps3 Capital Accord to deal with extreme market movements. Further, this work could start a new chapter in forecast combination research, as there is sparse literature (or even no literature) addressing the issue of how to combine ES forecasts. 

The future extension of this work can be considered as follows. On the one hand, the proposed joint combination method can be extended by adding a regularisation term (or even double regularisations) to further enhance the forecast accuracy. In addition, how to decide the value(s) of the regulisation parameter(s), such as time series cross-validation, can be a simultaneous extension. On the other hand, dependence modelling terms, such as copulas, can be considered in the risk modelling of cryptocurrency market, as it is well-known that the cryptocurrency market may exhibit higher potential correlations. For example, the movements of other currencies in the market may highly depend on the movement of Bitcoin. Additionally, interested researchers can explore the measurement equation of the cryptocurrency market.   

\bibliographystyle{agsm}
\bibliography{references.bib}

\appendix
\chapter{Data Analysis}\label{AppendixA}

\begin{figure}[H]
\begin{center}
\caption{Log Return Distributions}
\includegraphics[width=\textwidth,height=0.55\textheight]{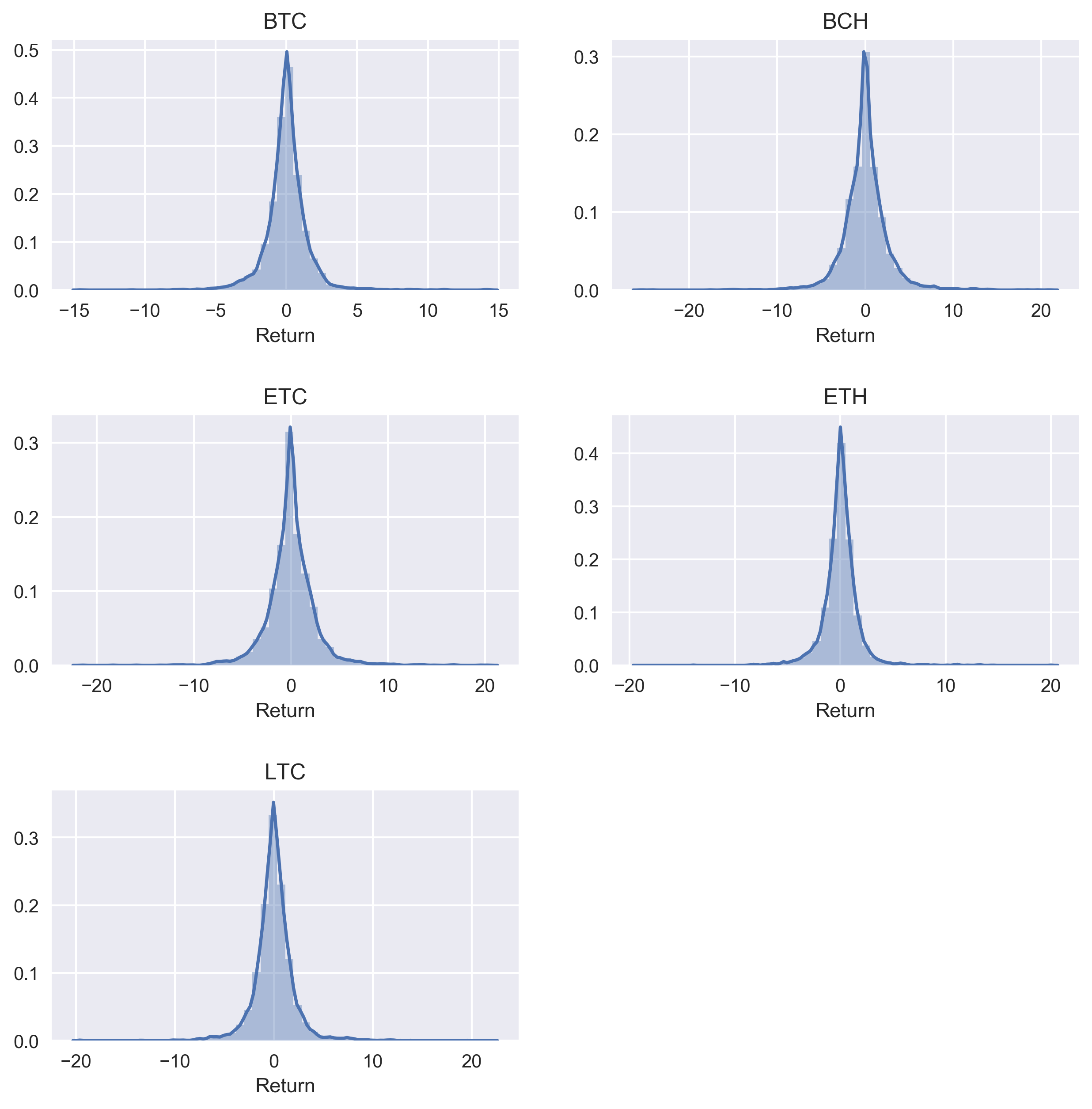}
\label{Distribution Plots}
\end{center}
\end{figure}

\begin{figure}[H]
\begin{center}
\caption{ACF and PACF of the Squared Returns}
\includegraphics[width=\textwidth,height=0.9\textheight]{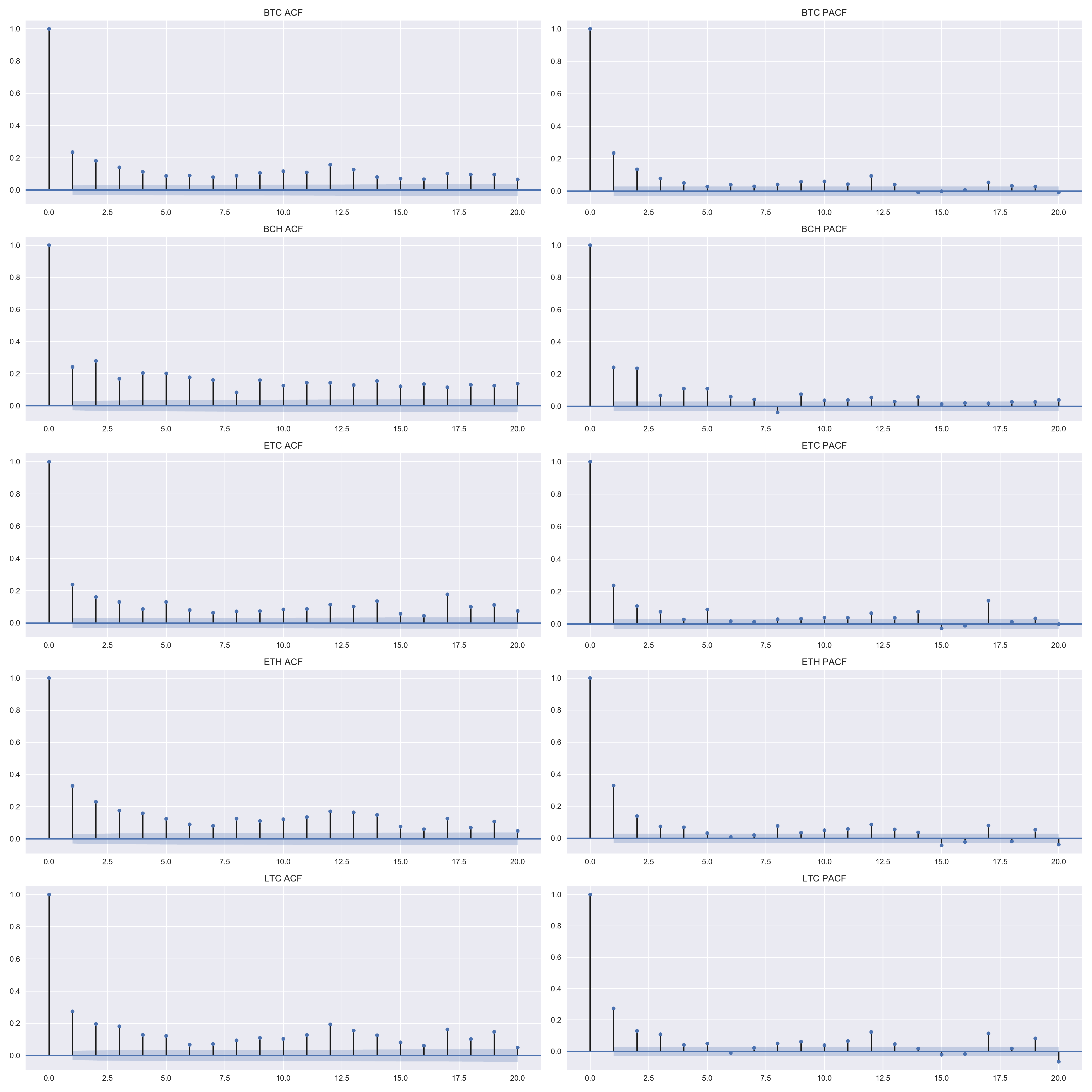}
\label{ACF and PACF}
\end{center}
\end{figure}

\begin{table}[H]
\centering
\caption {Descriptive Statistics Summary Table for Student's t-Distribution DOF}
\begin{tabular}{cccccc}
    \toprule
          & BTC   & BCH   & ETC   & ETH   & LTC \\
    \midrule
    count & 1200.000 & 1200.000 & 1200.000 & 1200.000 & 1200.000 \\
    mean  & 4.182 & 3.667 & 4.022 & 4.618 & 4.209 \\
    std   & 0.593 & 0.313 & 0.129 & 0.306 & 0.215 \\
    min   & 3.383 & 3.236 & 3.679 & 4.146 & 3.823 \\
    25\%  & 3.601 & 3.426 & 3.923 & 4.333 & 4.032 \\
    50\%  & 4.110 & 3.557 & 4.020 & 4.537 & 4.113 \\
    75\%  & 4.782 & 3.913 & 4.117 & 4.867 & 4.380 \\
    max   & 5.275 & 4.291 & 4.349 & 5.179 & 4.660 \\
\bottomrule
\bottomrule
\end{tabular}%
\label{t DOF}%
\end{table}

\begin{table}[H]
\centering
\caption{DOF of the Skewed Student's t-Distribution}
\begin{tabular}{cccccc}
    \toprule
          & BTC   & BCH   & ETC   & ETH   & LTC \\
    \midrule
    count & 1200.000 & 1200.000 & 1200.000 & 1200.000 & 1200.000 \\
    mean  & 4.215 & 3.669 & 4.022 & 4.694 & 4.209 \\
    std   & 0.625 & 0.312 & 0.130 & 0.325 & 0.215 \\
    min   & 3.383 & 3.237 & 3.674 & 4.198 & 3.822 \\
    25\%  & 3.604 & 3.428 & 3.921 & 4.377 & 4.033 \\
    50\%  & 4.130 & 3.558 & 4.019 & 4.607 & 4.114 \\
    75\%  & 4.843 & 3.909 & 4.117 & 4.967 & 4.381 \\
    max   & 5.382 & 4.289 & 4.352 & 5.274 & 4.657 \\
    \bottomrule
    \bottomrule
    \end{tabular}%
  \label{skt DOF}%
\end{table}

\chapter{Forecast Plots}\label{appendixB plot}
\section*{95\% VaR Forecasts}
\subsection*{BCH}
\begin{figure}[H]
\begin{center}
\caption{95\% VaR Forecasts for the BCH Market}
\includegraphics[width=\textwidth,height=0.3\textheight]{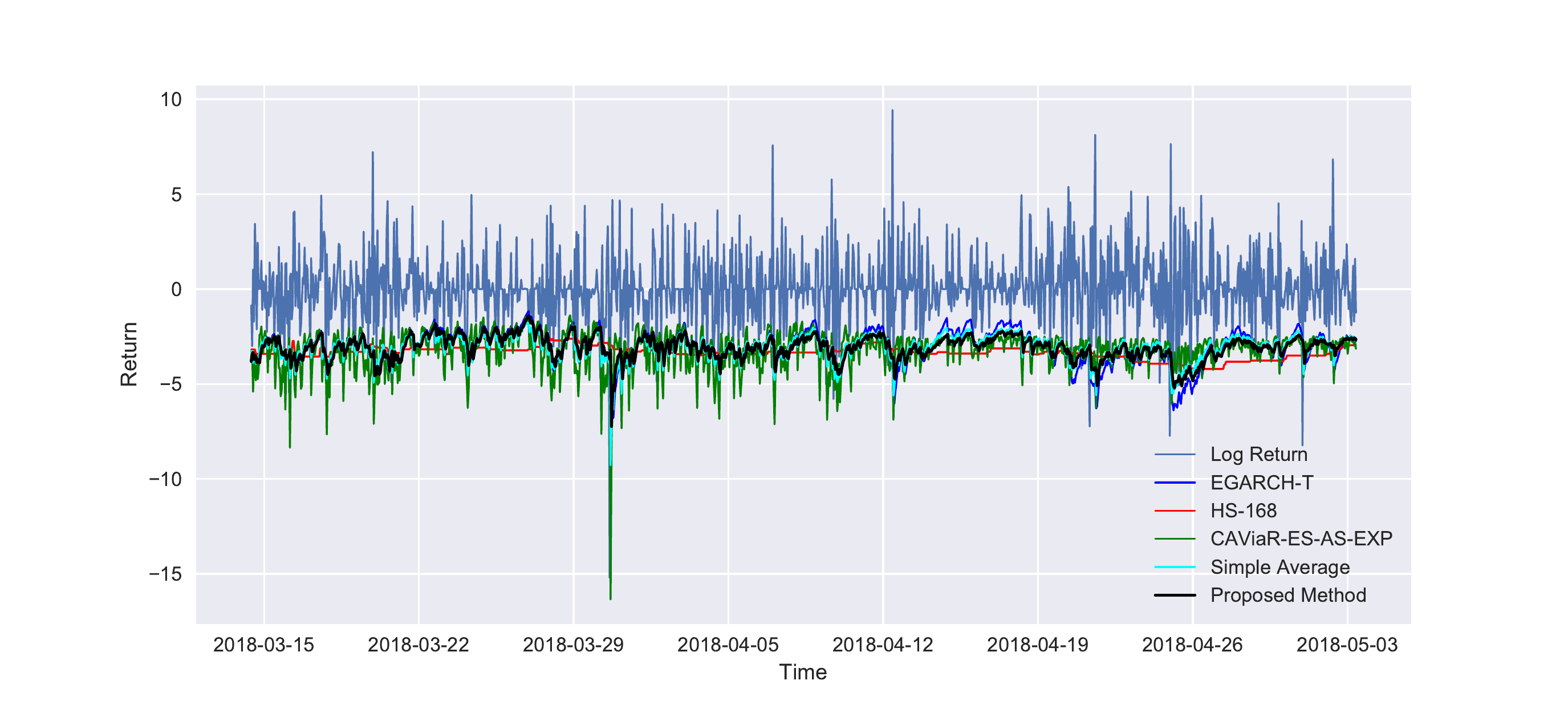}
\label{BCH_VaR_95}
\end{center}
\end{figure}

\begin{figure}[H]
\begin{center}
\caption{Individual Weights of the Combined VaR Forecasts at the 95\% Confidence Level for the BCH Market}
\includegraphics[width=\textwidth,height=0.3\textheight]{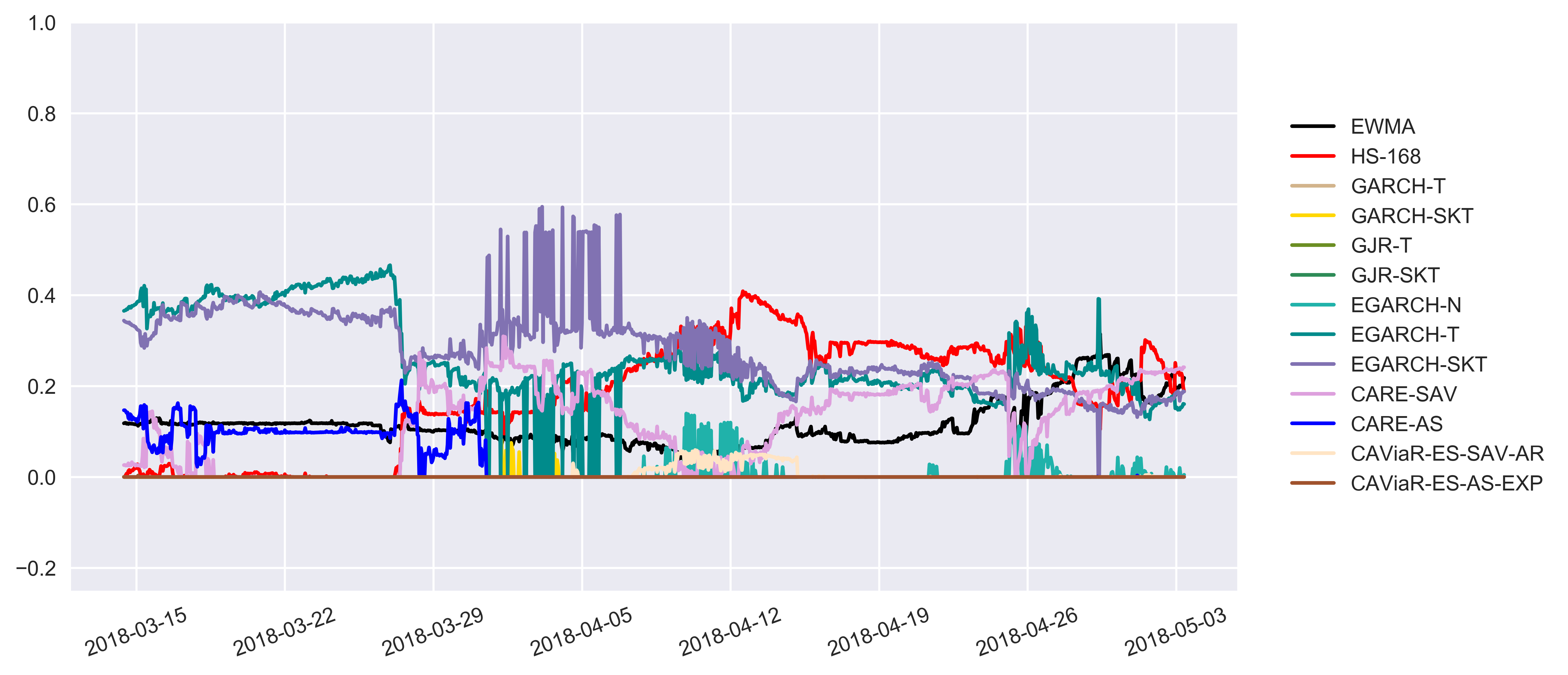}
\label{BCH_VaR_95_Weight}
\end{center}
\end{figure}

\subsection*{ETC}
\begin{figure}[H]
\begin{center}
\caption{95\% VaR Forecasts for the ETC Market}
\includegraphics[width=\textwidth,height=0.3\textheight]{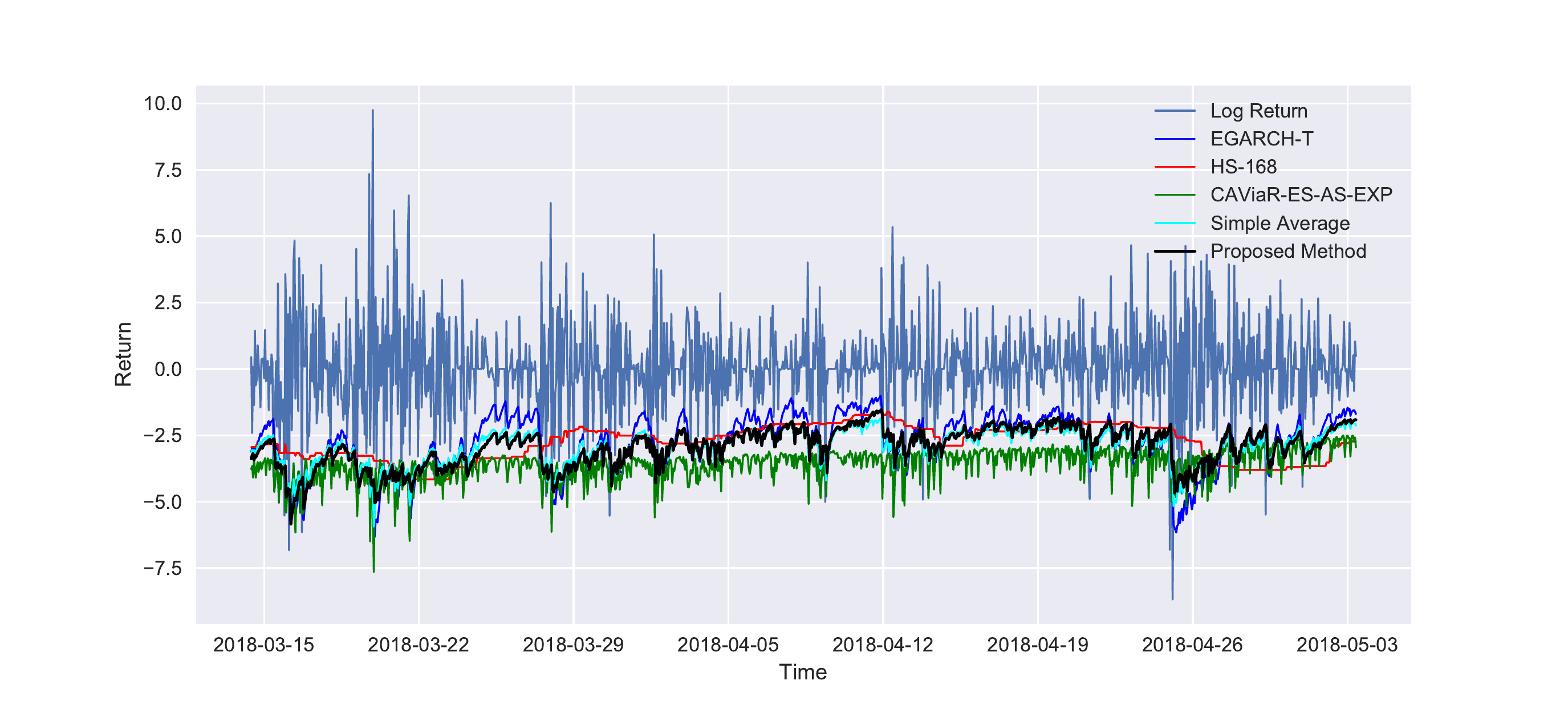}
\label{ETC_VaR_95}
\end{center}
\end{figure}

\begin{figure}[H]
\begin{center}
\caption{Individual Weights of the Combined VaR Forecasts at the 95\% Confidence Level for the ETC Market}
\includegraphics[width=\textwidth,height=0.3\textheight]{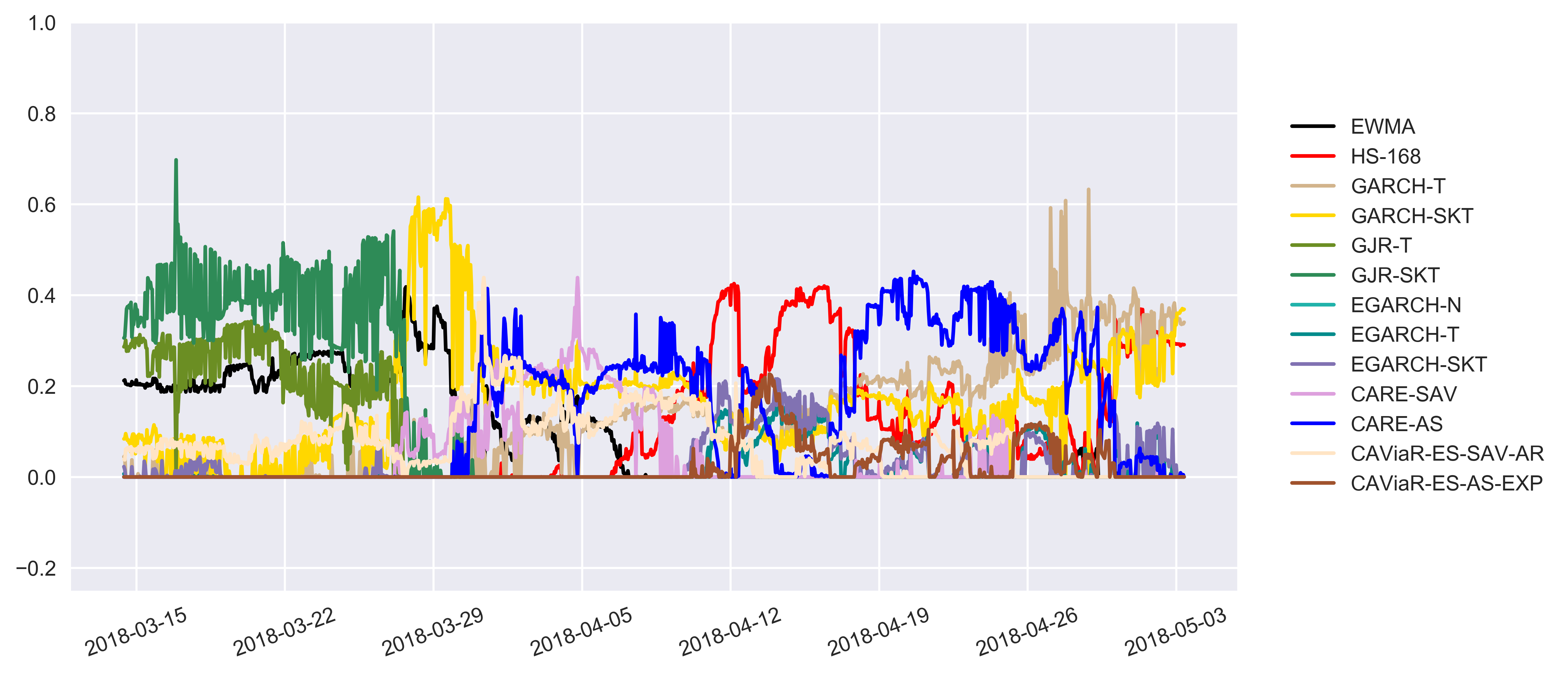}
\label{ETC_VaR_95_Weight}
\end{center}
\end{figure}

\subsection*{ETH}
\begin{figure}[H]
\begin{center}
\caption{95\% VaR forecasts for the ETH Market}
\includegraphics[width=\textwidth,height=0.3\textheight]{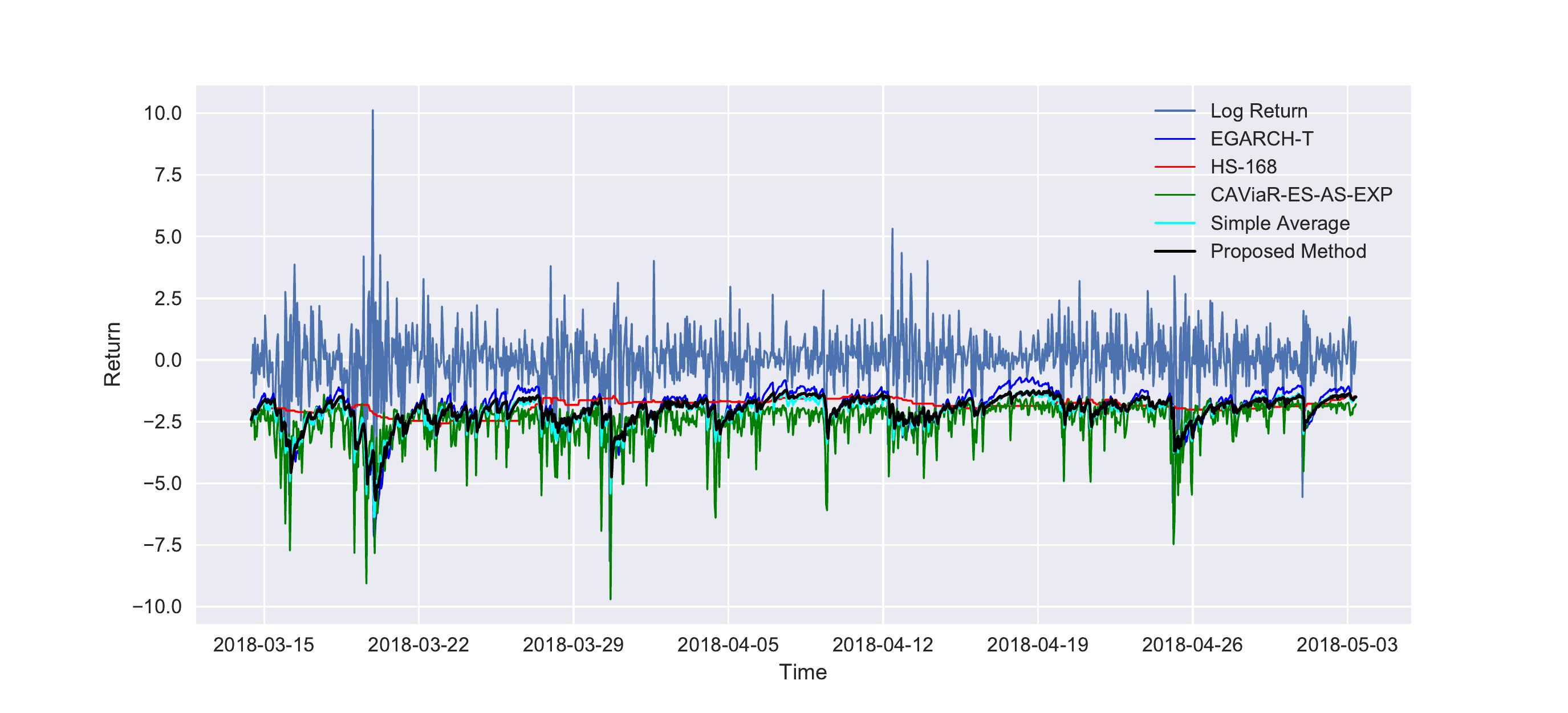}
\label{ETH_VaR_95}
\end{center}
\end{figure}

\begin{figure}[H]
\begin{center}
\caption{Individual Weights of the Combined VaR Forecasts at the 95\% Confidence Level for the ETH Market}
\includegraphics[width=\textwidth,height=0.3\textheight]{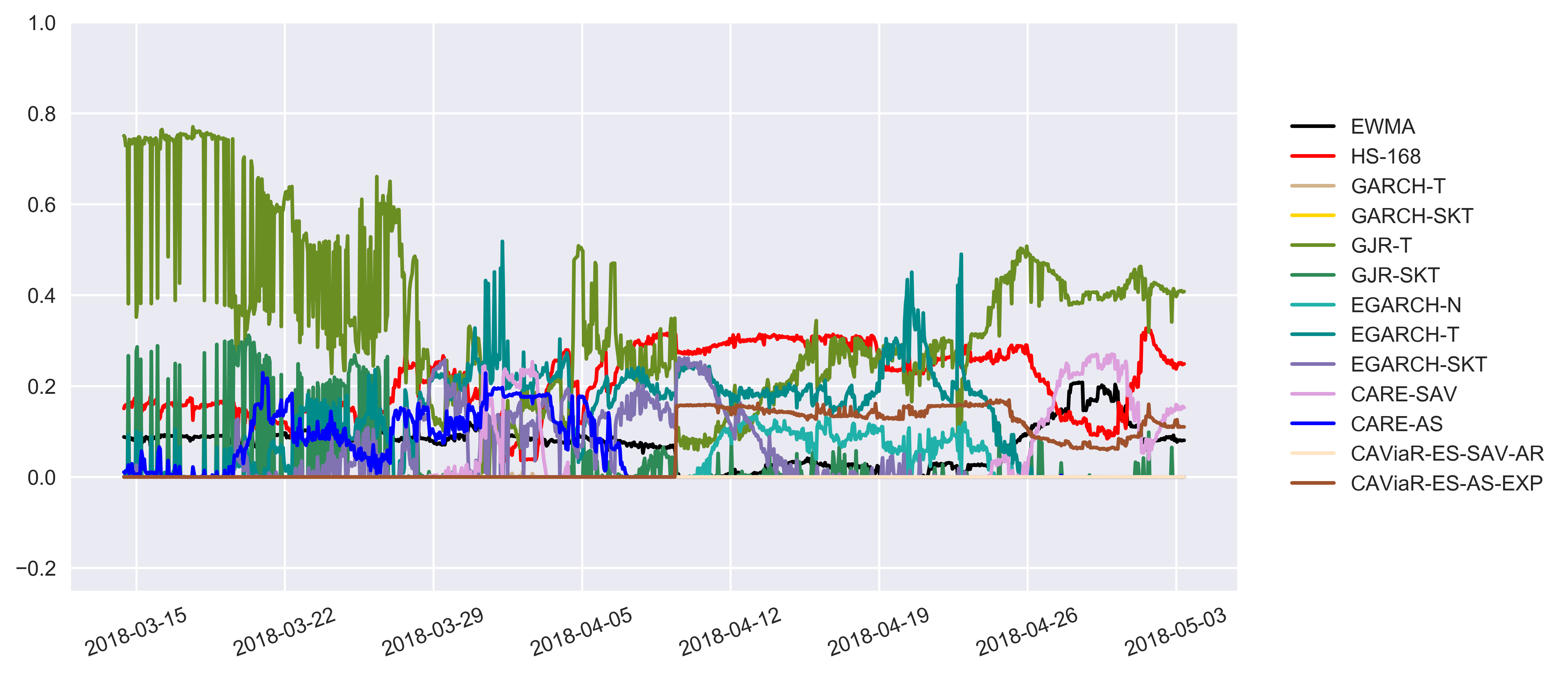}
\label{ETH_VaR_95_Weight}
\end{center}
\end{figure}

\subsection*{LTC}
\begin{figure}[H]
\begin{center}
\caption{95\% VaR Forecasts for the LTC Market}
\includegraphics[width=\textwidth,height=0.3\textheight]{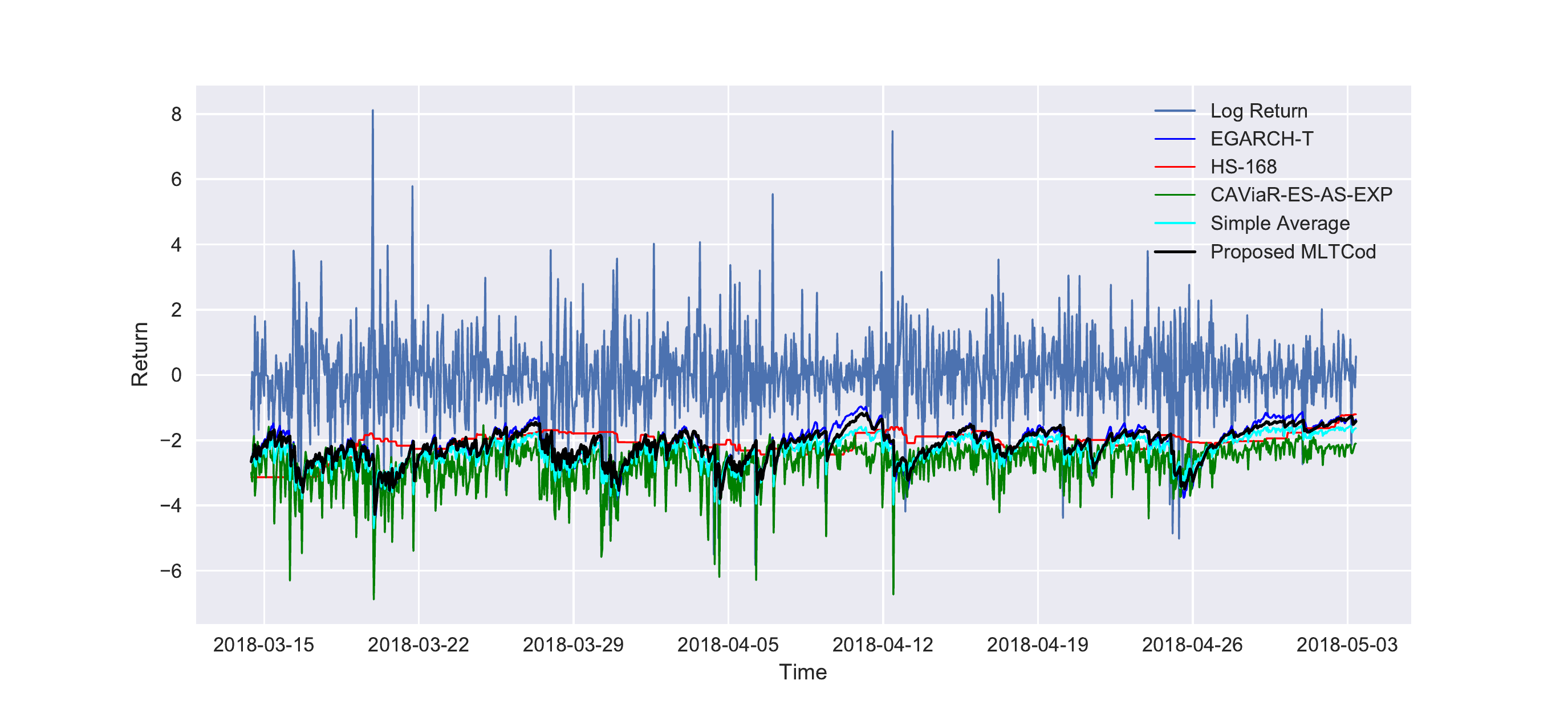}
\label{LTC_VaR_95}
\end{center}
\end{figure}

\begin{figure}[H]
\begin{center}
\caption{Individual Weights of the Combined VaR Forecasts at the 95\% Confidence Level for the LTC Market}
\includegraphics[width=\textwidth,height=0.3\textheight]{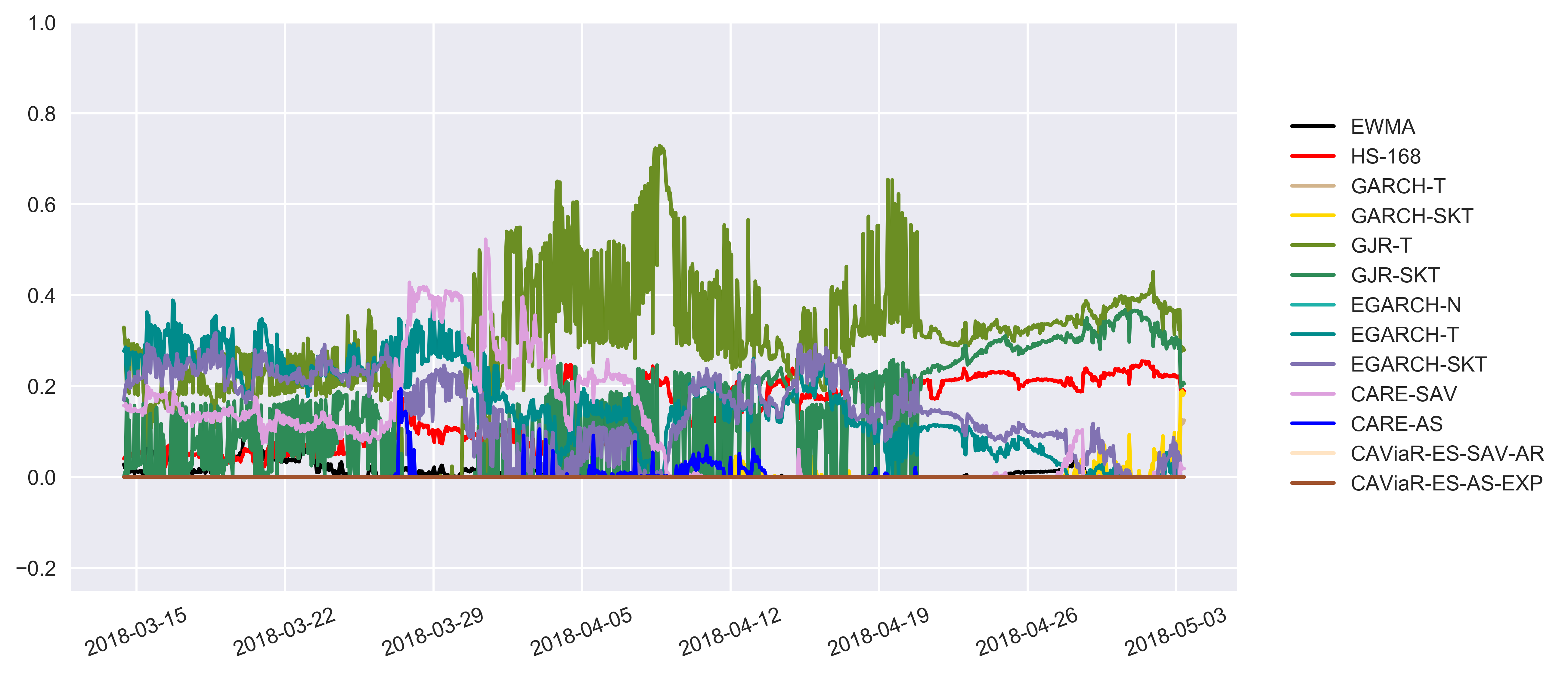}
\label{LTC_VaR_95_Weight}
\end{center}
\end{figure}

\section*{99\% VaR Forecasts}
\subsection*{BCH}
\begin{figure}[H]
\begin{center}
\caption{99\% VaR forecasts for the BCH Market}
\includegraphics[width=\textwidth,height=0.3\textheight]{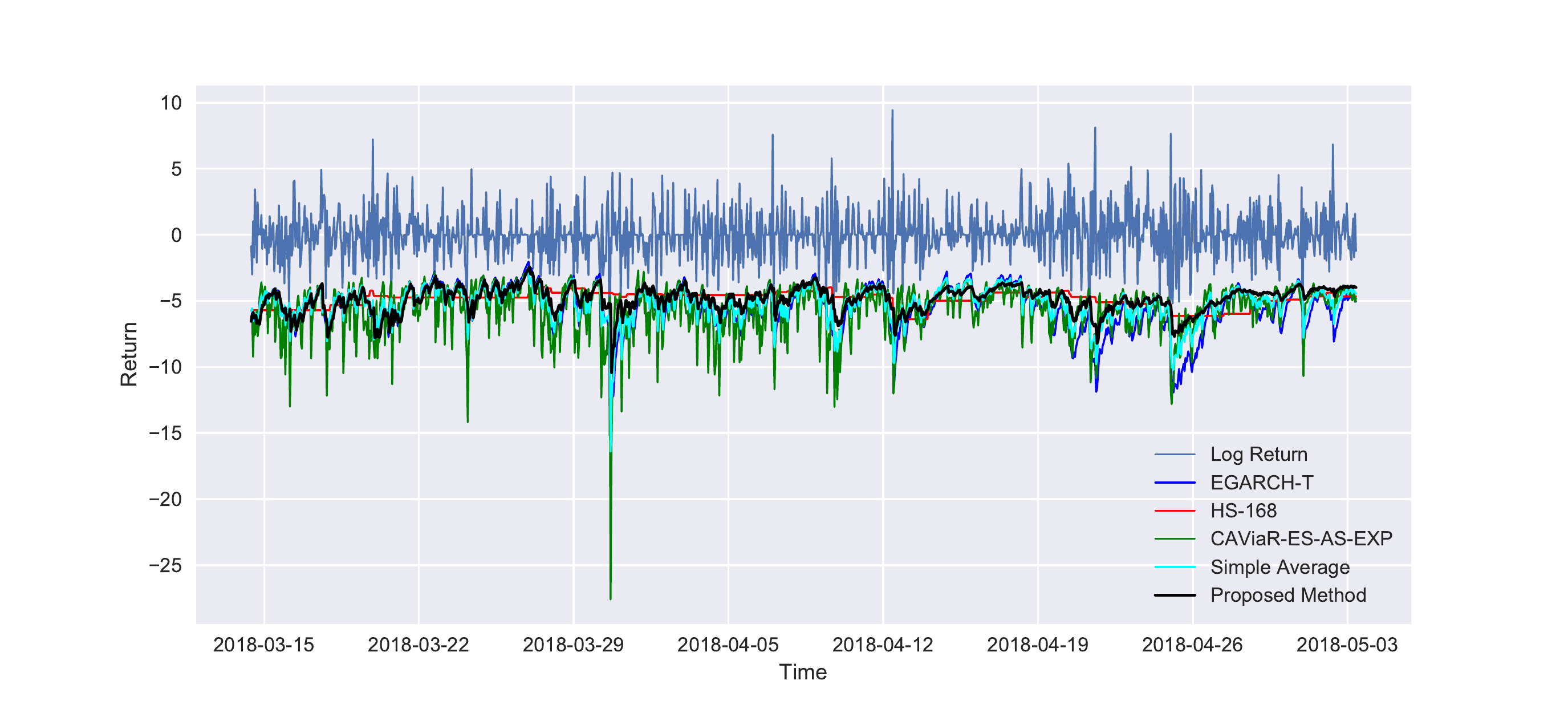}
\label{BCH_VaR_99}
\end{center}
\end{figure}

\begin{figure}[H]
\begin{center}
\caption{Individual Weights of the Combined VaR Forecasts at the 99\% Confidence Level for the BCH Market}
\includegraphics[width=\textwidth,height=0.3\textheight]{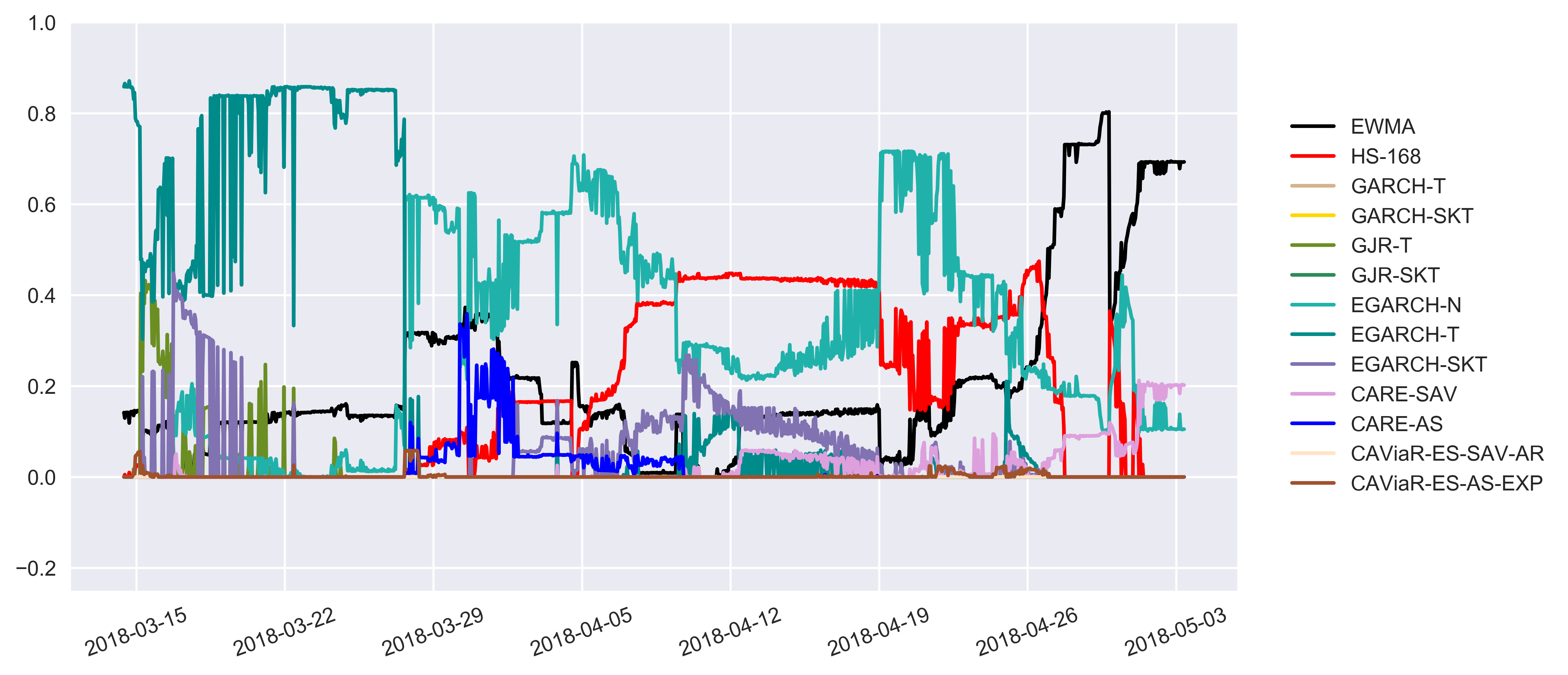}
\label{BCH_VaR_99_Weight}
\end{center}
\end{figure}

\subsection*{ETC}
\begin{figure}[H]
\begin{center}
\caption{99\% VaR Forecasts for the ETC Market}
\includegraphics[width=\textwidth,height=0.3\textheight]{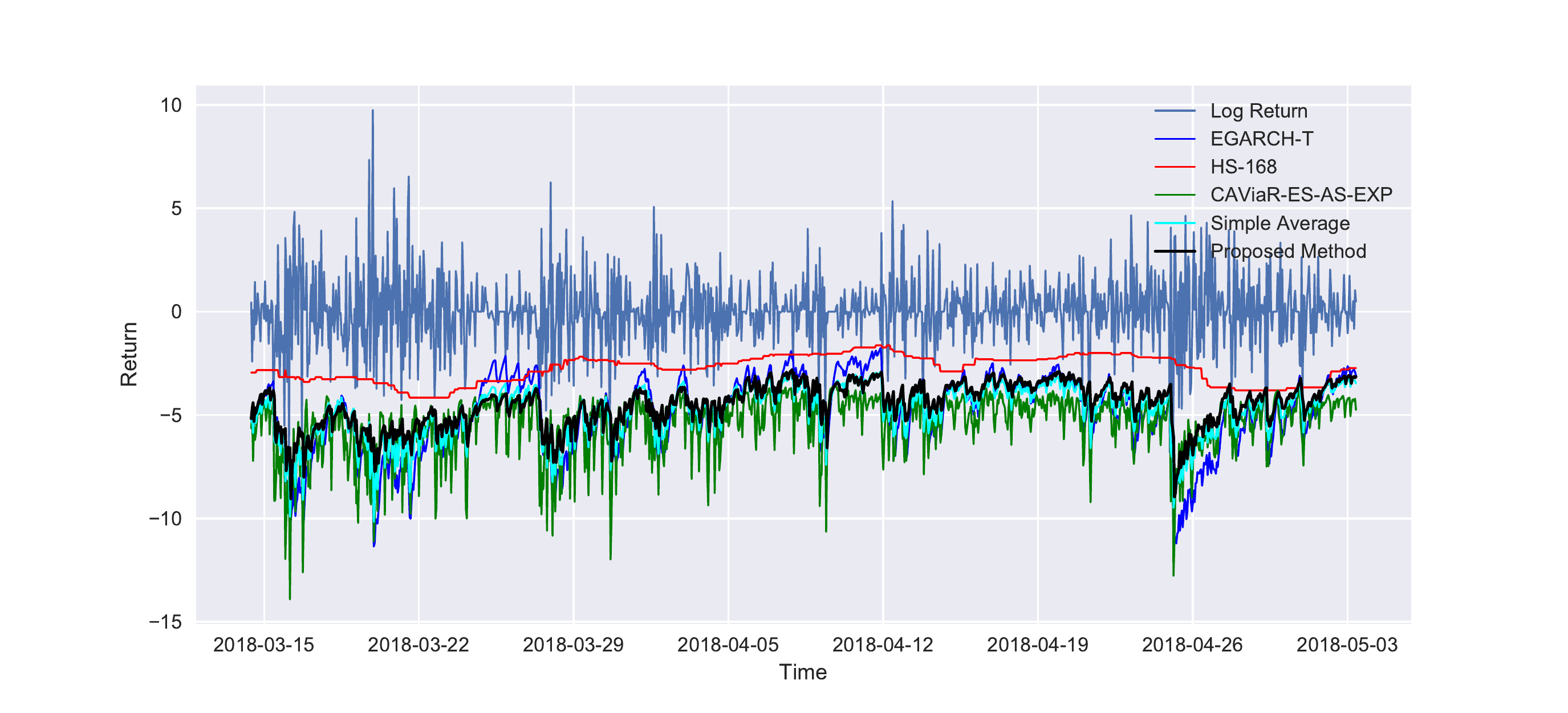}
\label{ETC_VaR_99}
\end{center}
\end{figure}

\begin{figure}[H]
\begin{center}
\caption{Individual Weights of the Combined VaR Forecasts at the 99\% Confidence Level for the ETC Market}
\includegraphics[width=\textwidth,height=0.3\textheight]{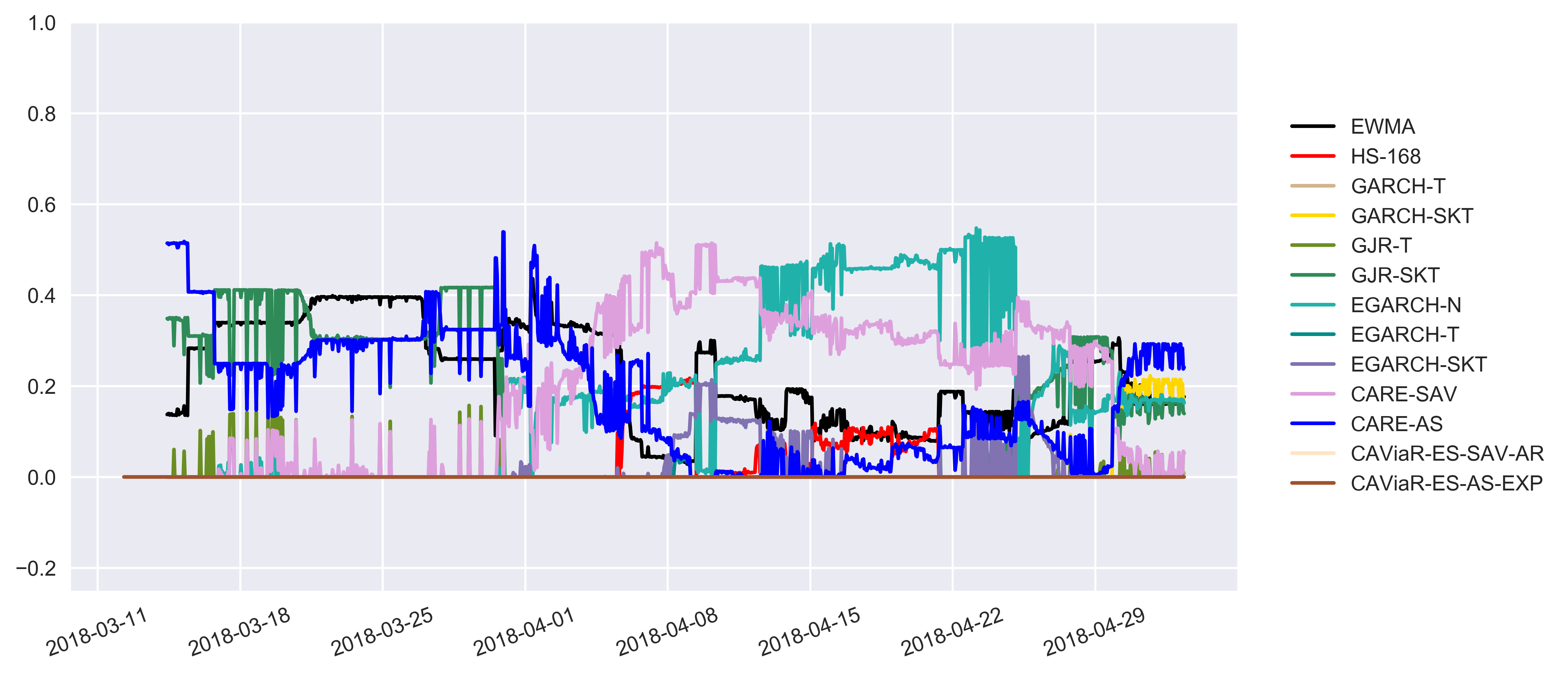}
\label{ETC_VaR_99_Weight}
\end{center}
\end{figure}

\subsection*{ETH}
\begin{figure}[H]
\begin{center}
\caption{99\% VaR Forecasts for the ETH Market}
\includegraphics[width=\textwidth,height=0.3\textheight]{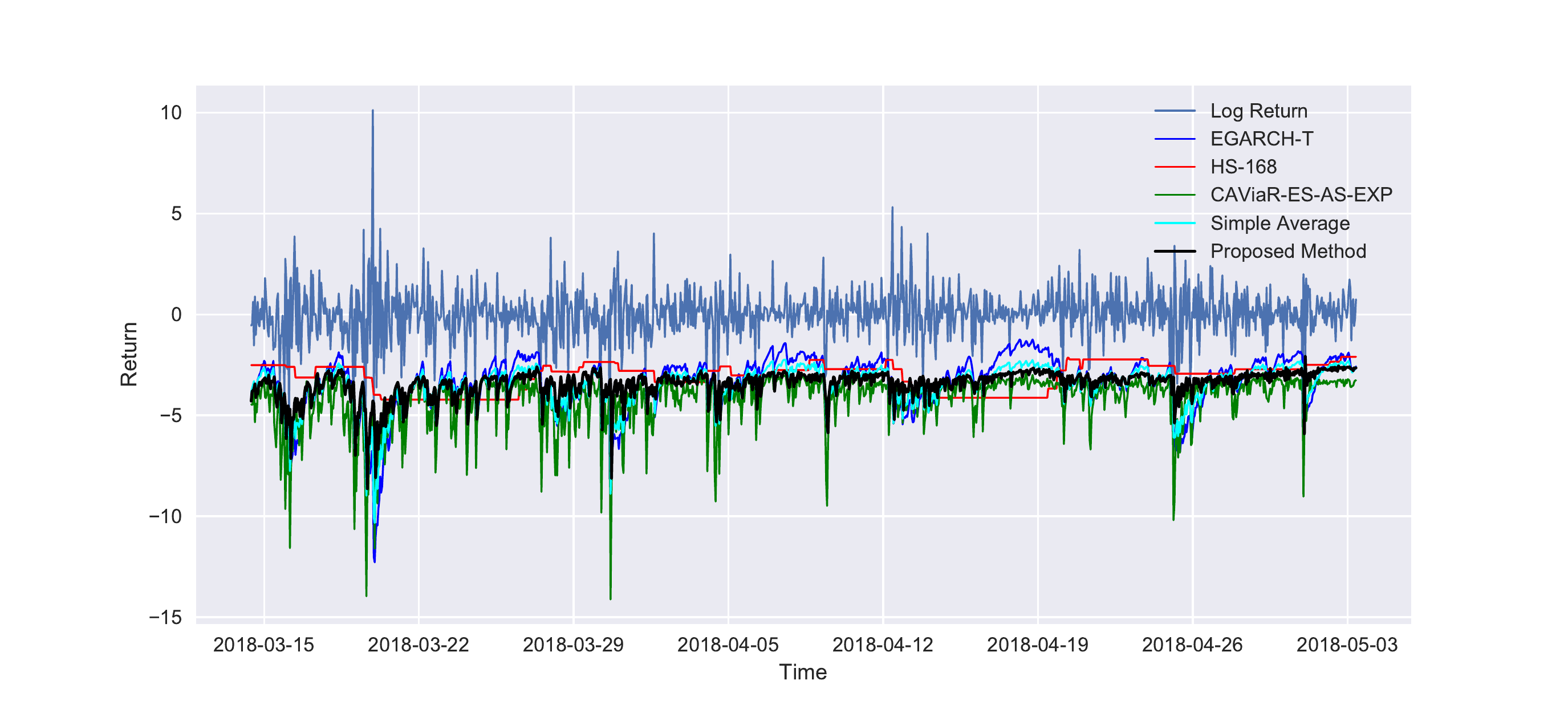}
\label{ETH_VaR_99}
\end{center}
\end{figure}

\begin{figure}[H]
\begin{center}
\caption{Individual Weights of the Combined VaR Forecasts at the 99\% Confidence Level for the ECH Market}
\includegraphics[width=\textwidth,height=0.3\textheight]{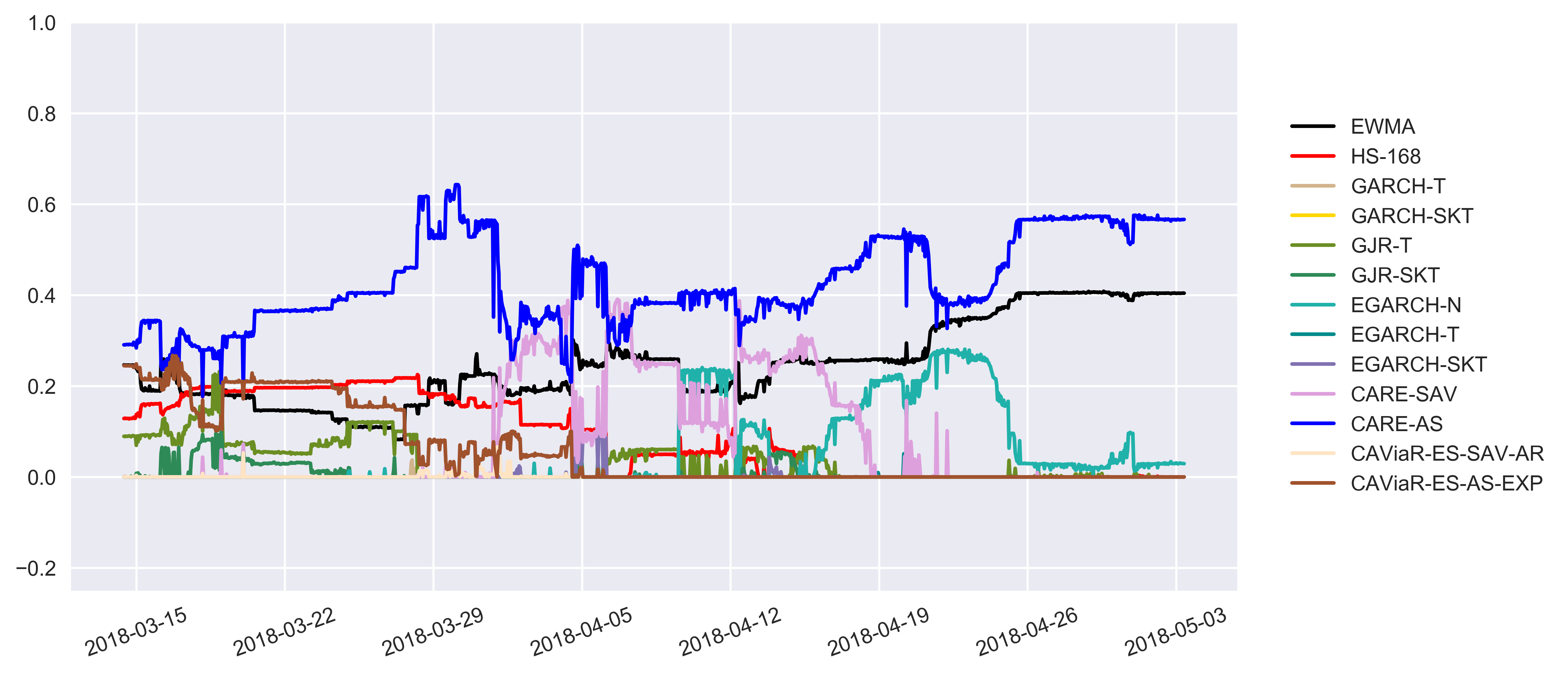}
\label{ETH_VaR_99_Weight}
\end{center}
\end{figure}

\subsection*{LTC}
\begin{figure}[H]
\begin{center}
\caption{99\% VaR Forecasts for the LTC Market}
\includegraphics[width=\textwidth,height=0.3\textheight]{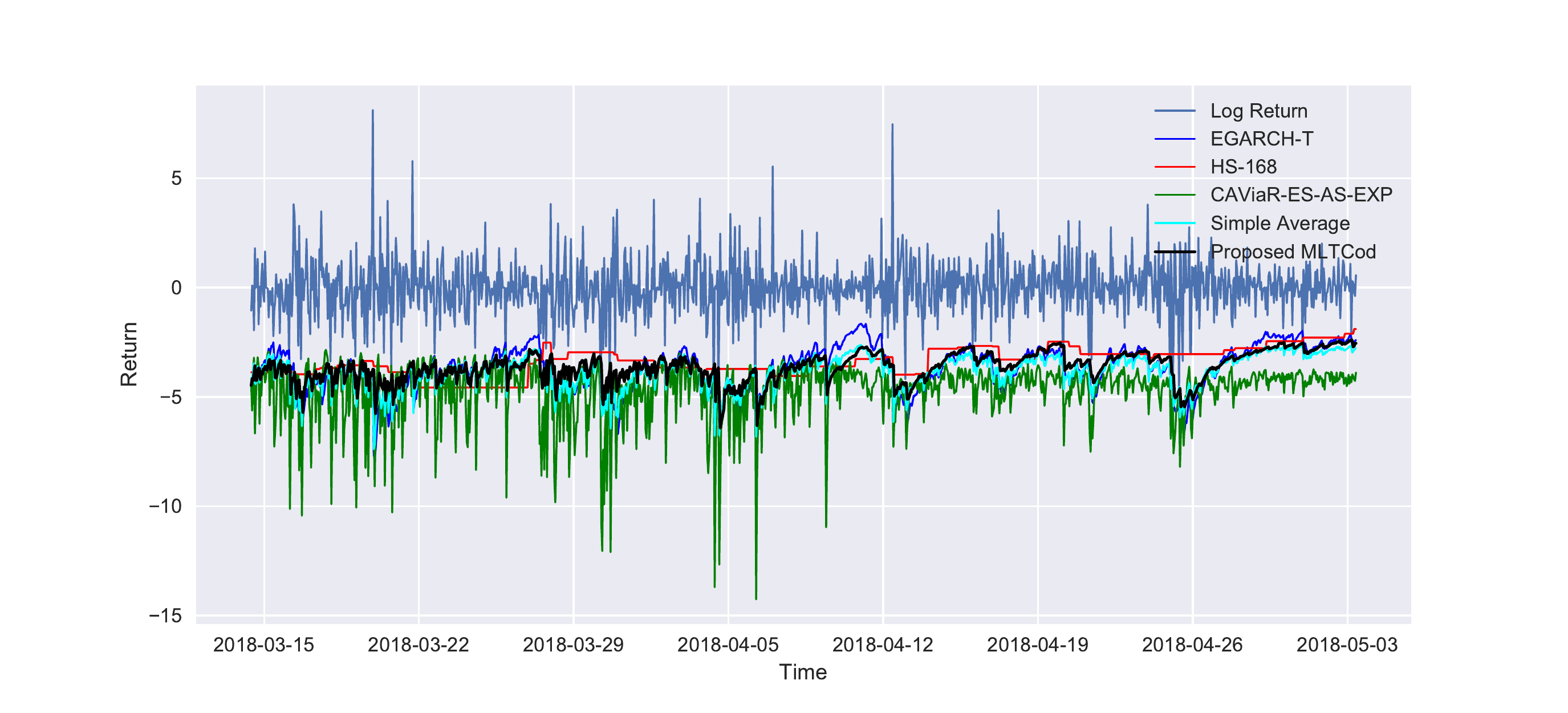}
\label{LTC_VaR_99}
\end{center}
\end{figure}

\begin{figure}[H]
\begin{center}
\caption{Individual Weights of the Combined VaR Forecasts at the 99\% Confidence Level for the LTC Market}
\includegraphics[width=\textwidth,height=0.3\textheight]{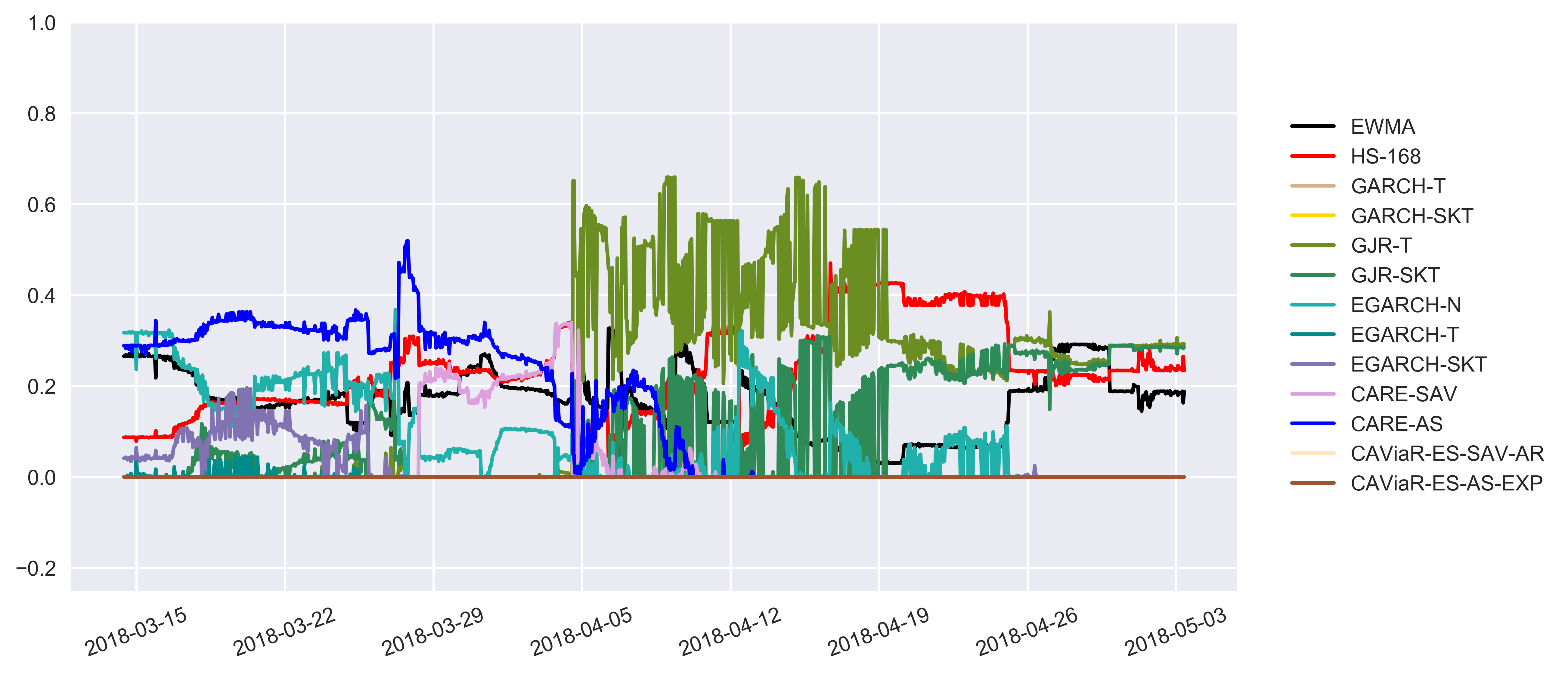}
\label{LTC_VaR_99_Weight}
\end{center}
\end{figure}

\section*{95\% ES Forecasts}

\subsection*{BCH}
\begin{figure}[H]
\begin{center}
\caption{95\% ES Forecasts for the BCH Market}
\includegraphics[width=\textwidth,height=0.3\textheight]{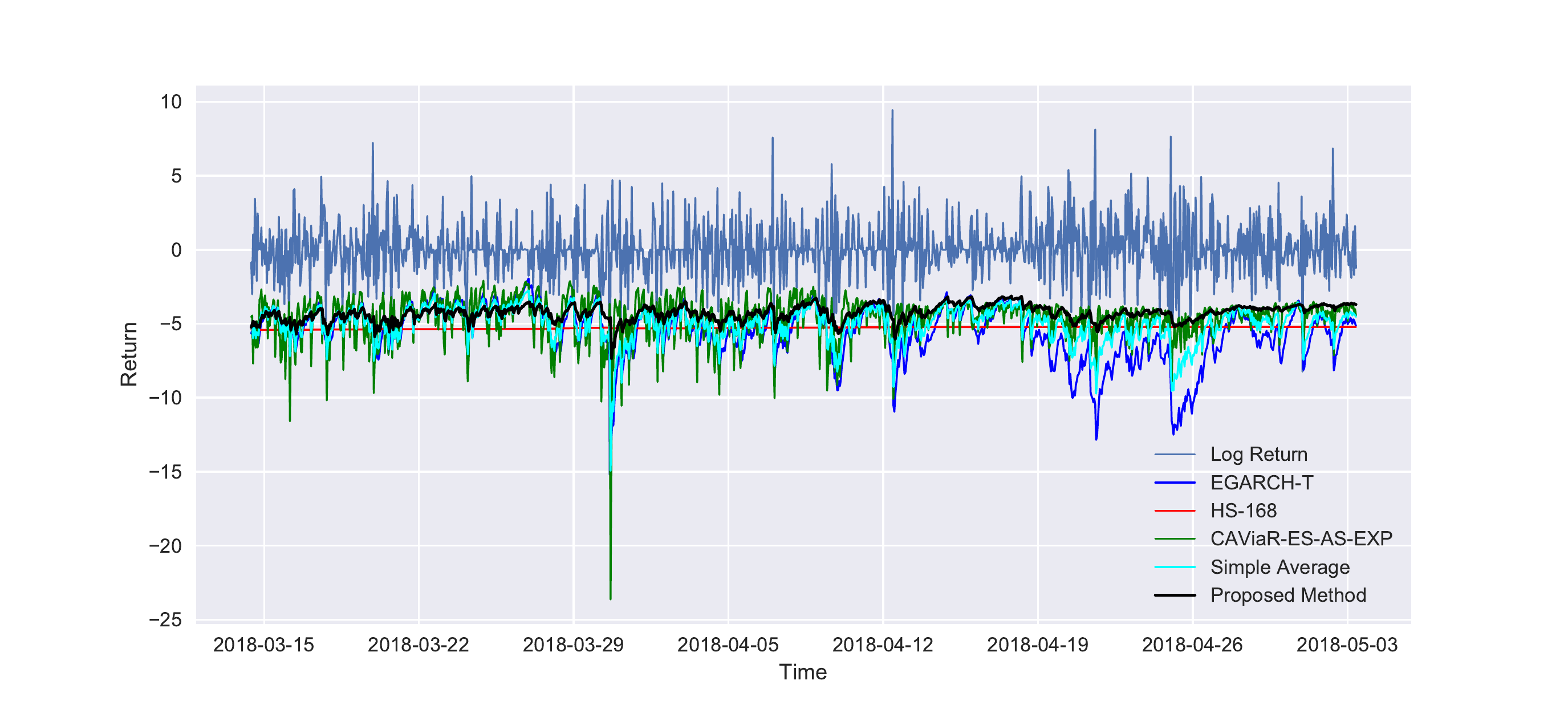}
\label{BCH_ES_95}
\end{center}
\end{figure}

\begin{figure}[H]
\begin{center}
\caption{Individual Weights of the Combined ES Forecasts at the 95\% Confidence Level for the BCH Market}
\includegraphics[width=\textwidth,height=0.3\textheight]{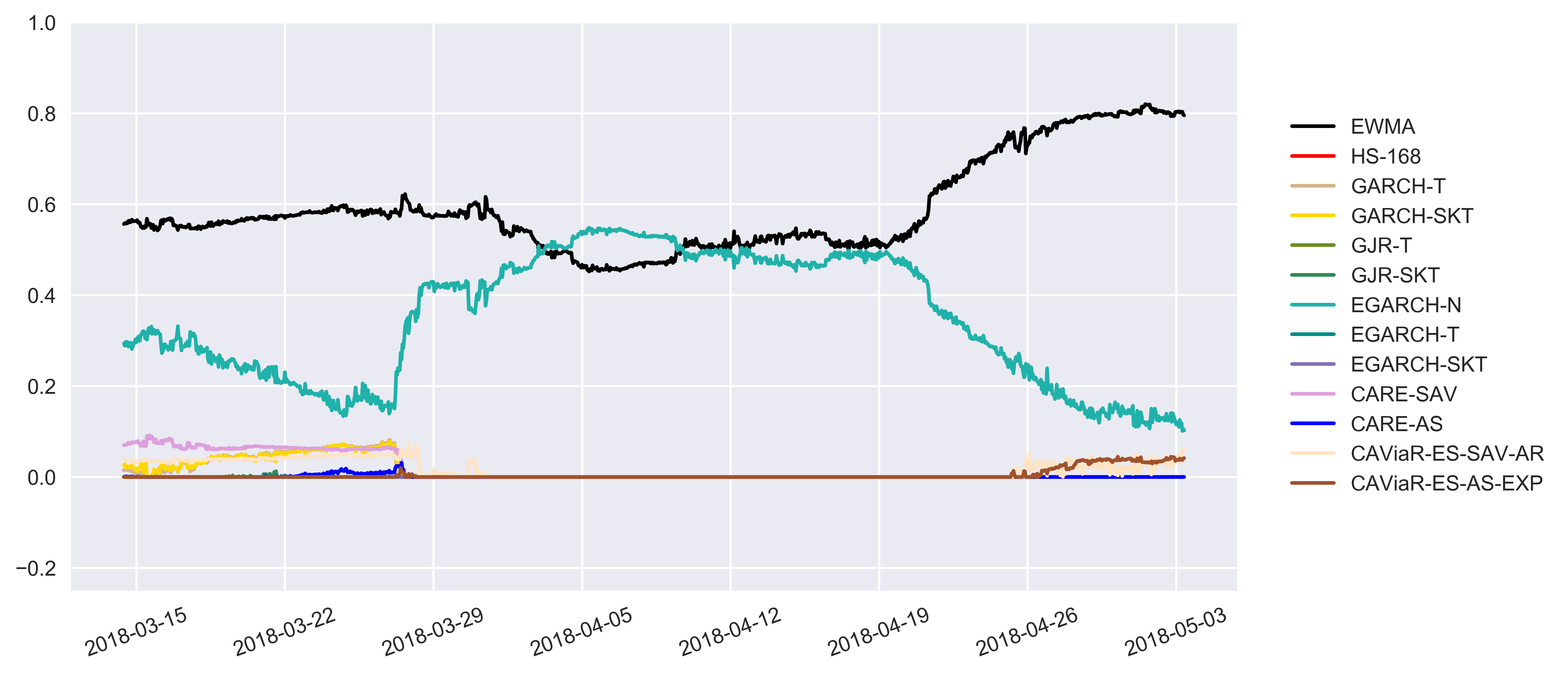}
\label{BCH_ES_95_Weight}
\end{center}
\end{figure}

\subsection*{ETC}
\begin{figure}[H]
\begin{center}
\caption{95\% ES Forecasts for the ETC Market}
\includegraphics[width=\textwidth,height=0.3\textheight]{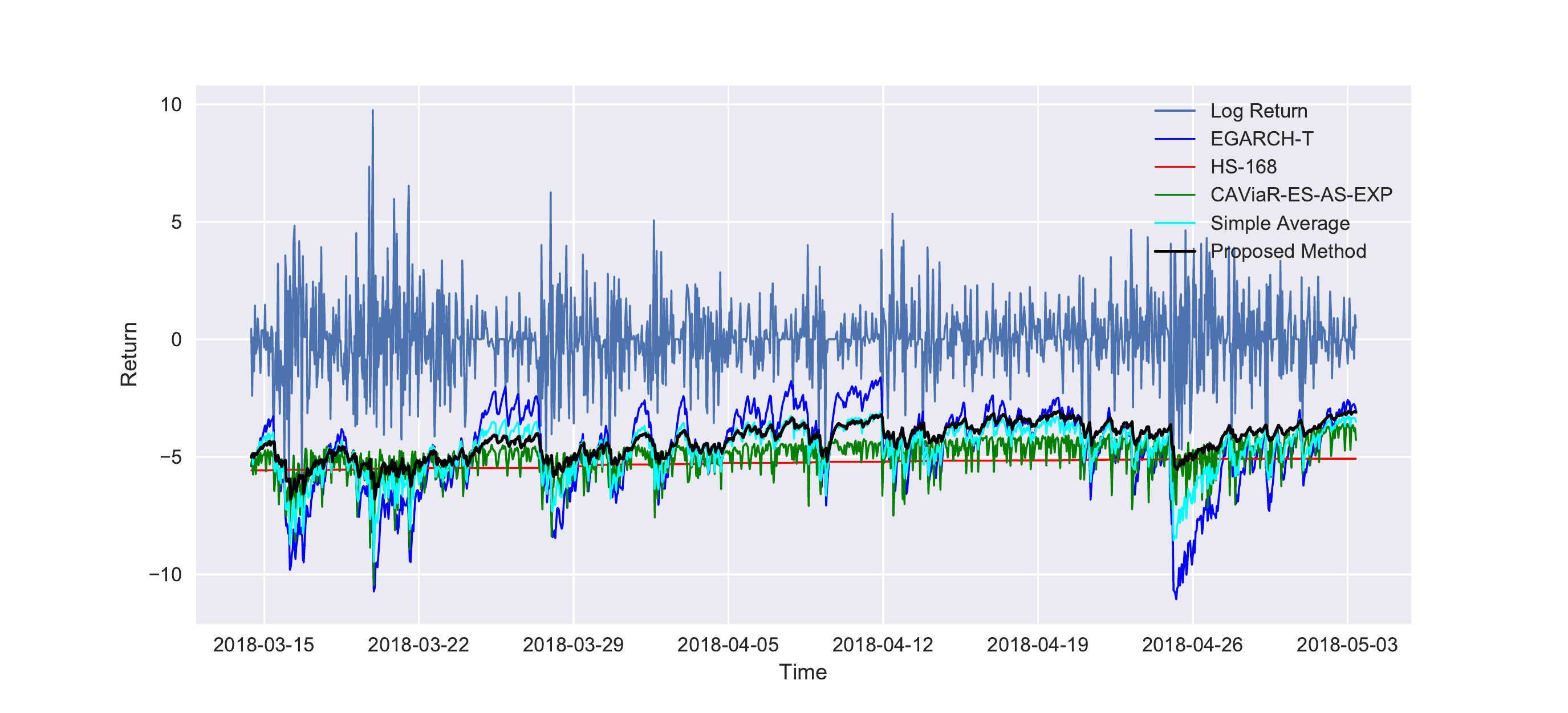}
\label{ETC_ES_95}
\end{center}
\end{figure}

\begin{figure}[H]
\begin{center}
\caption{Individual Weights of the Combined ES Forecasts at the 95\% Confidence Level for the ETC Market}
\includegraphics[width=\textwidth,height=0.3\textheight]{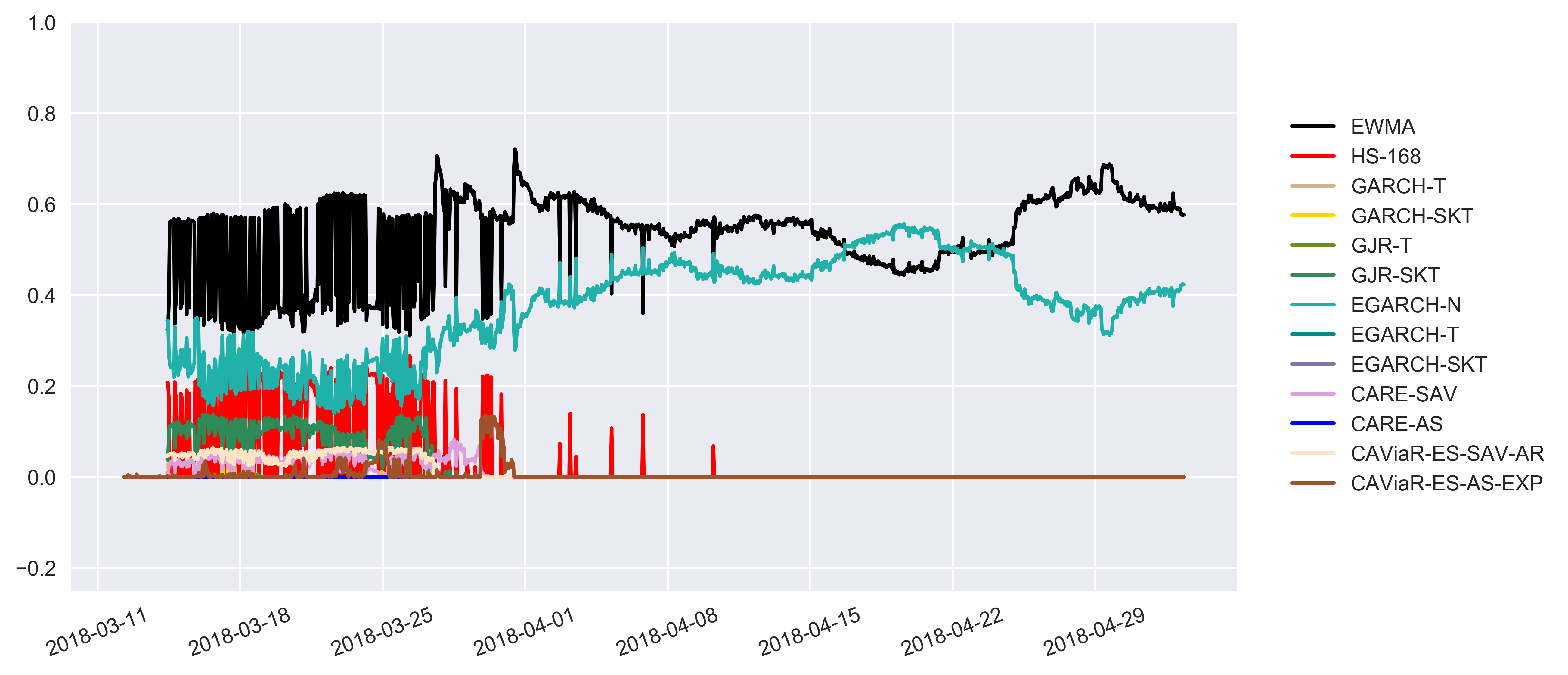}
\label{ETC_ES_95_Weight}
\end{center}
\end{figure}

\subsection*{ETH}
\begin{figure}[H]
\begin{center}
\caption{95\% ES Forecasts for the ETH Market}
\includegraphics[width=\textwidth,height=0.3\textheight]{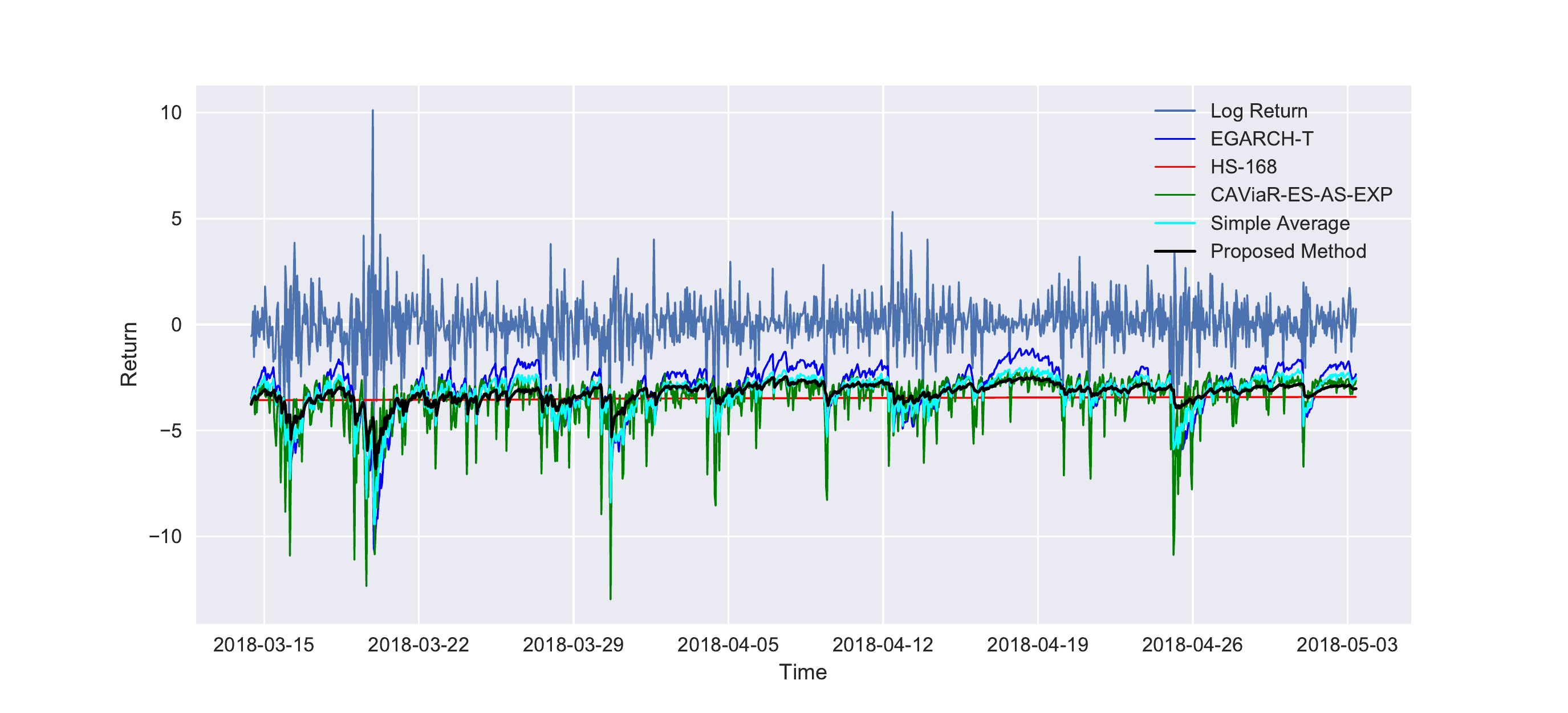}
\label{ETH_ES_95}
\end{center}
\end{figure}

\begin{figure}[H]
\begin{center}
\caption{Individual Weights of the Combined ES Forecasts at the 95\% Confidence Level for the LTC Market}
\includegraphics[width=\textwidth,height=0.3\textheight]{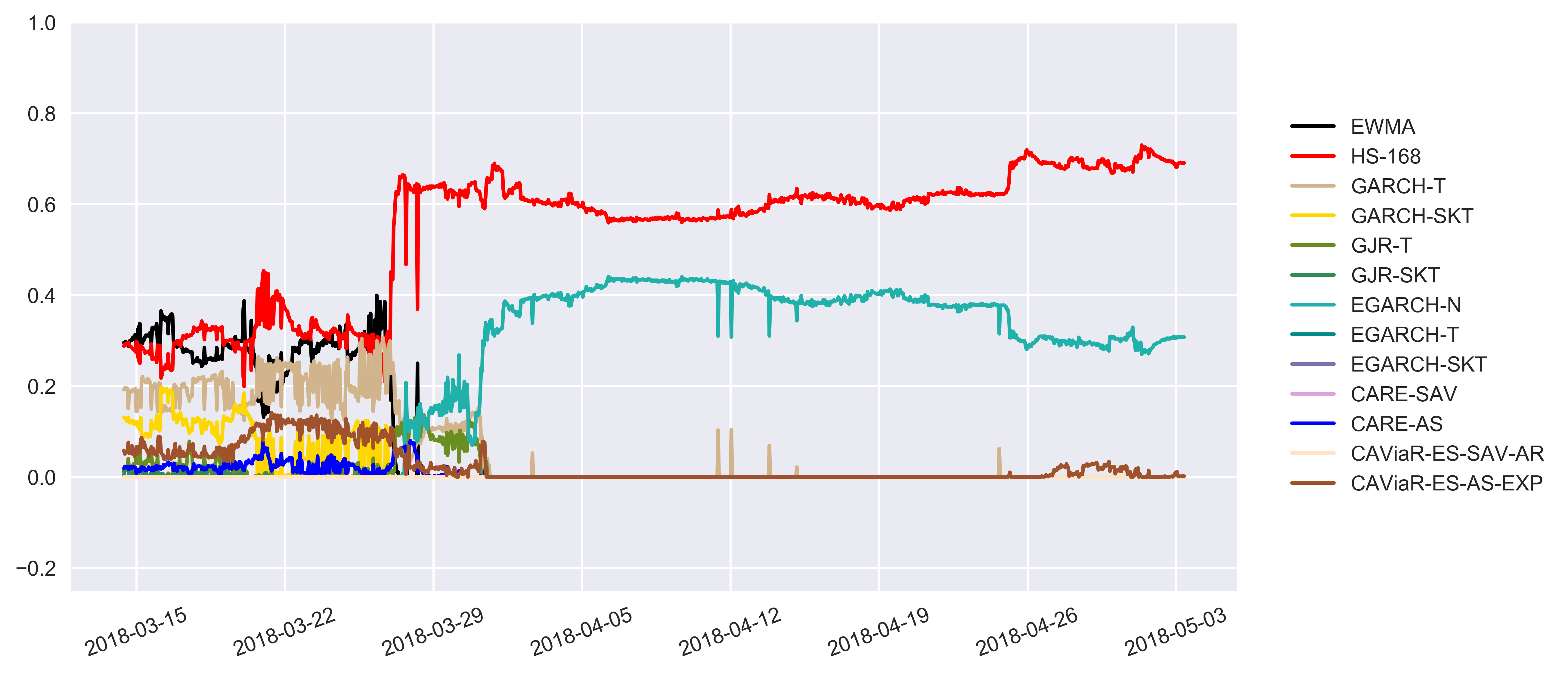}
\label{ETH_ES_95_Weight}
\end{center}
\end{figure}

\subsection*{LTC}
\begin{figure}[H]
\begin{center}
\caption{95\% ES Forecasts for the LTC Market}
\includegraphics[width=\textwidth,height=0.3\textheight]{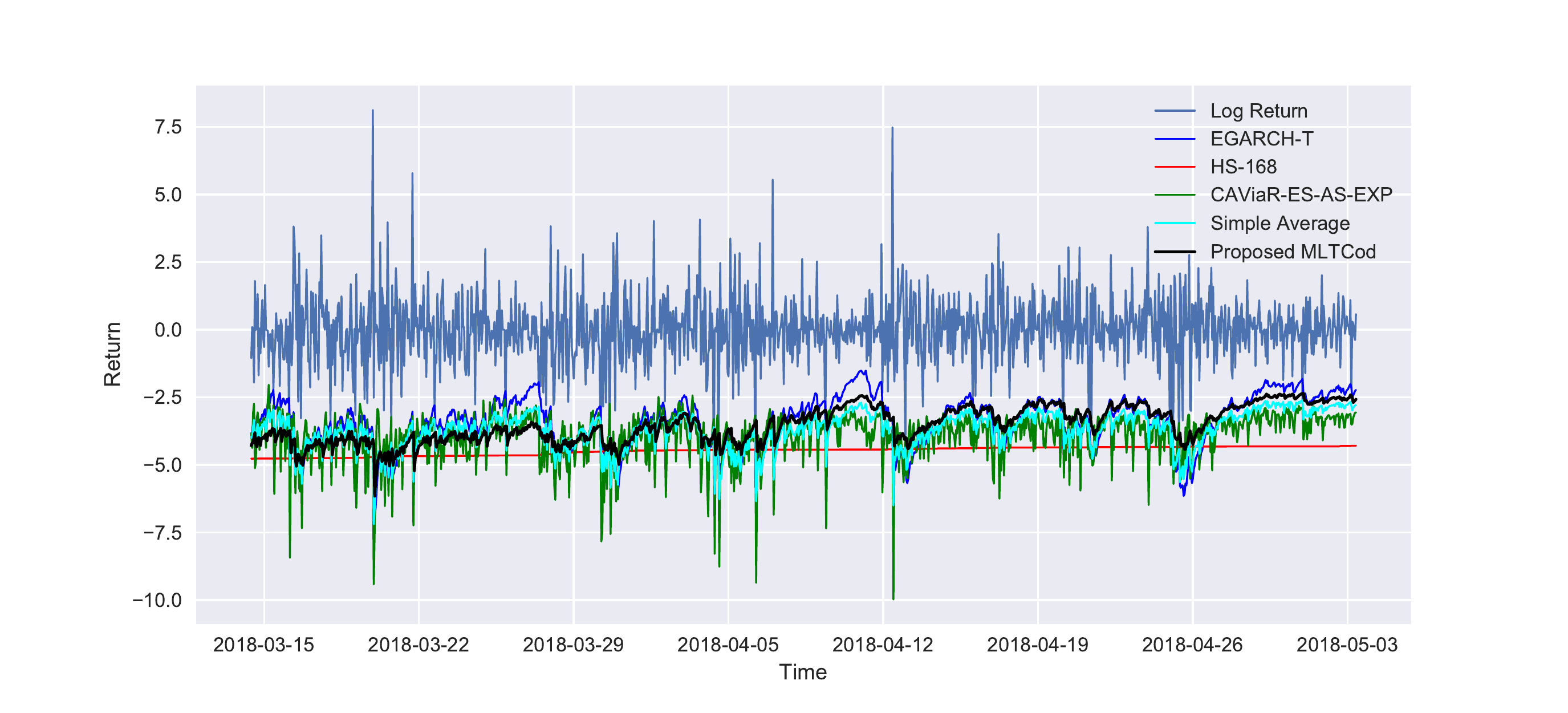}
\label{LTC_ES_95}
\end{center}
\end{figure}

\begin{figure}[H]
\begin{center}
\caption{Individual Weights of the Combined ES Forecasts at the 95\% Confidence Level for the LTC Market}
\includegraphics[width=\textwidth,height=0.3\textheight]{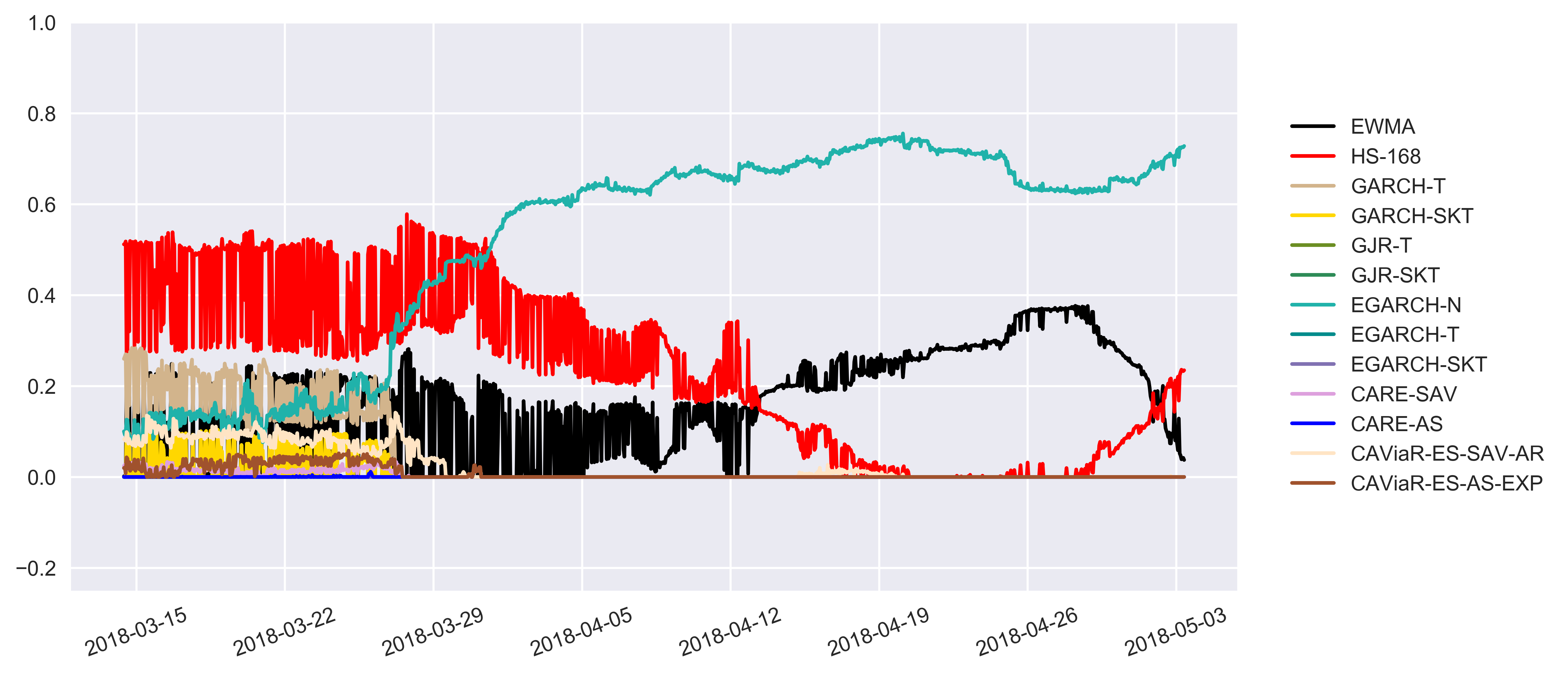}
\label{LTC_ES_95_Weight}
\end{center}
\end{figure}

\section*{99\% ES Forecasts}
\subsection*{BCH}
\begin{figure}[H]
\begin{center}
\caption{99\% ES Forecasts for the BCH Market}
\includegraphics[width=\textwidth,height=0.3\textheight]{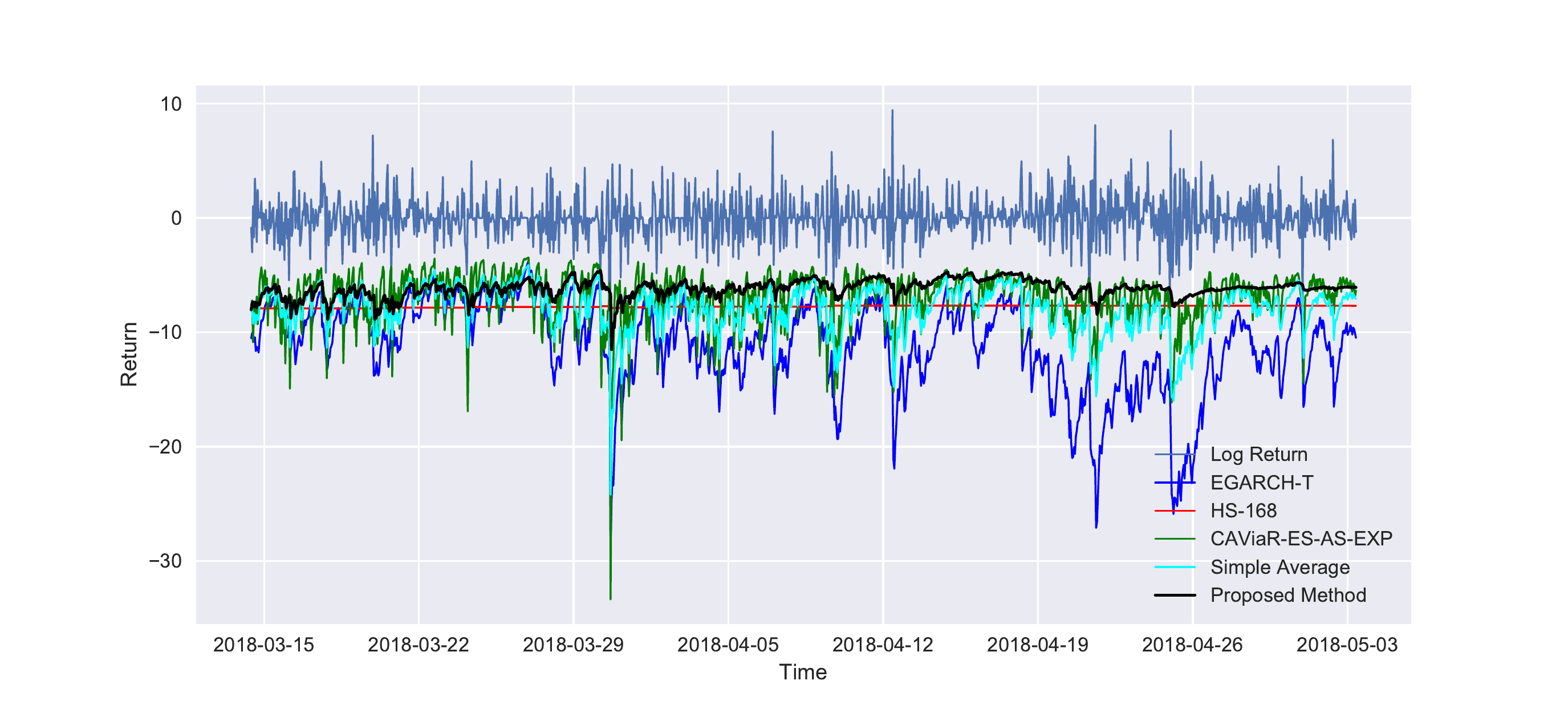}
\label{BCH_ES_99}
\end{center}
\end{figure}

\begin{figure}[H]
\begin{center}
\caption{Individual Weights of the Combined ES Forecasts at the 99\% Confidence Level for the BCH Market}
\includegraphics[width=\textwidth,height=0.3\textheight]{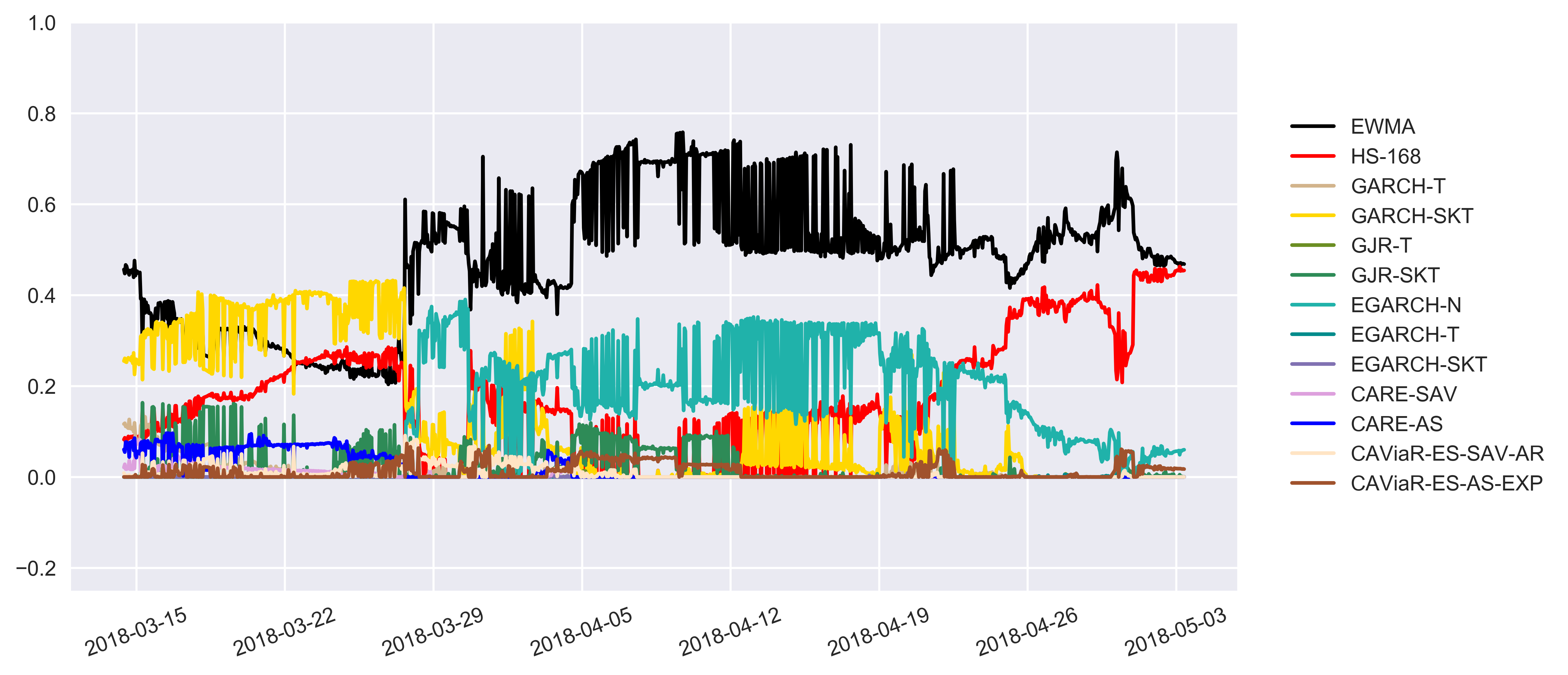}
\label{BCH_ES_99_Weight}
\end{center}
\end{figure}
\subsection*{ETC}
\begin{figure}[H]
\begin{center}
\caption{99\% ES Forecasts for the ETC Market}
\includegraphics[width=\textwidth,height=0.3\textheight]{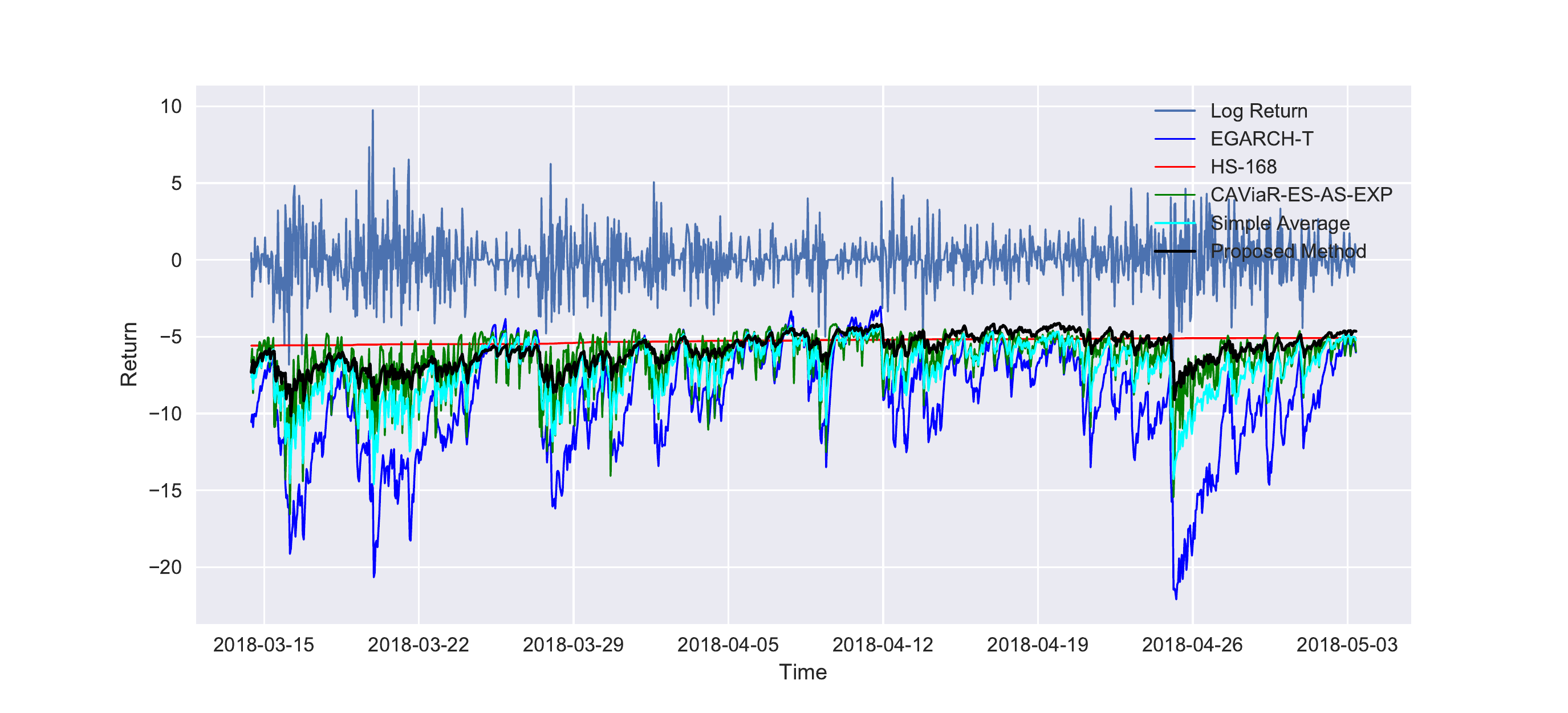}
\label{ETC_ES_99}
\end{center}
\end{figure}

\begin{figure}[H]
\begin{center}
\caption{Individual Weights of the Combined ES Forecasts at the 99\% Confidence Level for the ETC Market}
\includegraphics[width=\textwidth,height=0.3\textheight]{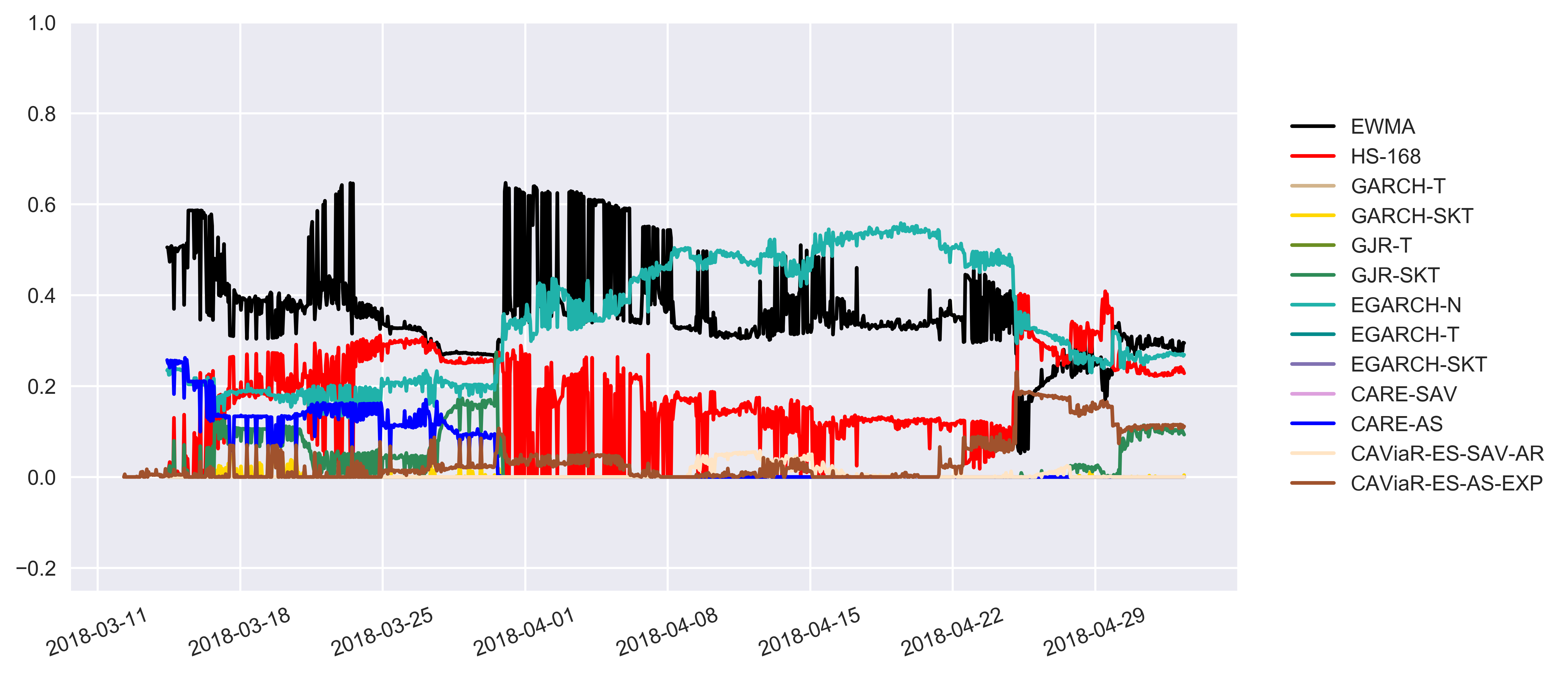}
\label{ETC_ES_99_Weight}
\end{center}
\end{figure}

\subsection*{ETH}

\begin{figure}[H]
\begin{center}
\caption{99\% ES Forecasts for the ETH Market}
\includegraphics[width=\textwidth,height=0.3\textheight]{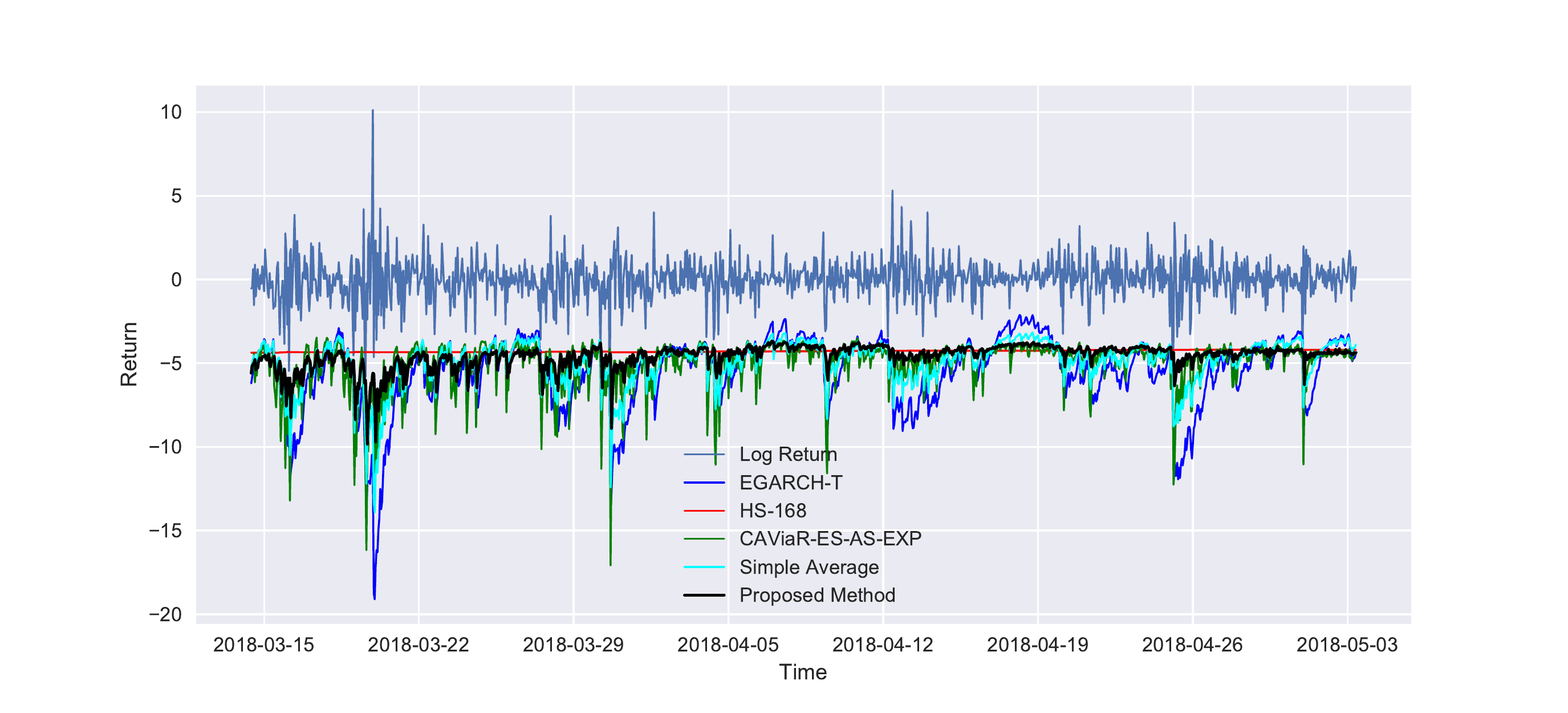}
\label{ETH_ES_99}
\end{center}
\end{figure}

\begin{figure}[H]
\begin{center}
\caption{Individual Weights of the Combined ES Forecasts at the 99\% Confidence Level for the ETH Market}
\includegraphics[width=\textwidth,height=0.3\textheight]{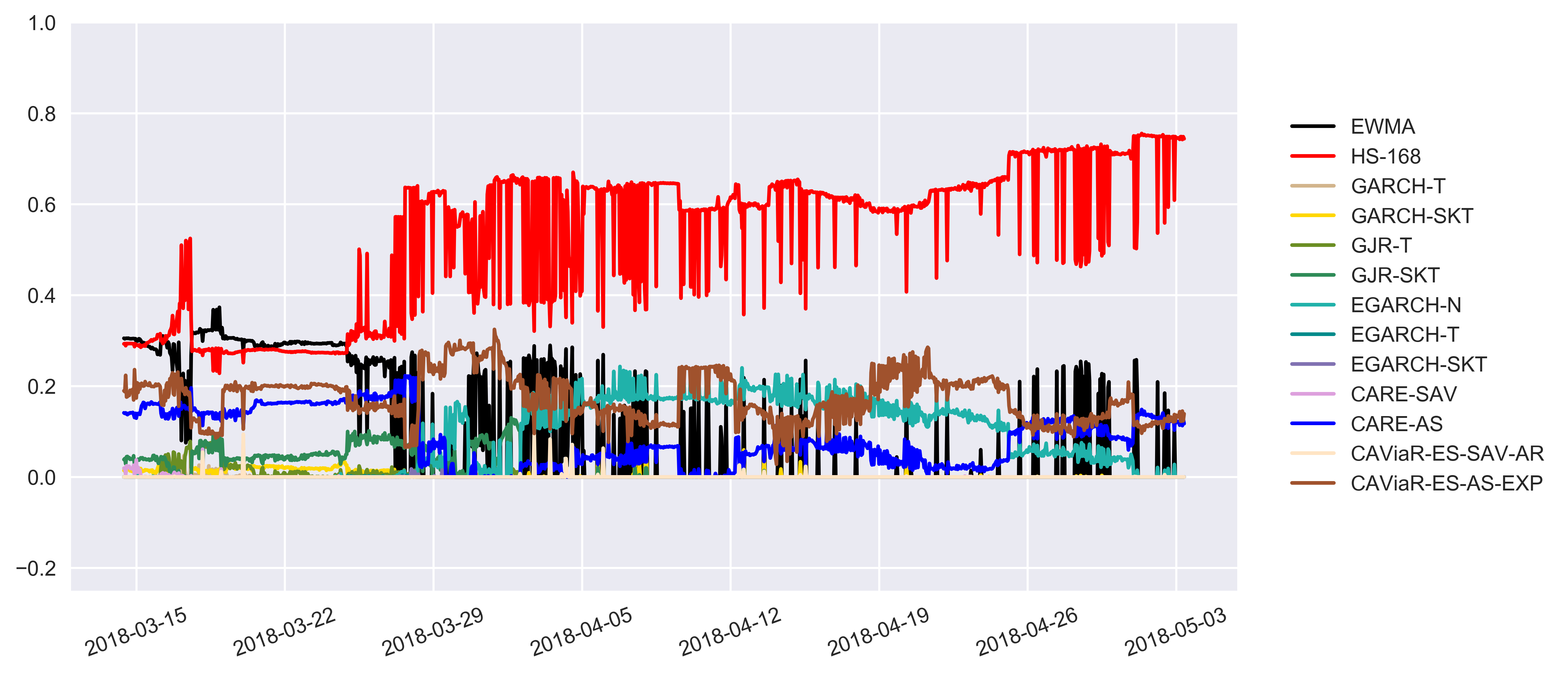}
\label{ETH_ES_99_Weight}
\end{center}
\end{figure}

\subsection*{LTC}
\begin{figure}[H]
\begin{center}
\caption{99\% ES Forecasts for the LTC Market}
\includegraphics[width=\textwidth,height=0.3\textheight]{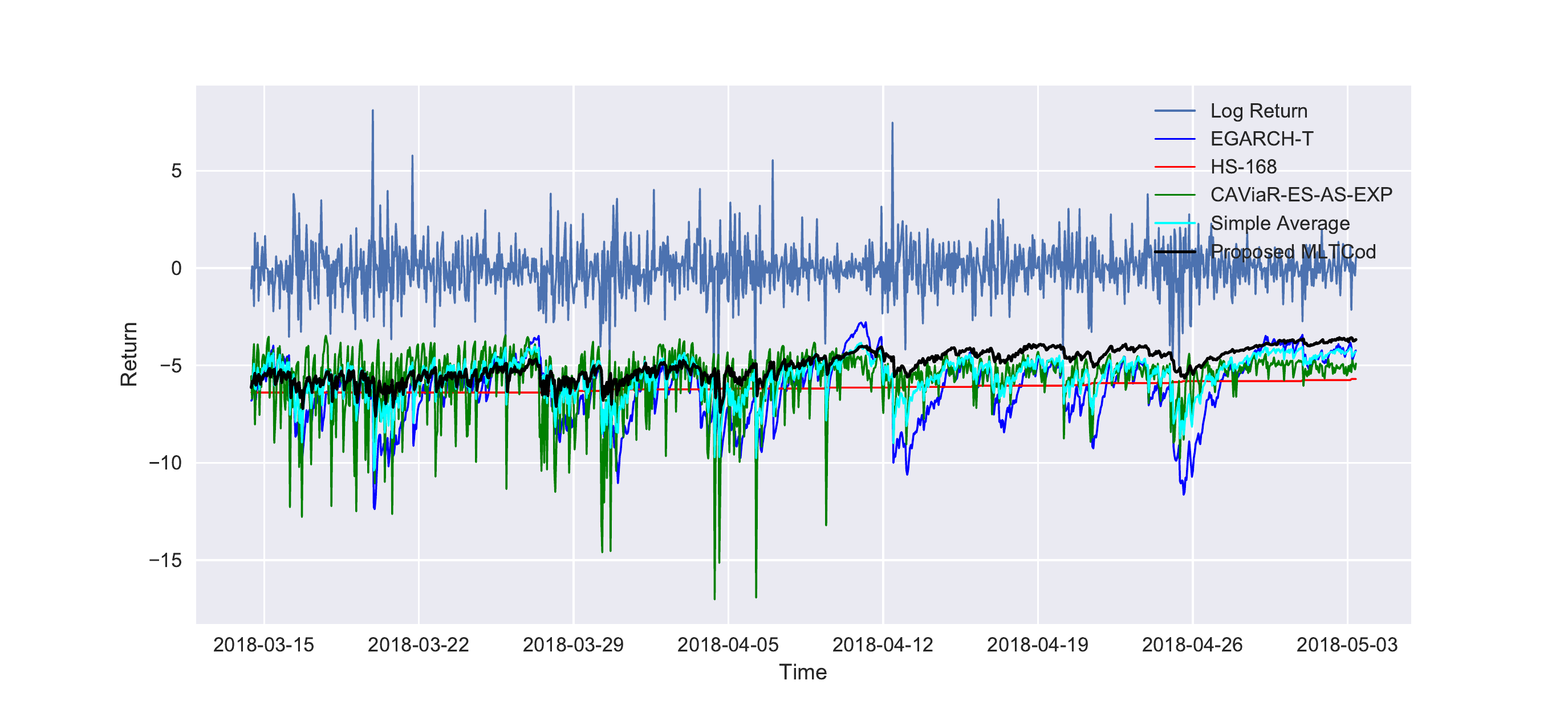}
\label{LTC_ES_99}
\end{center}
\end{figure}

\begin{figure}[H]
\begin{center}
\caption{Individual Weights of the Combined ES Forecasts at the 99\% Confidence Level for the LTC Marlet}
\includegraphics[width=\textwidth,height=0.3\textheight]{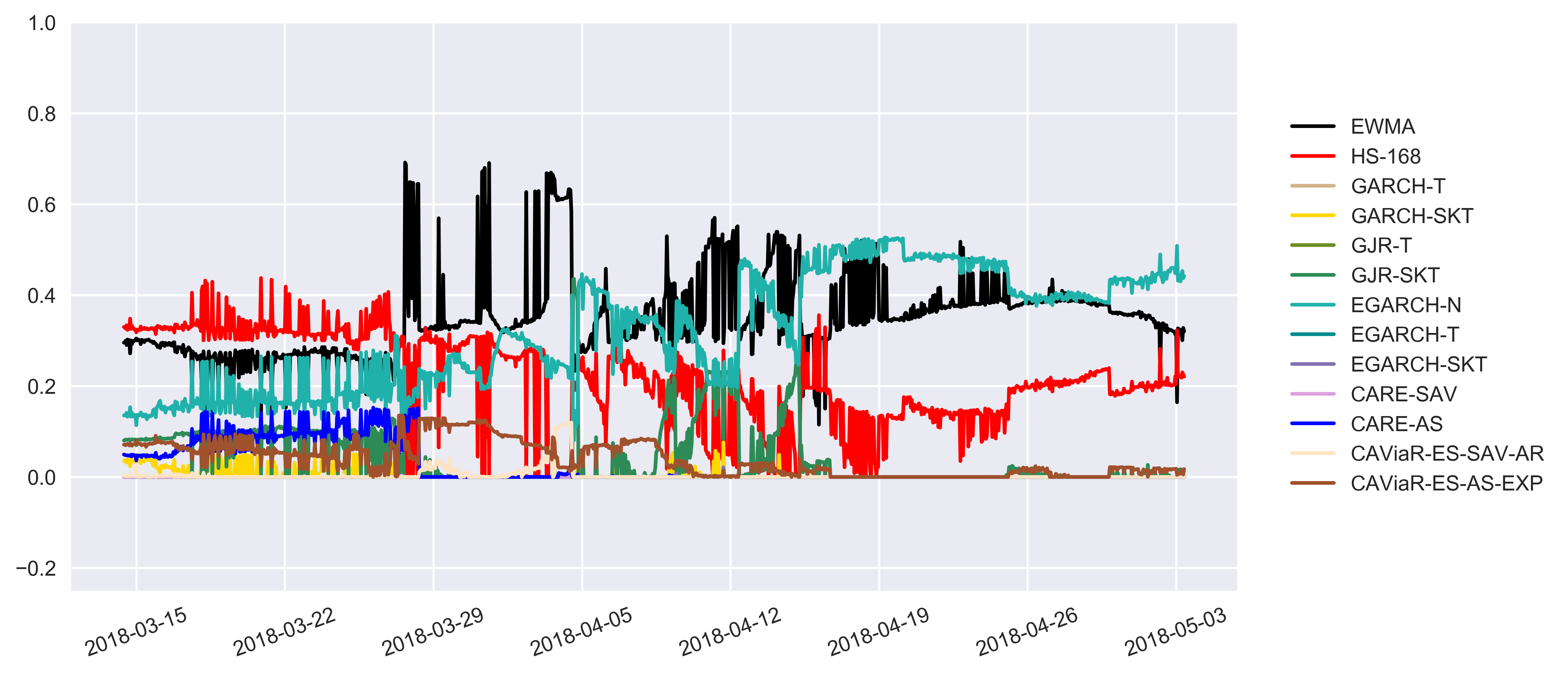}
\label{LTC_ES_99_Weight}
\end{center}
\end{figure}

\end{document}